\documentclass[12pt]{article}
\pdfoutput=1
\usepackage{tikz}
\usepackage{a4wide,epsfig,psfrag,amsmath,amssymb,cite,scalefnt}
\usepackage{color}
\usepackage{amsmath,comment,braket}
\usepackage{placeins}
\usepackage{slashed}

\graphicspath{ {figs/} }

\allowdisplaybreaks

\newcommand{\gsim}{\;\rlap{\lower 3.5 pt \hbox{$\mathchar \sim$}} \raise 1pt
 \hbox {$>$}\;}
\newcommand{\lsim}{\;\rlap{\lower 3.5 pt \hbox{$\mathchar \sim$}} \raise 1pt
 \hbox {$<$}\;}

\begin{document}

\title{\vskip-3cm{\baselineskip14pt
    \begin{flushleft}
      \normalsize FERMILAB-PUB-21-470-T\\
      \normalsize P3H-21-068\\
      \normalsize TTP21-034 \\
      \normalsize TU-1133
  \end{flushleft}}
  \vskip1.5cm
  Real corrections to Higgs boson pair production
  at NNLO in the large top quark mass limit
}

\author{
  Joshua Davies$^{a}$,
  Florian Herren$^{b}$,
  Go Mishima$^{c}$,
  Matthias Steinhauser$^{d}$
  \\[1mm]
  {\small\it $^a$ Department of Physics and Astronomy, University of Sussex,
    Brighton BN1 9QH, UK}
  \\[1mm]
  {\small\it $^b$ Fermi  National  Accelerator  Laboratory, Batavia,  IL,  60510,  USA}
  \\[1mm]
  {\small\it $^c$ Department of Physics, Tohoku University, Sendai, 980-8578 Japan}
  \\[1mm]
  {\small\it $^d$Institut f{\"u}r Theoretische Teilchenphysik, Karlsruhe Institute of Technology (KIT)}\\
  {\small\it Wolfgang-Gaede Stra\ss{}e 1, 76128 Karlsruhe, Germany}
}
  
\date{}

\maketitle

\thispagestyle{empty}

\begin{abstract}

  In this paper we consider the next-to-next-to-leading order total cross
  section of Higgs boson pair production in the large top quark mass limit and
  compute four expansion terms in $1/m_t^2$. Good convergence is observed
  below the top quark threshold, which makes our results a valuable input for
  approximation methods which aim for next-to-next-to-leading order
  corrections over the whole kinematic range.  We present details on various
  steps of our calculation; in particular, we provide results for three- and
  four-particle phase-space master integrals and describe in detail the
  evaluation of the collinear counterterms.

\end{abstract}

\thispagestyle{empty}

\sloppy


\newpage


\section{Introduction}

After the discovery of the Higgs boson~\cite{ATLAS:2012yve,CMS:2012qbp}
one of the major tasks of the Large Hadron Collider (LHC) at CERN, and in
particular the High-Luminosity LHC, is to investigate the scalar sector of the
Standard Model (SM) of particle physics.  In principle, within the SM all
parameters are known: the scalar potential contains two parameters, the Higgs
boson mass ($m_H$) and the vacuum expectation value ($v$). Both are available
to high precision~\cite{ParticleDataGroup:2020ssz} and their combination
determines the triple and quartic Higgs boson couplings via
$\lambda= m_H^2/(2v^2)\approx 0.13$.  However it is important to verify this
value of $\lambda$ (and thus the structure of the scalar sector)
independently, therefore experimental measurements are crucial.

There are two promising approaches which are sensitive to the value of
$\lambda$. Firstly there is double Higgs boson production, where the dominant
production process is via gluon fusion. The coupling $\lambda$ already enters
at leading order (LO), providing a good sensitivity. On the experimental side,
however, a measurement of this cross section with a reasonable precision is
very challenging (see, e.g., Refs.~\cite{ATLAS:2020jgy,CMS:2020tkr}).  A
second process through which information about $\lambda$ can be obtained is
single Higgs boson production.  The LHC is anticipated to be able to measure
this cross section with a precision at the percent level.  In the theory
predictions, however, $\lambda$ only enters via quantum corrections at
next-to-leading order (NLO), reducing the sensitivity to its value in this
process.

In this paper we consider double Higgs boson production at 
next-to-next-to-leading order (NNLO)
in the large-$m_t$ limit. Special emphasis is put on the 
real-radiation corrections which to date are only known
in the infinite top quark mass limit. We compute finite
$1/m_t$ corrections up to order $1/m_t^6$. We describe
technical aspects of our calculation in detail, which might be useful
for other processes.

Leading order corrections to
$gg\to HH$ were first considered more than 30 years
ago~\cite{Glover:1987nx,Plehn:1996wb}.  At NLO the infinite top quark mass
limit is known from~\cite{Dawson:1998py} and finite $1/m_t$ corrections
were considered in~\cite{Grigo:2013rya,Degrassi:2016vss}. At NLO various
further approximations have been constructed. Among them is an expansion in
the high-energy limit~\cite{Davies:2018ood,Davies:2018qvx}, for small
transverse momentum of the Higgs bosons~\cite{Bonciani:2018omm} and around the
top quark threshold~\cite{Grober:2017uho}.  In the latter, the large-$m_t$
expansions of the form factors were combined with information from
the threshold expansion using conformal mapping and Pad\'e approximation; this
produces good approximations of the virtual NLO corrections
up to a di-Higgs invariant mass of about $700$~GeV.

Exact NLO corrections for the real-radiation contribution were
first obtained in Ref.~\cite{Maltoni:2014eza}.
The complete exact NLO results are known from
Refs.~\cite{Borowka:2016ehy,Borowka:2016ypz,Baglio:2018lrj}, where they are
computed using numerical methods. In contrast to
the approximations discussed above, which are in general available in analytic from, these
numerical computations are quite expensive in terms of CPU time.
It is therefore advantageous still to use 
approximations in the region of the phase space where they are valid.
For example in Ref.~\cite{Davies:2019dfy} the exact results from
Refs.~\cite{Borowka:2016ehy,Borowka:2016ypz} were combined with the
high-energy expansion of Refs.~\cite{Davies:2018ood,Davies:2018qvx}. The
CPU-time expensive calculations were only necessary for relatively small
values of the Higgs transverse momentum, say below
$p_T\approx 200$~GeV, and the fast evaluation of the analytic
high-energy expansions could be used for the remaining phase space.

The dependence on the renormalization scheme of the top quark mass has been
discussed in Ref.~\cite{Baglio:2020wgt}. Sizeable uncertainties were
observed for large values of the di-Higgs invariant mass.  To reduce this
uncertainty a NNLO calculation is necessary.  A full NNLO calculation is
currently out of reach, so it is important to construct approximations
valid in different region of the phase space.  In the region where the
large-$m_t$ expansion---the topic of this paper---is valid the scheme
uncertainty is small. It is nevertheless useful to have such corrections at
hand since they serve as valuable input for approximation methods such as, e.g., the one
described in Ref.~\cite{Grober:2017uho}.

At NNLO, only approximations in the large-$m_t$ limit are
available so far. In the infinite-$m_t$ limit the cross section has been computed in
Refs.~\cite{deFlorian:2013jea,deFlorian:2013uza,Grigo:2014jma}
and finite $1/m_t$ expansion terms for the virtual corrections are
available from~\cite{Grigo:2015dia,Davies:2019xzc}.
In Ref.~\cite{Grazzini:2018bsd} a NNLO approximation has been constructed
which incorporates, in addition to the exact NLO corrections and the
infinite-$m_t$ results at NNLO, the exact expression for the double-real
radiation contributions at NNLO.

Even N$^3$LO corrections are available for $gg\to HH$,
although only in the infinite top quark mass limit. The results presented in
Refs.~\cite{Chen:2019lzz,Chen:2019fhs} are based on ingredients computed
in Refs.~\cite{Anastasiou:2016cez,Mistlberger:2018etf,Spira:2016zna,Gerlach:2018hen,Banerjee:2018lfq}.

In the present work we consider the $1/m_t$ corrections at
NNLO.  The virtual corrections to the form factors have been computed in
\cite{Davies:2019xzc} including terms to order $1/m_t^8$. In
Section~\ref{sec::virt} we describe our procedure to obtain the corresponding contribution
to the cross section to this order.  The main purpose of this paper is to
complement the virtual corrections by the corresponding real-radiation
contributions.  We compute all contributions up to order $1/m_t^6$ and the
channels which involve quarks to $1/m_t^8$.

\begin{figure}[t]
  \begin{center}
    \begin{tabular}{cccccc}
        \hspace{-2mm}\includegraphics[width=0.15\textwidth]{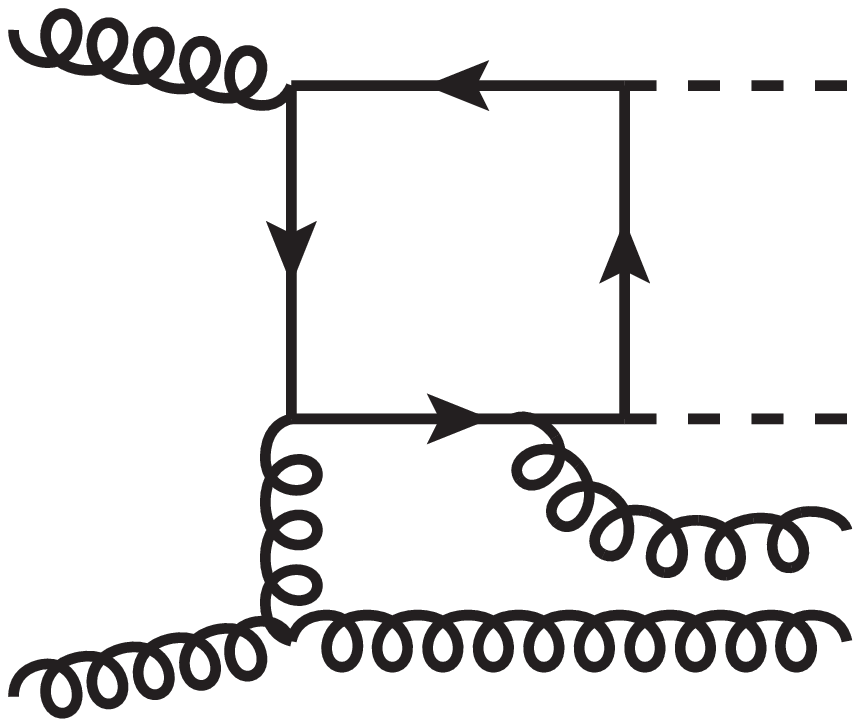}
      &
        \hspace{-2mm}\includegraphics[width=0.15\textwidth]{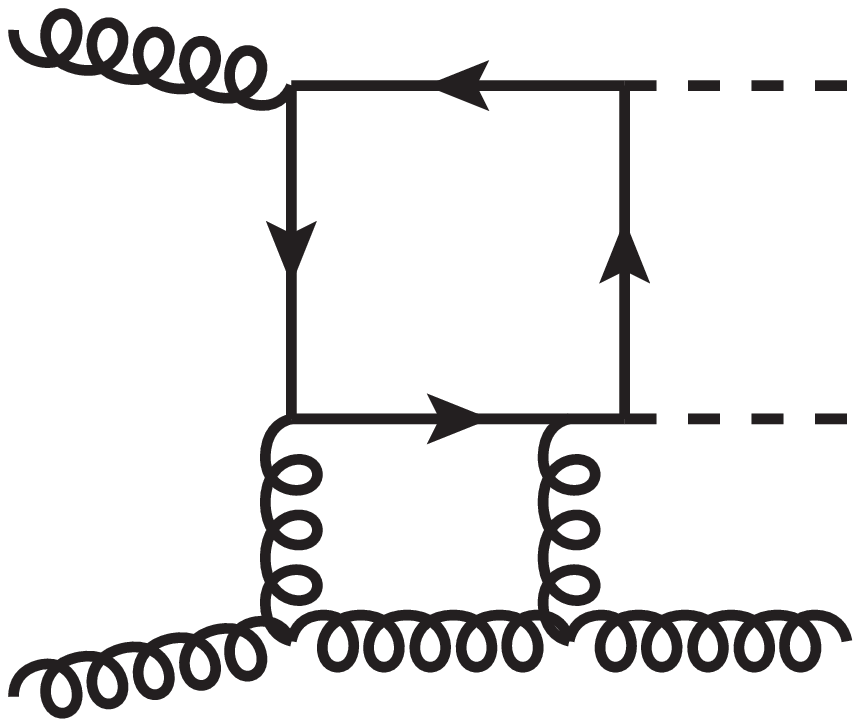}
      &
        \hspace{-2mm}\includegraphics[width=0.15\textwidth]{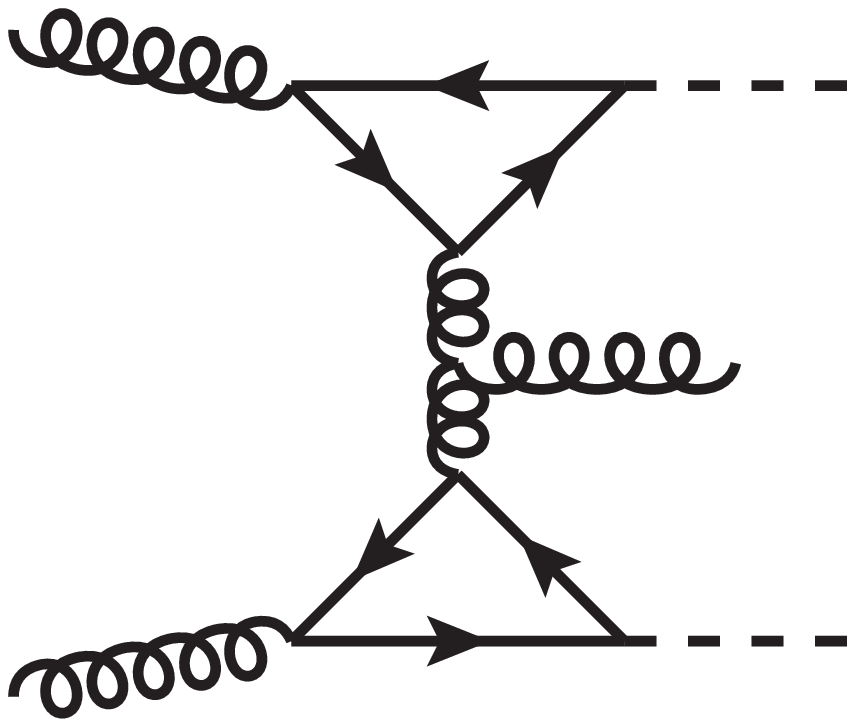}
      &
        \hspace{-2mm}\includegraphics[width=0.15\textwidth]{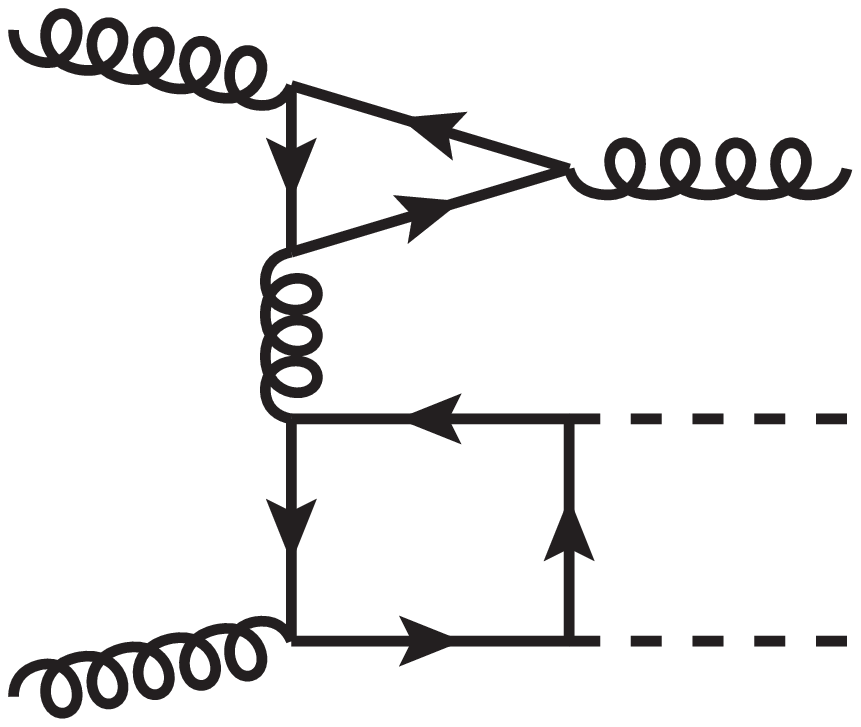}
      & 
        \hspace{-2mm}\includegraphics[width=0.15\textwidth]{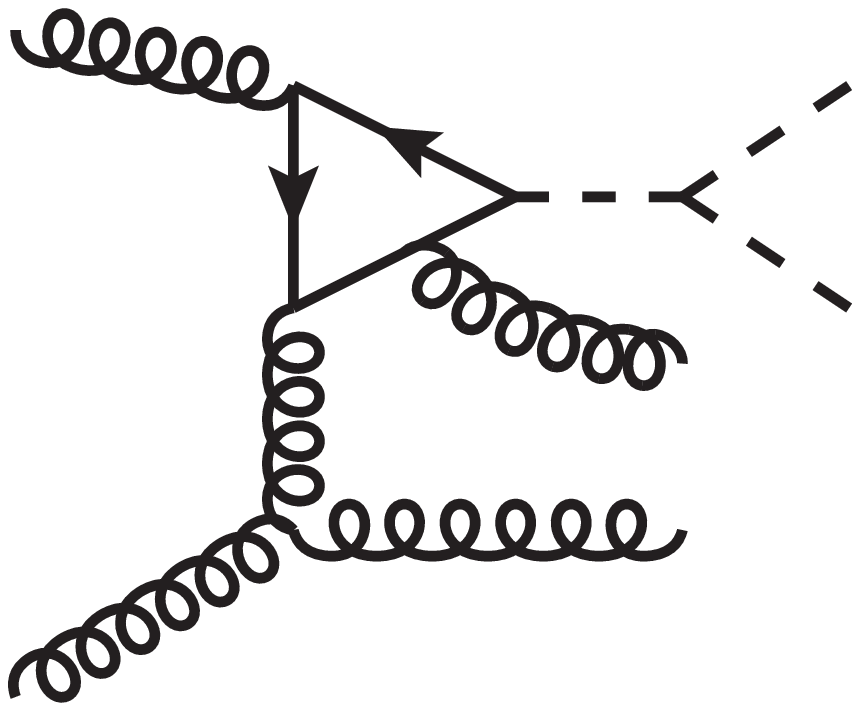}
      &
        \hspace{-2mm}\includegraphics[width=0.15\textwidth]{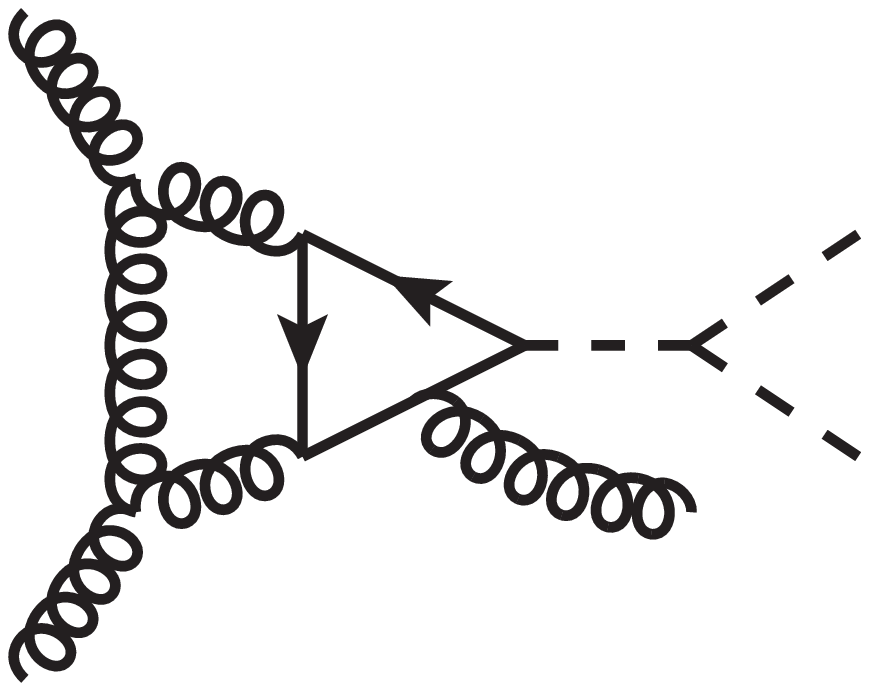}
        \\
      (a) & (b) & (c) & (d) & (e) & (f)
    \end{tabular}
    \caption{\label{fig::gghh}Sample Feynman diagrams contributing to the
      real radiation. Contributions such as those shown in
      (c) lead to $n_h^3$ contributions which have already been
      computed in Ref.~\cite{Davies:2019xzc}. 
      The $n_h^3$ contributions of (d) contain a top quark loop
      without a Higgs coupling and have not been computed in
      Ref.~\cite{Davies:2019xzc}; they are considered here.}
  \end{center}
\end{figure}

\begin{figure}[t]
  \begin{center}
    \begin{tabular}{cccc}
        \hspace{-2mm}\includegraphics[width=0.24\textwidth]{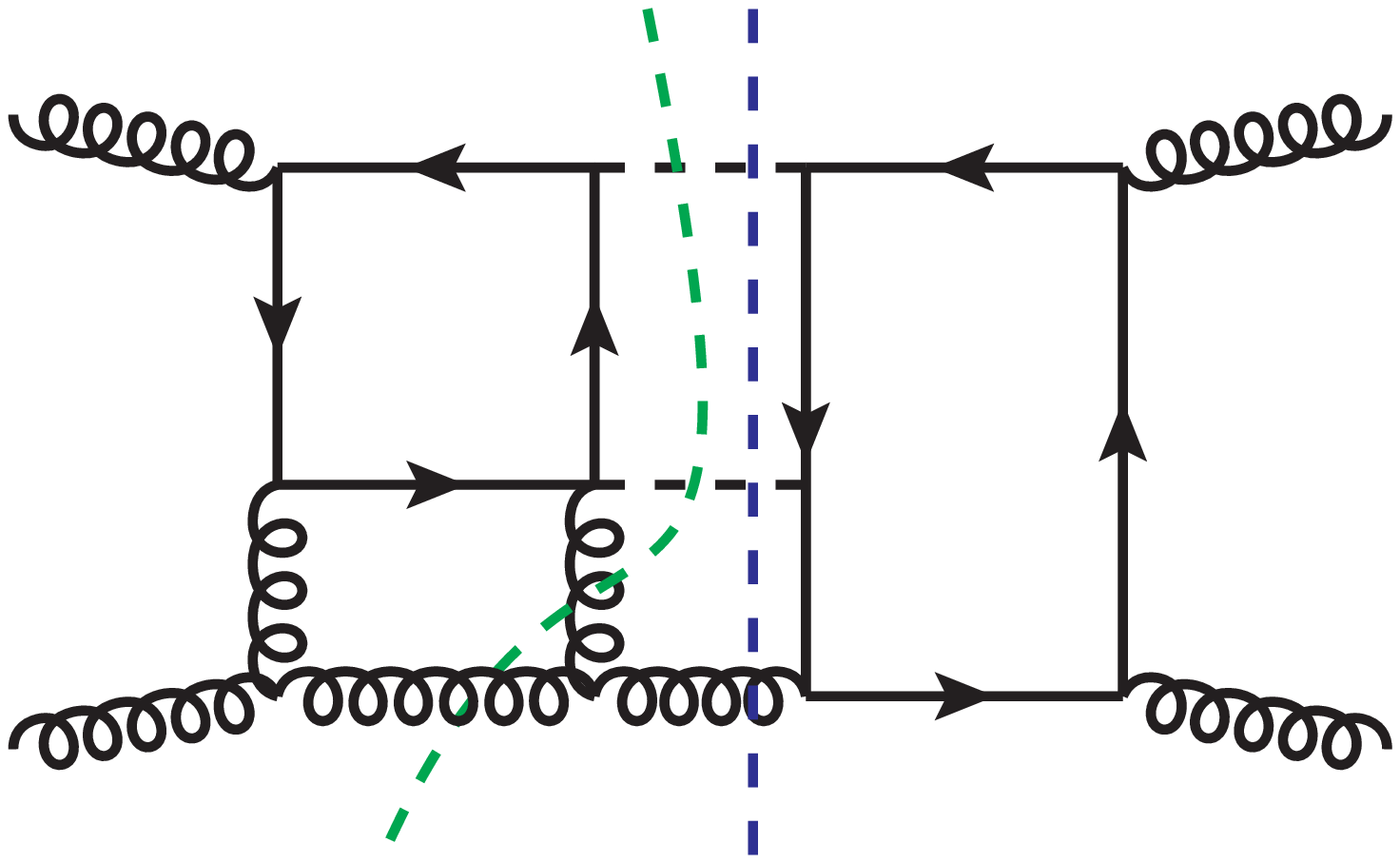}
      &
        \hspace{-2mm}\includegraphics[width=0.24\textwidth]{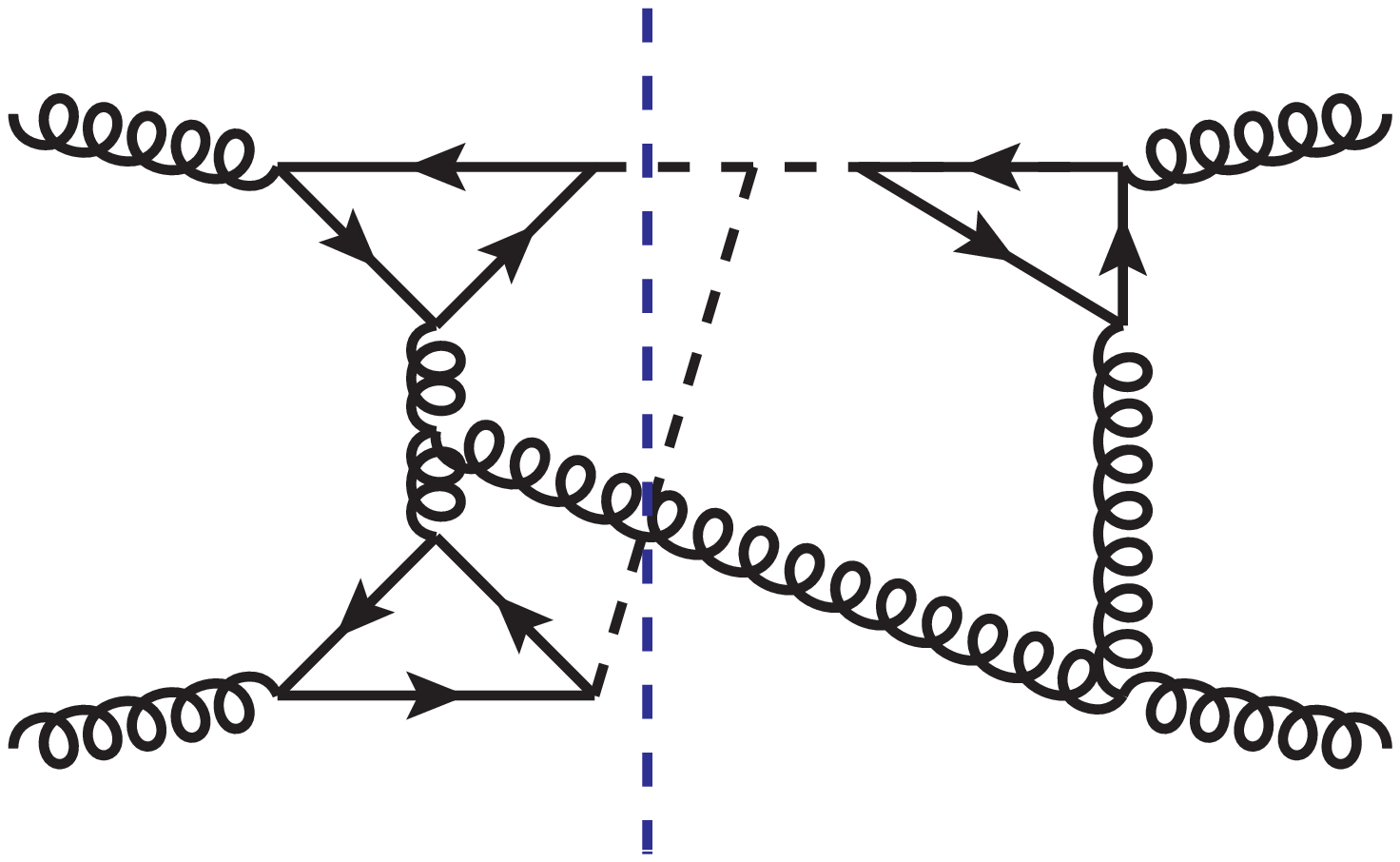}
      &
        \hspace{-2mm}\includegraphics[width=0.24\textwidth]{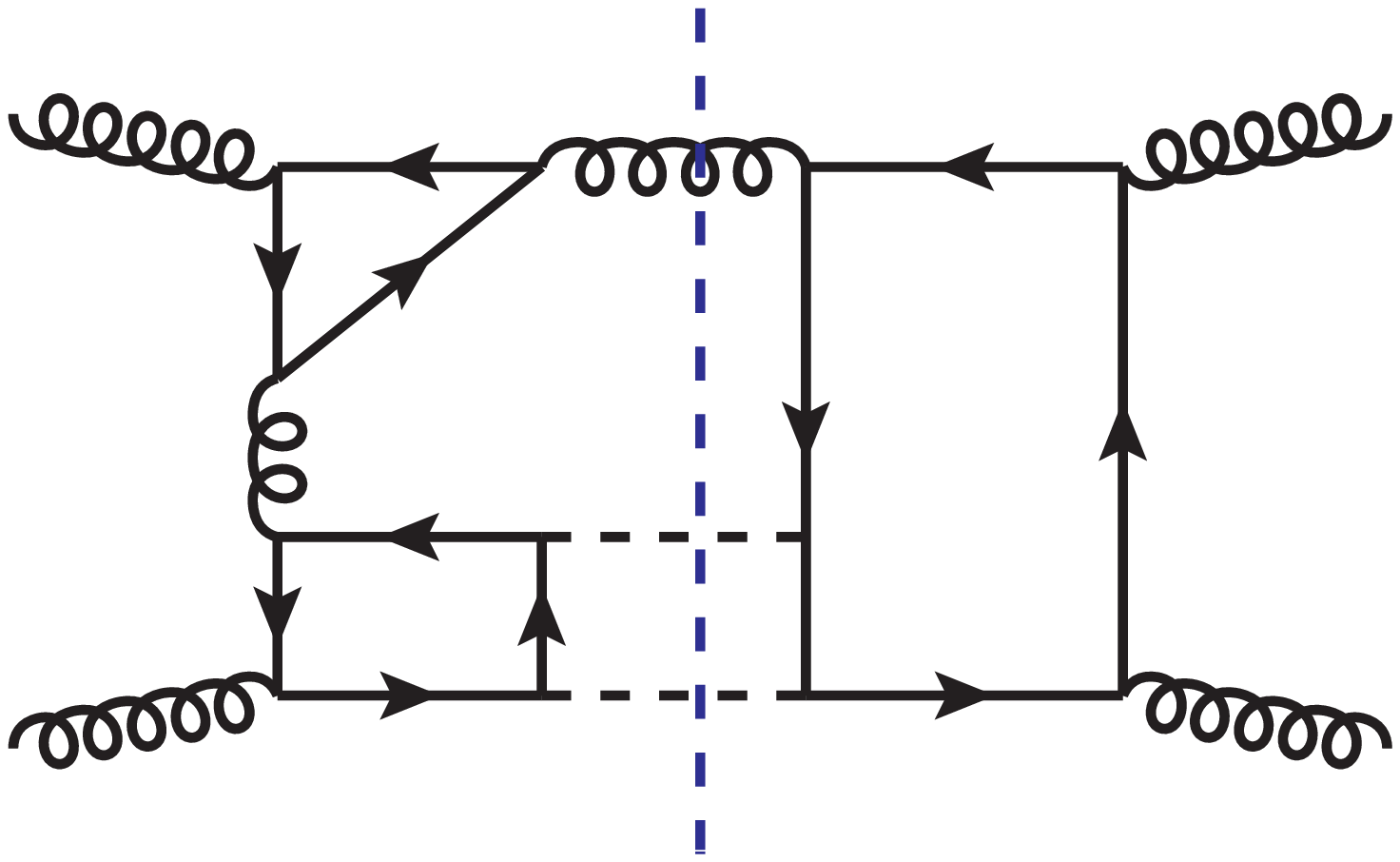}
      &
        \hspace{-2mm}\includegraphics[width=0.24\textwidth]{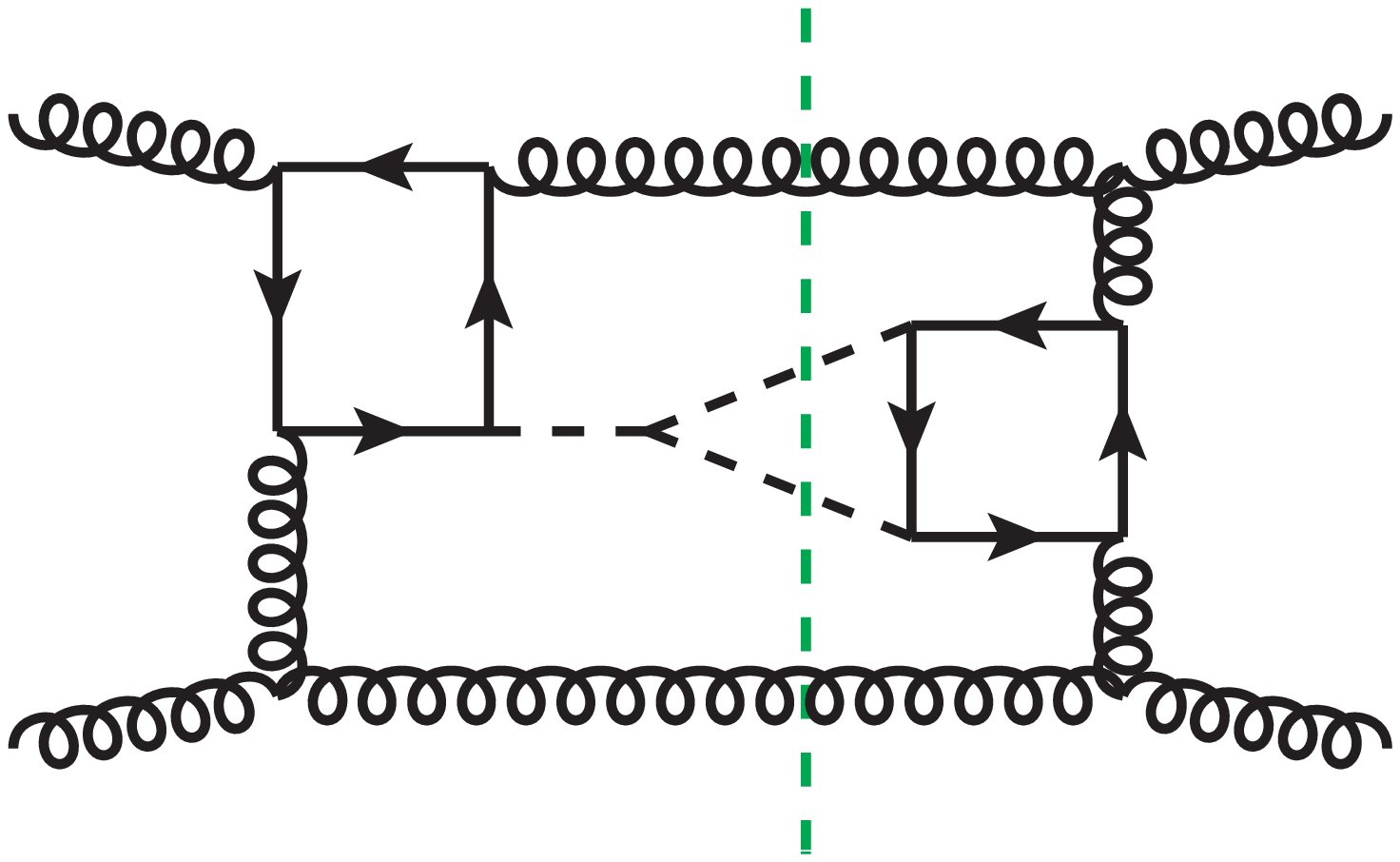}
        \\
      (a) & (b) & (c) & (d)
    \end{tabular}
    \caption{\label{fig::gghhgg}Sample Feynman diagrams in the forward-scattering
      kinematics. Three- and four-particle cuts are shown by blue
      and green dashed lines, respectively. The $n_h^3$ contributions as shown in (b) have
      already been considered in~\cite{Davies:2019xzc} but those in (c) have not; they
      are considered here.}
  \end{center}
\end{figure}

The real-radiation contribution can be divided into one-loop $2\to4$
processes and two-loop $2\to3$ processes; sample Feynman diagrams are shown in
Fig.~\ref{fig::gghh}. In the following we denote these contributions as
``real-real'' and ``real-virtual'', respectively.  We are interested in the
total cross section which is conveniently obtained using the optical
theorem applied to the forward-scattering amplitude (see Fig.~\ref{fig::gghhgg}). 
At NNLO this leads to
five-loop diagrams. However, in the large-$m_t$ limit one observes a
factorization into $n$-loop tadpole and $(5-n)$-loop so-called {\it
  phase-space integrals}, where in our case we have $n=2$ or $n=3$.
The tadpole integrals are well studied in the literature, however the phase-space
integrals are not;  we discuss in detail the various integral families, their
reduction to master integrals and provide analytic results.

We want to mention that a part of the real-radiation contribution, where the two Higgs
bosons couple to different top quark loops (cf. Fig.~\ref{fig::gghh}(c)) have
been computed in Ref.~\cite{Davies:2019xzc}. In the forward-scattering picture of
Fig.~\ref{fig::gghhgg} such contributions have three closed top quark loops
and we denote them the ``$n_h^3$ contribution''. Note that there are further
$n_h^3$ contributions which have a closed loop with only gluon couplings (as shown in
Fig.~\ref{fig::gghh}(c)). Such terms are not included in Ref.~\cite{Davies:2019xzc},
but are computed in this paper.

The remainder of the paper is organized as follows: in the next section we
discuss the individual parts of our calculation. This concerns in particular
the setup used for the computation of the real-radiation corrections including
the asymptotic expansion and the reduction to phase-space master
integrals. Furthermore, we discuss the ultraviolet and
collinear counterterms to subtract the divergences from initial-state
radiation. Section~\ref{sec::masters} is dedicated to the phase-space master
integrals. We provide details on the transformation of the system of
differential equations to $\epsilon$ form and on the computation of the
boundary conditions in the soft limit. We discuss our analytic and numerical
results in Section~\ref{sec::res} and summarize our findings in
Section~\ref{sec::conclusion}. In the appendix we provide useful additional
material such as explicit formulae used for the computation of the collinear
\mbox{counterterms}, the integrands of the phase-space master integrals, NNLO virtual
corrections to the channel $q\bar{q}\to HH$ and NNLO virtual corrections
involving four closed top quark loops.  Furthermore, we describe in detail our
approach to obtain the leading $1/m_t$ term for double Higgs production from
the analytic expressions of the single-Higgs production cross section.


\section{\label{sec:tech}Computation}


\subsection{Partonic channels and generation of amplitudes}
\label{sec::sub::ampgen}

We write the perturbative expansion of the partonic cross
sections as
\begin{eqnarray}
  \sigma_{ij\to HH + X}(s,\rho) 
  &=& 
      \delta_{ig} \delta_{jg} \sigma_{gg}^{(0)} 
      + \sigma_{ij}^{(1)} 
      + \sigma_{ij}^{(2)} \ldots \,.
      \label{eq::sig}
\end{eqnarray}
where the superscripts (0), (1) and (2)
stand for the LO, NLO and NNLO contributions,
$s$ is the centre-of-mass energy and $\rho$ is defined
in Eq.~(\ref{eq::def_rho}).
At NLO and NNLO we further subdivide the
cross sections according to the number of
closed top quark loops and write
\begin{eqnarray}
  \sigma_{ij}^{(1)} &=& \sigma_{ij}^{(1),n_h^2} 
                        + \delta_{ig} \delta_{jg} \sigma_{gg}^{(1),n_h^3}
                        \,,
                        \nonumber\\
  \sigma_{ij}^{(2)} &=& \sigma_{ij}^{(2),n_h^2}
                        + \sigma_{ij}^{(2),n_h^3}
                        + \delta_{ig} \delta_{jg} \sigma_{gg}^{(2),n_h^4}
                        \,.
                        \label{eq::sig_nh2_nh3}
\end{eqnarray}
Note that the superscripts $n_h^k$ counts the 
number of closed top quark loops which have at least one
coupling to a Higgs boson.
The contributions $\sigma_{gg}^{(1),n_h^3}$ and $\sigma_{ij}^{(2),n_h^3}$
have been computed in Ref.~\cite{Davies:2019xzc}.
In this paper we concentrate on $\sigma_{ij}^{(2),n_h^2}$ which can be decomposed as
\begin{eqnarray}
  \sigma_{ij}^{(2),n_h^2} 
  &=&
      \sigma_{ij,\rm virt}^{(2),n_h^2}
      + \sigma_{ij,\rm real}^{(2),n_h^2}
      + \sigma_{ij,\rm coll}^{(2),n_h^2}
      \,,
      \label{eq::sig_tot}
\end{eqnarray}
where the individual contributions are discussed in the remainder of this Section.
Most emphasis is put on $\sigma_{ij,\rm real}^{(2),n_h^2}$ since
the complicated calculation for $\sigma_{ij,\rm virt}^{(2),n_h^2}$ was
performed in~\cite{Davies:2019xzc} and for $\sigma_{ij,\rm coll}^{(2),n_h^2}$
``only'' convolutions of lower-order cross sections and splitting functions
have to be computed. $\sigma_{gg}^{(2),n_h^4}$ is discussed in Appendix~\ref{app::nh4}.

We compute the cross section in the large-$m_t$ limit. It is thus
convenient to introduce the variable
\begin{eqnarray}
  \rho &=& \frac{m_H^2}{m_t^2}\,,
  \label{eq::def_rho}
\end{eqnarray}
such that after the expansion in $1/m_t$ the coefficients
of $\rho^n$ depend on $x=m_H^2/s$. 
The analytic calculation of the phase-space integrals
is performed in an expansion around the production threshold
of the two Higgs bosons, for which it is convenient to define
\begin{eqnarray}
  \delta &=& 1-4x\,,
\end{eqnarray}
such that the threshold corresponds to $\delta=0$.

To compute $\sigma_{ij,\rm real}^{(2),n_h^2}$ we first
generate the forward-scattering amplitudes for the relevant
partonic channels, applying the optical theorem to obtain the real-radiation
contributions.  These amplitudes involve two external momenta $q_1$ and $q_2$,
with $s=(q_1+q_2)^2$.  One has to consider the $gg$, $gq$, $q\bar{q}$, $qq$
and $q q^\prime$ partonic channels, where $q$ and $q^\prime$ denote two
different light quark flavours.  Note that both quark and anti-quark
contributions have to be taken into account, i.e., $gq$ and $qq$
also stand for the $g\bar{q}$ and $\bar{q}\bar{q}$ contributions,
respectively. Similarly in $q q^\prime$, all quark-quark, quark--anti-quark and
anti-quark--anti-quark contributions have to be considered, where in each case
the quark flavours are different. For all channels one obtains
the same result after replacing a quark by an anti-quark and vice versa.
We also compute channels with one or both external gluons replaced
with ghosts ($c$) and anti-ghosts ($\bar{c}$), i.e., in addition to 
the channels listed above we have $gc$, $g\bar{c}$, 
$cc$,
$c\bar{c}$,
$\bar{c}\bar{c}$,
$cq$
and $\bar{c}q$.
This allows us to sum over the gluon polarizations according to
\begin{eqnarray*}
  \sum_{\lambda} \varepsilon^{(\lambda),*}_\mu(q_1)
  \varepsilon^{(\lambda)}_\nu(q_2) 
  &=& -g_{\mu\nu}
  \,,
\end{eqnarray*}
where $\lambda$ runs from 0 to 3.

We generate the amplitudes using \texttt{qgraf}~\cite{Nogueira:1991ex},
however this program does not allow one to filter only diagrams which admit a
cut through the required final-state particles (here two Higgs bosons and one
or more gluons or light fermion pairs), as depicted in Fig.~\ref{fig::gghhgg}
by the green and blue dashed lines.  We thus generate all possible
forward-scattering diagrams and post-process them using
\texttt{gen}~\cite{Pak_gen} which is able to filter the \texttt{qgraf} output
based on our required cuts.

\begin{table}
  \begin{center}
    \begin{tabular}{|c|c|c|c|}
      \hline 
      Channel & \texttt{qgraf} diagrams 
      & \texttt{gen}-filtered diagrams 
      & building block diagrams \\  
      \hline 
      $gg$          & 16,631,778 & 160,154 & 4,612 \\ 
      $gc$          & 1,671,006  & 5,426   & 336 \\
      $c\bar{c}$    & 406,662    & 3,879   & 243 \\
      $cc$          & (not considered)  & (not considered) & 8 \\
      $gq$          & 1,671,006  & 5,426   & 336 \\
      $cq$          & (not considered)   & (not considered) & 8 \\
      $q\bar{q}$    & 406,662    & 3,879   & 243 \\
      $qq$          & (not considered)     & (not considered) & 8 \\
      $qq^\prime$   & 203,331    & 34      & 4 \\
      \hline 
    \end{tabular}
    \caption{The number of ``full'' diagrams generated by \texttt{qgraf} and
      after filtering with \texttt{gen}, compared to the number of ``building
      block'' diagrams, for each channel. For the $cc$, $cq$ and $qq$ channels
      we have only applied the ``building block'' approach.}
    \label{tab::channel-dias}
  \end{center}
\end{table}

Due to the large \texttt{qgraf} output (for the $gg$ channel, 16.6 million
diagrams, 42~GB in total) it proved necessary to separate the diagrams into
subsets containing particular numbers of top quark, light quark, cut- and
un-cut Higgs lines which could be filtered by \texttt{gen} separately, in
parallel. After filtering, the number of diagrams remaining is greatly reduced
(for the $gg$ channel, to a total of 160 thousand diagrams, 416~MB in total). In
Table~\ref{tab::channel-dias} we show the numbers of diagrams for all 
channels which we compute.


\subsection{Real radiation: asymptotic expansion and building blocks}

After producing the filtered set of Feynman diagrams, we consider the
large-$m_t$ limit and apply an asymptotic expansion for $m_t^2 \gg s,m_H^2$.
This expresses each diagram as a sum of products of so-called ``hard subgraphs''
(which have to be expanded in their small quantities: masses and momenta) and
``co-subgraphs'' which are obtained from the original diagrams by shrinking
the subgraph lines to a point. The rules which must be applied to determine
all of the relevant subgraphs of a diagram are implemented in the software
\texttt{exp}~\cite{Harlander:1997zb,Seidensticker:1999bb}, which produces
\texttt{FORM}~\cite{Ruijl:2017dtg} code to perform the expansion.

\begin{figure}[t]
  \begin{center}
    \begin{tabular}{cccc}
        \hspace{-2mm}\includegraphics[width=0.24\textwidth]{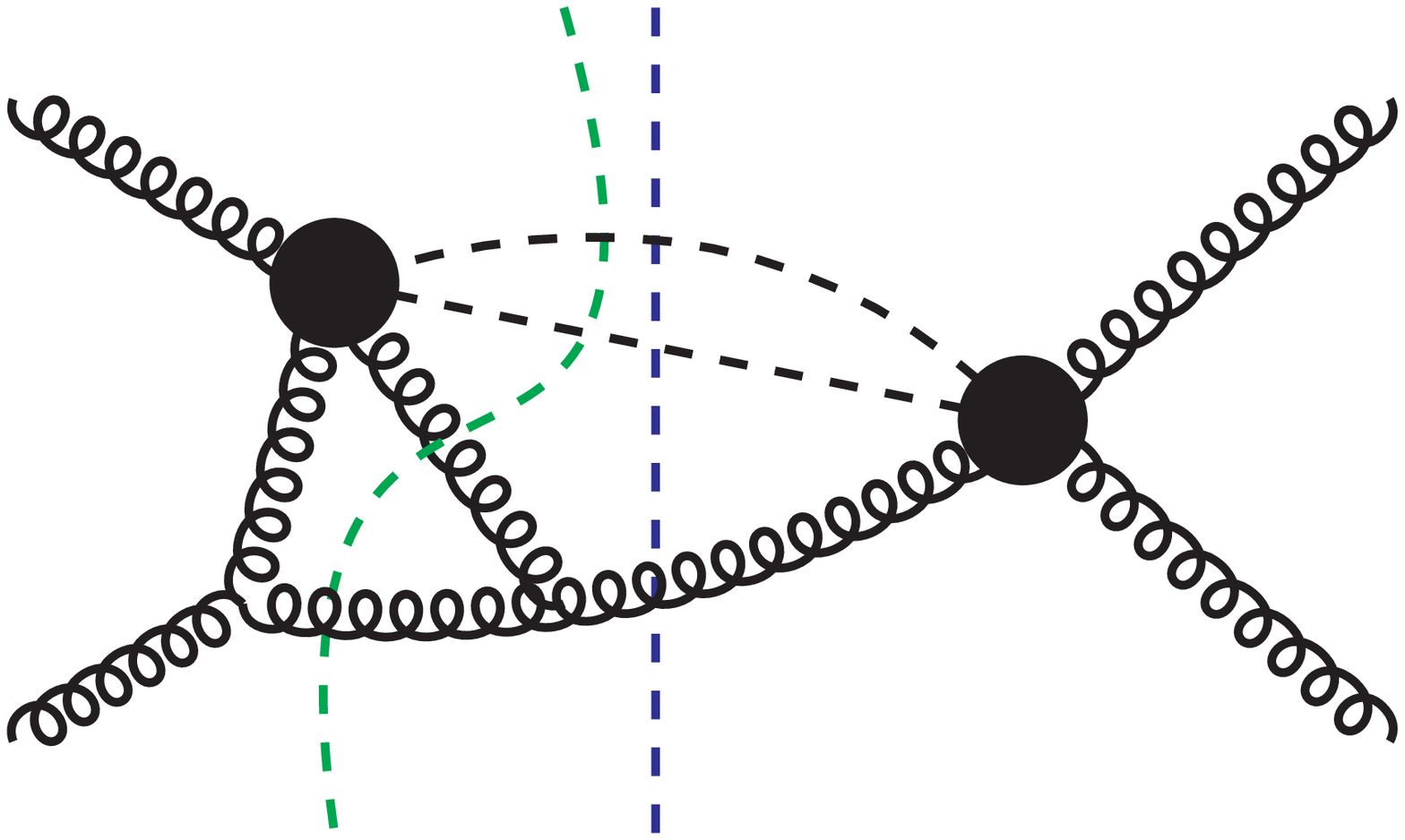}
      &
        \hspace{-2mm}\includegraphics[width=0.24\textwidth]{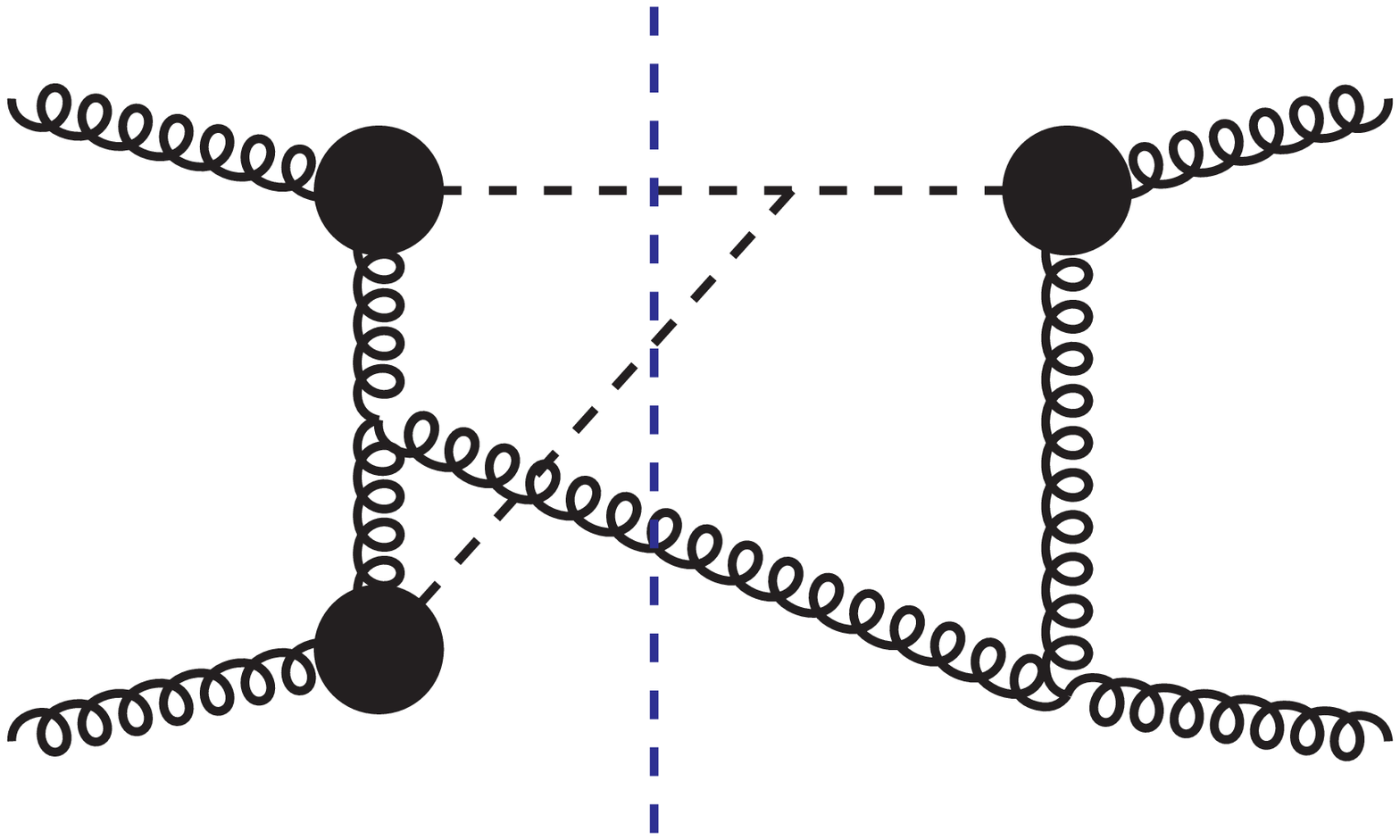}
      &
        \hspace{-2mm}\includegraphics[width=0.24\textwidth]{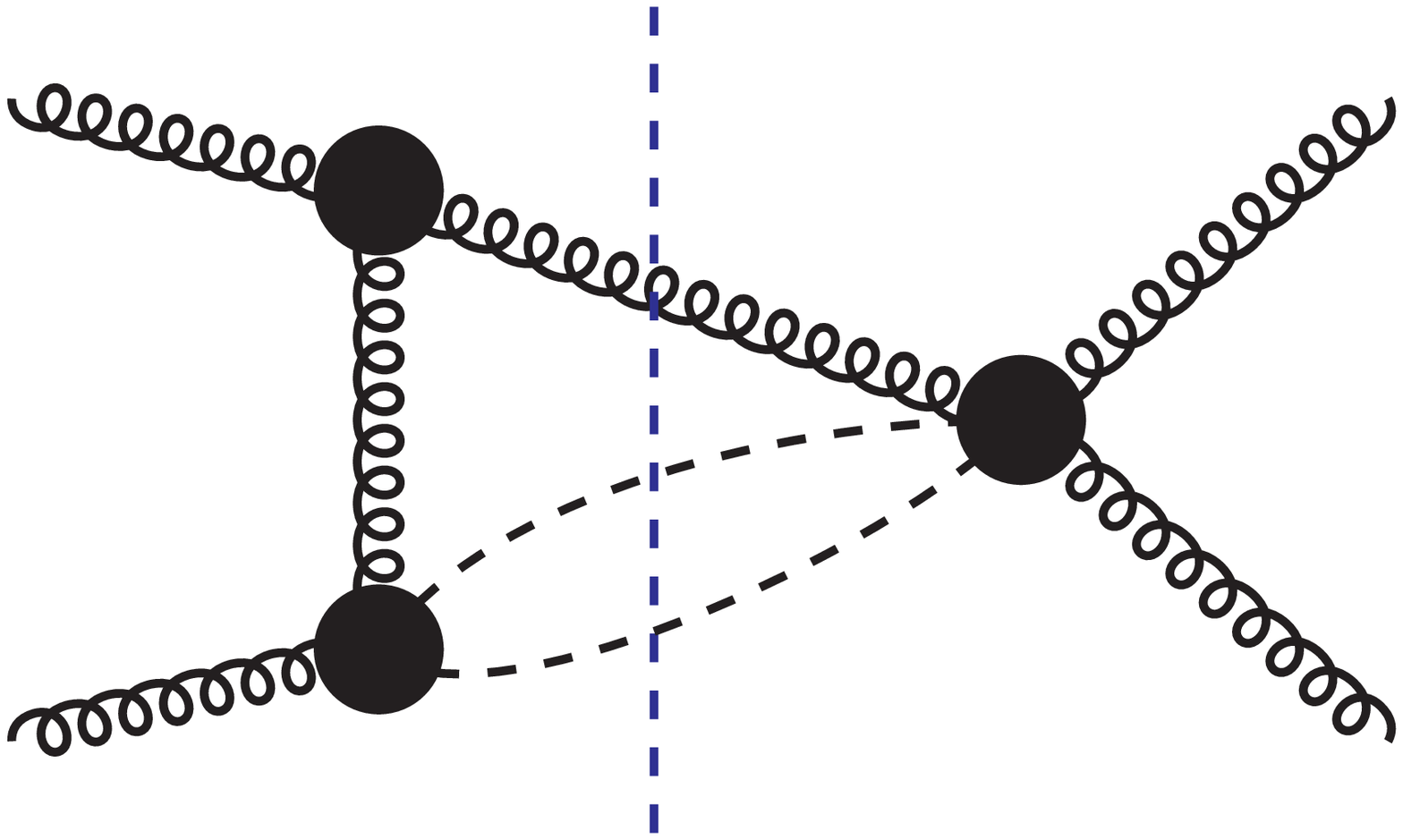}
      &
        \hspace{-2mm}\includegraphics[width=0.24\textwidth]{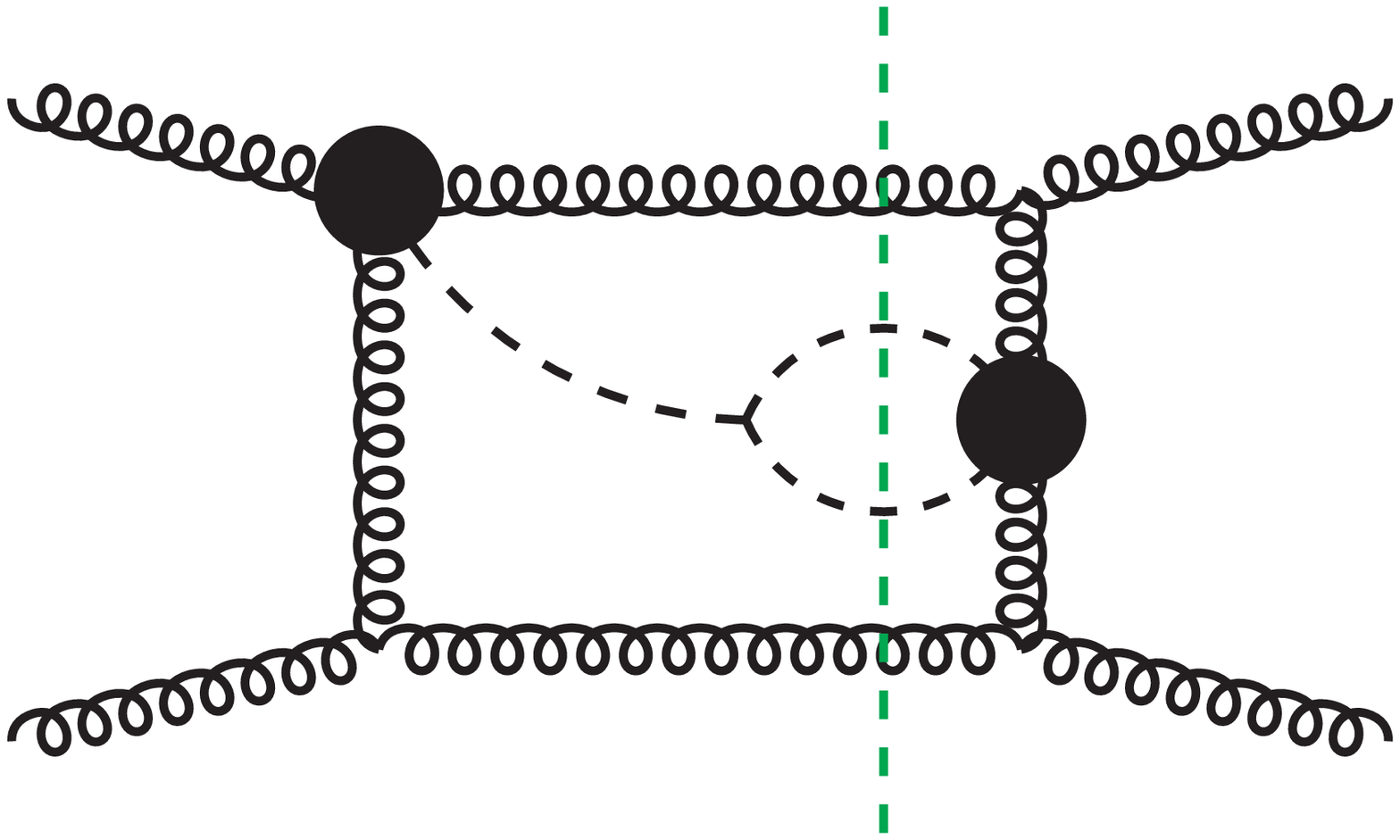}
      \\
      (a) & (b) & (c) & (d)
    \end{tabular}
  \caption{\label{fig::gghhgg_eff} The forward-scattering diagrams of Fig.~\ref{fig::gghhgg}, with black
  circles representing the expanded hard subgraphs in the large-$m_t$ expansion. Three- and four-particle
  cuts are shown by blue and green dashed lines, respectively.}
  \end{center}
\end{figure}

The subgraphs generated by the asymptotic expansion are, in this case, one-
and two-loop one-scale vacuum integrals; this class of integrals is well
studied, and it is possible to compute tensor integrals which contract with
external momenta or line momenta of the co-subgraph. Such routines are
implemented in \texttt{MATAD}~\cite{Steinhauser:2000ry}. The co-subgraphs of
the expansion lead to two- and three-loop forward-scattering integrals (or
``phase-space'' integrals) which depend on $s$ and $m_H^2$. The two-loop
phase-space integrals involve three-particle cuts whereas the three-loop
phase-space integrals involve both three- and four-particle cuts and thus
contribute to both the real-virtual and real-real contributions.  From
the diagrams in Fig.~\ref{fig::gghhgg} one obtains the diagrams in
Fig.~\ref{fig::gghhgg_eff} where the black circles represent the expanded
subgraphs.

In principle, the computation can now be performed using the method and tools
described above and in Section~\ref{sec::sub::ampgen}, that is,
\begin{enumerate}
\item generate the diagrams with \texttt{qgraf} and filter for valid cuts with
  \texttt{gen},
\item use \texttt{q2e/exp} to identify subgraphs and generate \texttt{FORM} code,
\item compute the asymptotic expansion with \texttt{FORM} and \texttt{MATAD},
\end{enumerate}
yielding expressions in terms of the remaining phase-space integrals which
must still be computed. In practice however, this approach is very
computationally challenging due to the enormous number of five-loop
forward-scattering diagrams generated and the complexity of expanding each of
those diagrams due to the large number of propagators present. We use this
method to compute in full only the leading contribution to the large-$m_t$
expansion (terms of order $1/m_t^0$) in order to cross-check an alternative
approach described below. For the channels $gq$, $q\bar{q}$ and $c\bar{c}$, we
additionally perform this cross-check to order $1/m_t^2$ in the large-$m_t$
expansion.

\begin{figure}
  \begin{center}
    \begin{tabular}{cc}
      \includegraphics[width=0.45\textwidth]{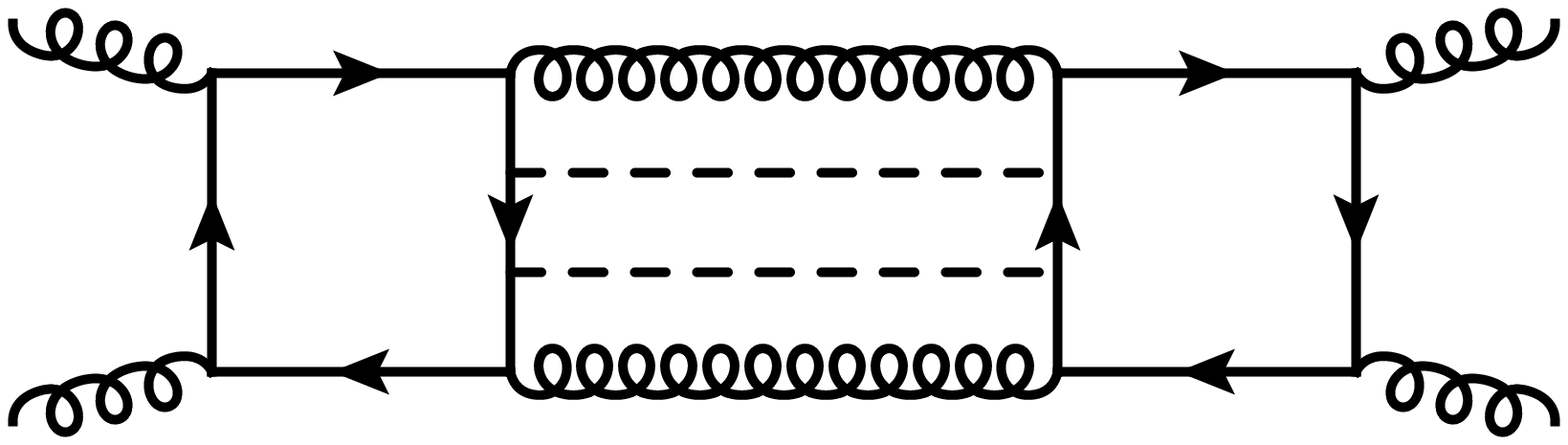}
      &
        \includegraphics[width=0.45\textwidth]{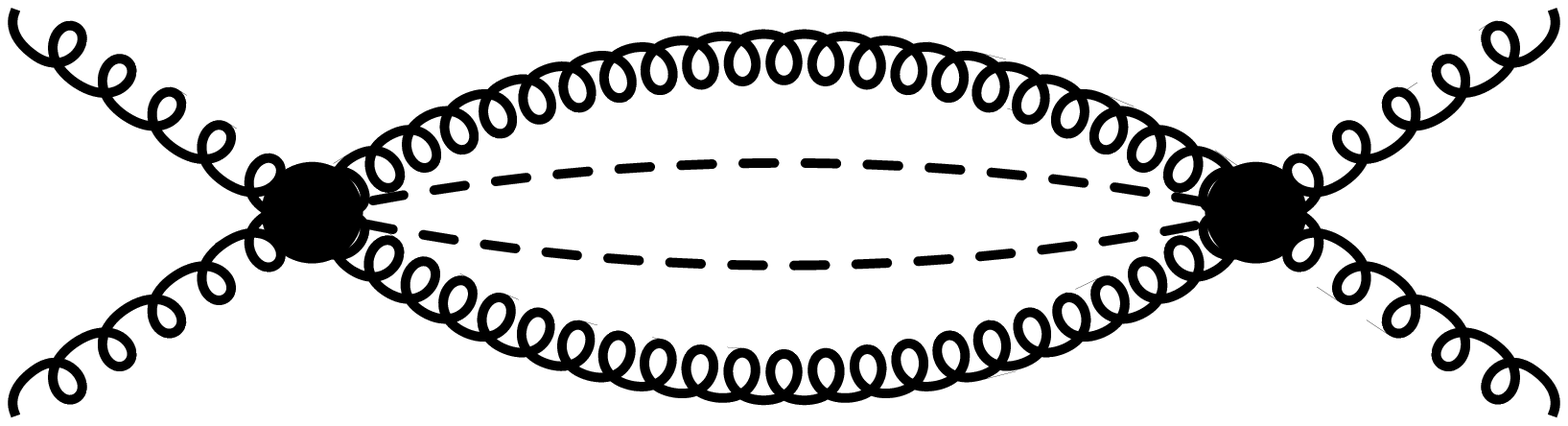}
        \\
      (a) & (b)
    \end{tabular}
    \caption{(a) An example ``full'' Feynman diagram, which contributes to the
      real-real-virtual corrections of $gg\to HH$. There are 3600 such diagrams.
      (b) The equivalent ``building-block'' diagram, of which there is only one.
    }
    \label{fig::BBexample}
  \end{center}
\end{figure}

Many of the diagrams have identical subgraphs, which are expanded in the same
small parameters. This suggests that it is beneficial to pre-compute the
subgraphs, expanding them to the required order only once and storing the
results. We can then generate two- and three-loop forward-scattering
amplitudes with ``effective vertices'' in place of the subgraphs, of which
there are comparatively very few (for the $gg$ channel we generate 4612
diagrams, compared to the 160 thousand diagrams discussed in
Section~\ref{sec::sub::ampgen}. In Table~\ref{tab::channel-dias} we show the
numbers of diagrams for all channels). As an example, in
Fig.~\ref{fig::BBexample} we show diagrams which involve the 4-gluon--2-Higgs
effective vertex; there are 3600 ``full'' diagrams, one of which is shown in
Fig.~\ref{fig::BBexample}(a), and just one effective-vertex diagram shown in
Fig.~\ref{fig::BBexample}(b). This subset of diagrams has been discussed in
Ref.~\cite{Davies:2019esq}.  In the course of the calculation, the effective
vertices are replaced with the pre-expanded subgraphs. We call this the
``building block approach''.

\begin{figure}[t]
  \begin{center}
    \begin{tabular}{ccc}
		\includegraphics[width=0.25\textwidth]{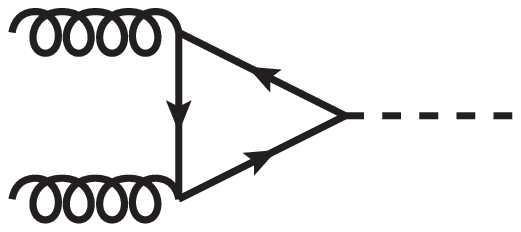} &
		\includegraphics[width=0.25\textwidth]{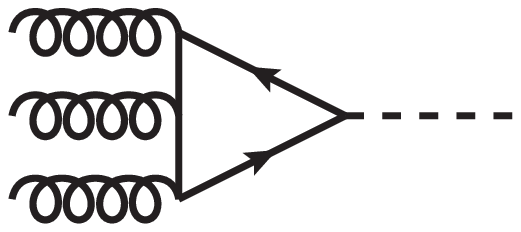} &
		\includegraphics[width=0.25\textwidth]{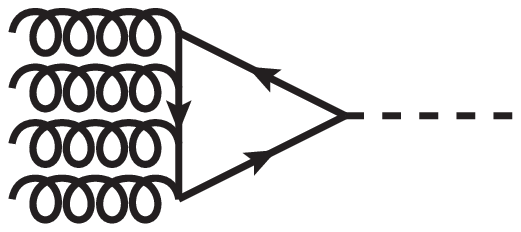} \\
		\includegraphics[width=0.25\textwidth]{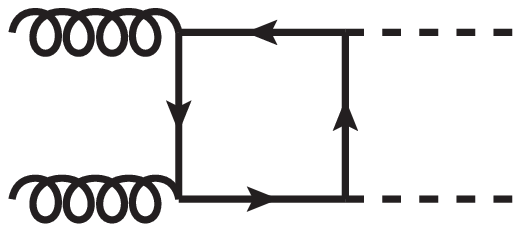} &
		\includegraphics[width=0.25\textwidth]{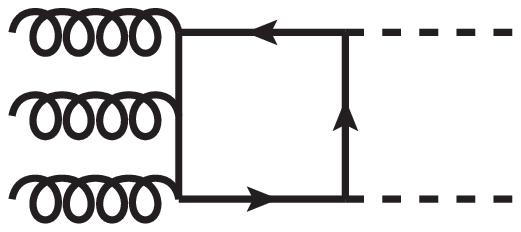} &
		\includegraphics[width=0.25\textwidth]{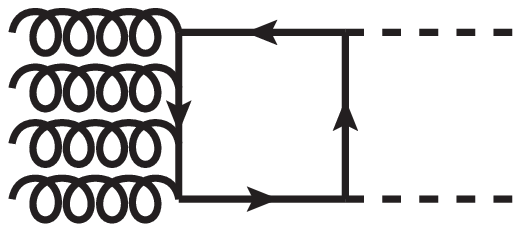} \\
    \end{tabular}
  \end{center}
  \caption{Sample Feynman diagrams contributing to the building blocks.}
  \label{fig::bb}
\end{figure}

In practice, we adopt something of a hybrid approach in order to avoid having
to deal with two-loop building blocks which are challenging to
compute. This means that we implement the
following one-loop building blocks: $ggH$, $gggH$, $ggggH$, $ggHH$, $gggHH$,
$ggggHH$, shown graphically in Fig.~\ref{fig::bb}. We compute them to order
$1/m_t^8$ with all external momenta off shell, and dependence on the open
external Lorentz and colour indices retained. The resulting expressions range
between 84~kB for $ggH$ and 164~MB for $ggggHH$. This means that they have to
be inserted into the phase-space diagrams carefully, to obtain acceptable
performance. Initially the coefficients of each term of the $1/m_t^2$
expansion are inserted only as placeholder symbols, while the rest of the
diagram is computed. Once the expression is written in terms of a linear
combination of phase-space integrals, the placeholder symbols are replaced
by using \texttt{Term} environments in \texttt{FORM} to sort the coefficients of the
phase-space integrals and $1/m_t$ expansion coefficients individually, so
as not to blow up the intermediate size of the whole expression.

For the building blocks involving up to three gluons, the Lorentz and colour
structures factorise and can thus be inserted separately into the computation
of the kinematic and colour parts of the diagrams, as is required by the
structure of our setup to compute the diagrams.  For the four-gluon building
blocks this is no longer the case, so we proceed in the following way. First,
we simply compute the expansion of the building blocks to arrive at a result
in the form
\begin{equation}
  \label{eq::BggggHHunfac}
  B_{ggggHH}^{\mu\nu\rho\sigma,\,abcd} = \sum_i
  K_i^{\mu\nu\rho\sigma}\left(m_t,\{p_k\}\right) \: C_i^{abcd} 
\end{equation}
where the factors $K_i$ contain the kinematic dependence. $\{p_k\}$ is the set
of external momenta of the building block. The colour structures are given by
the $C_i$ factors.  We now define the symbol $\delta_i$ such that
\begin{equation}
	\label{eq::LorentzColourDelta}
	\delta_i \otimes \delta_j = \left\{
		\begin{array}{ll}
			1 & i = j \\
			0 & i \neq j \\
		\end{array}
	\right. ,
\end{equation}
which allows us to re-write Eq.~(\ref{eq::BggggHHunfac}) as
\begin{equation}
  B_{ggggHH}^{\mu\nu\rho\sigma,\,abcd} =
  \left( \sum_i K_i^{\mu\nu\rho\sigma}\left(m_t,\{p_k\}\right) \delta_i \right)
  \otimes
  \left (
    \sum_j \delta_j C_j^{abcd}
  \right ).
\end{equation}
We can thus compute the Lorentz and colour parts of diagrams involving this
building block separately, and finally multiply them according to
Eq.~(\ref{eq::LorentzColourDelta}). If a diagram contains two insertions of
this building block (such as the diagram shown in Fig.~\ref{fig::BBexample})
we introduce two independent sets of the $\delta$ symbols.

\begin{figure}[t]
  \begin{center}
    \begin{tabular}{cc}
      \includegraphics[width=0.3\textwidth]{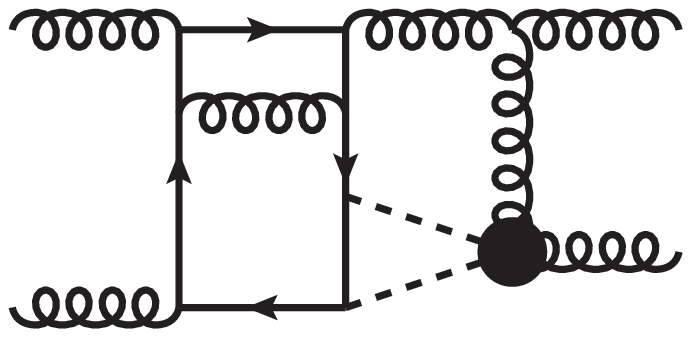} &
      \includegraphics[width=0.3\textwidth]{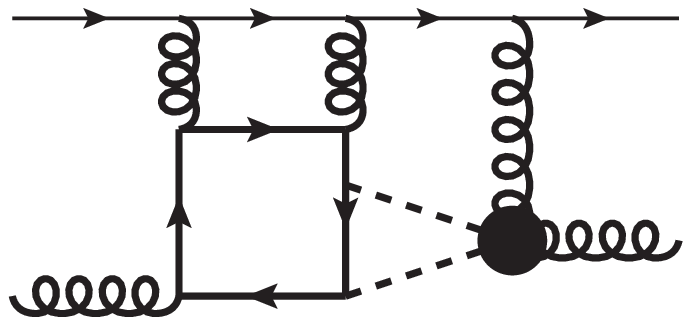} \\
    \end{tabular}
  \end{center}
  \caption{Examples for diagrams for which we perform a direct expansion of
    the two-loop tadpole subgraph, so as to avoid having to compute two-loop
    building blocks.}
  \label{fig::tad2l-examples}
\end{figure}

For diagrams for which the top quark dependence can not be completely
constructed from these building blocks (i.e. diagrams which would require
two-loop building blocks), we can generate diagrams with at least one (one-loop)
building block vertex and treat the remaining top quark propagators explicitly
with \texttt{exp}. This means that at most, we have to directly expand the
subgraphs of four-loop forward-scattering diagrams. Two examples of such
diagrams are shown in Fig.~\ref{fig::tad2l-examples}.  For some diagrams of
this type (such as the second example) \texttt{exp} also generates a one-loop
subgraph which we discard, since this case is already included in the building
block computation.

\subsection{Real radiation: partial fraction decomposition and IBP reduction}

After integrating out the top quark loops we end up with 
two- and three-loop phase-space integrals.
It is useful to distinguish three different 
sets of integral families according to the number of loops and the 
cuts through two Higgs bosons and massless particles
(see also Fig.~\ref{fig::fams_IBP} below):
\begin{itemize}
\item Two-loop families with one or two three-particle cuts. These appear in the
  real corrections at NLO and real-virtual corrections involving three top quark
  loops at NNLO.
\item
  Three-loop families with a three-particle cut. These appear in the remaining
  real-virtual corrections at NNLO.
\item
  Three-loop families with one or two four-particle cuts. These produce the
  real-real corrections at NNLO.
\end{itemize}
In total 16 two-loop families, 28 three-loop families with a three-particle cut
and 64 four-particle cut families are  required to map all diagrams. 

\begin{figure}[t]
  \includegraphics[width=\textwidth]{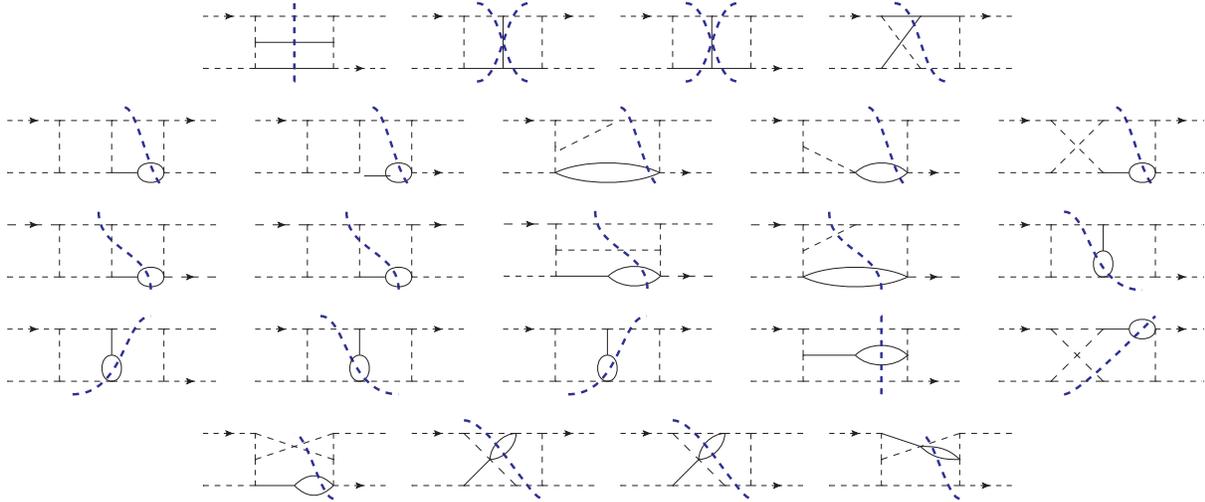}
  \caption{\label{fig::fams_IBP}Minimal set of integral families, obtained by
    {\tt LIMIT}, for the two-loop (first
    line), three-loop real-virtual (second line) and three-loop real-real
    (third to fifth lines) phase-space contributions. The arrow on external
    lines denotes the momentum $q_1$ and the blue dashed lines represent the
    cuts.}
\end{figure}   

We generate the four-point amplitudes with three independent
external momenta, $q_1$, $q_2$ and $q_3$.  After using the forward-scattering
kinematics, i.e. $q_3=-q_2$, some of the propagators become linearly
dependent; a partial fraction decomposition is needed to obtain proper
input for the IBP reduction. For this purpose we have developed the program
{\tt LIMIT}~\cite{Herren:2020ccq} which automates this step of our
calculation. This program is based on ideas presented in Ref.~\cite{Pak:2011xt}.

In the following we describe the workflow of {\tt LIMIT}. First,
{\tt LIMIT} identifies linear relations among the propagators of a given
family, derives relations to perform a partial fraction decomposition, and
generates {\tt FORM} routines to apply these relations to the
integrals. Each initial, linearly dependent, family may yield
multiple linearly independent families.

Next, families with multiple cuts are split into multiple families
with one cut each, since each integral with two cuts corresponds to
the sum of two integrals with one cut. Splitting the families in
this way is necessary because {\tt LiteRed}~\cite{Lee:2012cn,Lee:2013mka}
can only handle families with a single cut. Now we use {\tt LiteRed} 
to identify vanishing sectors of the families and to derive symmetry relations
for the non-vanishing sectors.  Based on this, {\tt LIMIT} then generates {\tt FORM}
code to nullify vanishing integrals and apply the symmetry relations,
reducing the number of terms present in each amplitude.

Finally, {\tt LIMIT} identifies equivalent families based on their graph
polynomials and groups them into classes. For each class, {\tt LiteRed} is
employed to derive relations for translating integrals of each family to
integrals of a single representative family.  As not all linearly independent
families have the same number of propagators, {\tt LIMIT} then tries to embed
classes of families with fewer lines into classes of families with more
lines. The rules for mapping integrals into this minimal number of
representative families are translated into {\tt FORM} code and applied to the
amplitudes.

By applying {\tt LIMIT} to the three sets of integral families mentioned above,
we obtain 4, 5 and 14 integral families, respectively.
Their graphical representation is shown in Fig.~\ref{fig::fams_IBP}.
For more details we refer to~\cite{Herren:2020ccq}.

\begin{figure}[t]
  \begin{center}
  \includegraphics[width=0.8\textwidth]{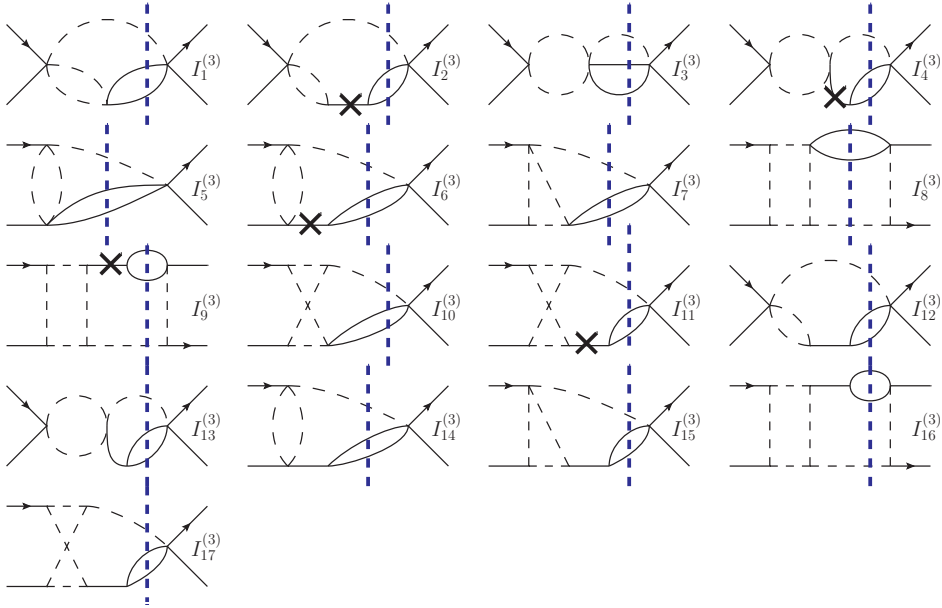}
  \caption{\label{fig::3l3p}Three-loop three-particle cut master
    integrals. Solid and dashed lines denote massive and massless lines,
    respectively. The blue dashed line symbolizes the cut and crosses denote
    propagators raised to negative power.}
  \end{center}
\end{figure}

\begin{figure}[t]
  \includegraphics{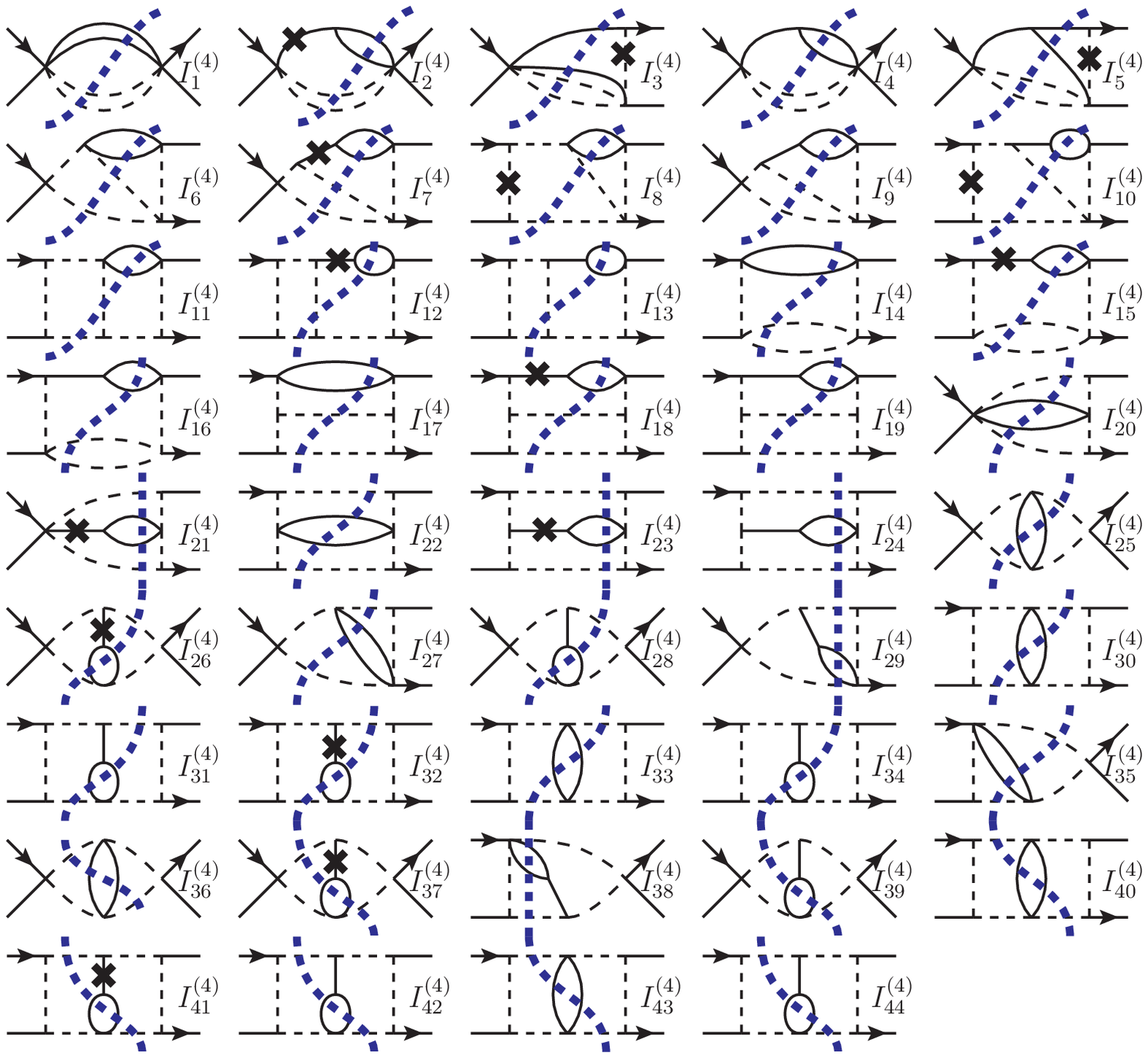}
  \caption{\label{fig::3l4p}Three-loop four-particle cut master
    integrals. Solid and dashed lines denote massive and massless lines,
    respectively. The blue dashed line symbolizes the cut and crosses denote
    propagators raised to negative power.}
\end{figure}

\begin{figure}[t]
  \includegraphics{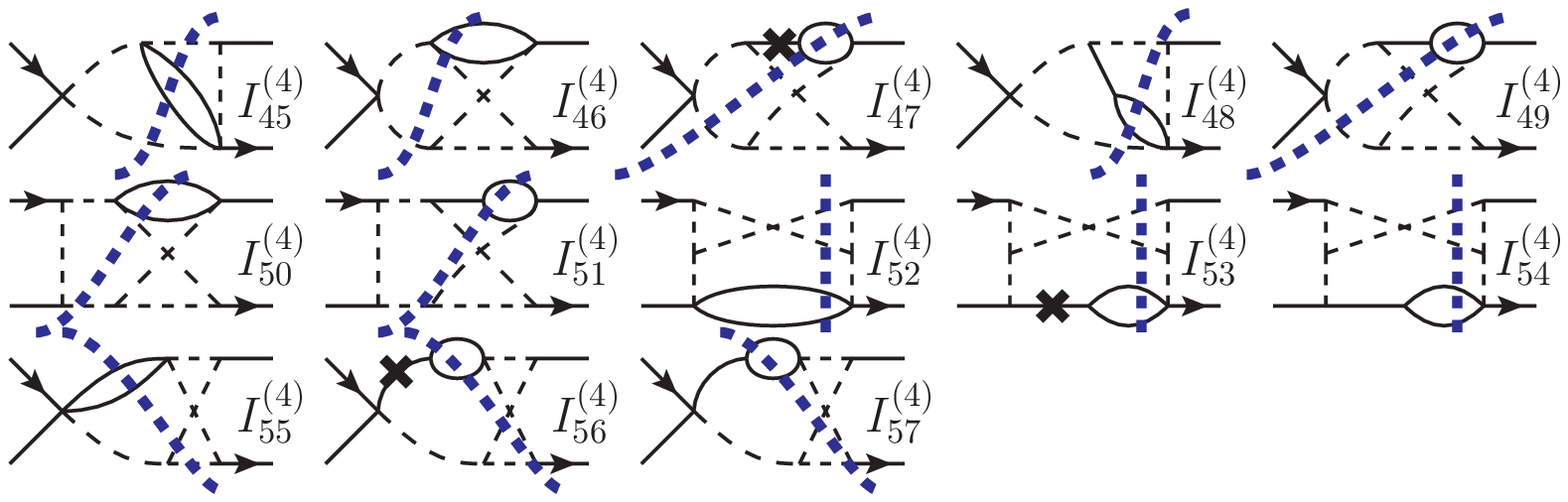}
  \caption{\label{fig::3l4p2}Continuation of Fig.~\ref{fig::3l4p}.}
\end{figure}

Using {\tt LiteRed} we relate equivalent sectors of families in the three
different classes and perform an IBP reduction to master integrals. In total
$\mathcal{O}\left(500\,000\right)$ integrals need to be reduced, which takes
{\tt LiteRed} less than two days.
The two-loop reduction leads to 16 master integrals with a three-particle
cut. They have already been studied in Ref.~\cite{Davies:2019xzc} where the
$n_h^3$ contributions were computed where both Higgs bosons couple to
different top-quark loops.
We furthermore obtain 17 three-loop three-particle cut and 57 four-particle
cut master integrals. They are shown in Fig.~\ref{fig::3l3p}, as well as
Figs.~\ref{fig::3l4p} and~\ref{fig::3l4p2}, respectively.

Details on the computation of the master integrals
are provided in Section~\ref{sec::masters}.
After inserting the results for the master integrals we obtain an expansion
for each Feynman diagram in $1/m_t$ and $\epsilon$.  For the real-radiation
contribution we observe poles of order $1/\epsilon^4$.


\subsection{Ultraviolet counterterms}

The NNLO real-radiation amplitude receives a one-loop
counterterm contribution induced by the 
NLO corrections, which are required to order $\epsilon$.
To be more precise, we have to renormalize
$\alpha_s$, $m_t$ and the (external) gluon field. The corresponding
renormalization constants are given by
\begin{eqnarray}
  Z_{\alpha_s} &=& 1 + \frac{\alpha_s}{4\pi\epsilon}\left(-\frac{11}{3}C_A +
                   \frac{4}{3}n_f T_f\right)  ,
                   \,\nonumber\\
  Z_m &=& 1 - \frac{3\alpha_s}{4\pi\epsilon}C_F\,,
          \,\nonumber\\
  Z_3^{\mathrm{OS}} &=& \frac{1}{\zeta_3^0} = 1 -
                        \frac{\alpha_s}{3\epsilon\pi}{T_f} n_h \Bigg[1 +
                        \epsilon\ln\frac{\mu^2}{m_t^2}
                        +
                        \epsilon^2\left(\frac{\zeta_2}{2}+\frac{1}{2}\ln^2\frac{\mu^2}{m_t^2}
                        \right)\nonumber\\
               &&\mbox{}+
                  \epsilon^3\left(-\frac{\zeta_3}{3}+\frac{\zeta_2}{2}\ln\frac{\mu^2}{m_t^2}
                  +\frac{1}{6}\ln^3\frac{\mu^2}{m_t^2}\right)+\mathcal{O}(\epsilon^4)\Bigg],
\end{eqnarray}
where $\mu$ is the renormalization scale and $\alpha_s = \alpha_s(\mu)$.
We first perform the renormalization in the full theory (i.e., we
have $n_f=6=n_l+n_h=5+1$). Afterwards, we perform a decoupling from
$\alpha_s^{(6)}$ to $\alpha_s^{(5)}$.  Note that we renormalize the top quark
mass first in the $\overline{\mathrm{MS}}$ scheme; the transition to
the on-shell scheme in the final result is straightforward.


\subsection{Collinear counterterms}

The NNLO collinear counterterms are obtained
from the convolution of the LO and NLO cross sections 
$\sigma_{gg}^{(0)}$ and $\sigma_{ij}^{(1)}$
and the gluon or quark splitting functions.
In general such a convolution has the form
\begin{eqnarray}
  \frac{\sigma_{ij}^{\rm bare}(x)}{x} 
  &=& \sum_{k,l} 
      \frac{\sigma_{{kl}}(z)}{z}
      \otimes \Gamma_{ki}
      \otimes \Gamma_{lj}\,,
      \label{eq::sig_Gam}
\end{eqnarray}
with
\begin{eqnarray}
  (f \otimes g)(z) &=& \int_0^1 {\rm d}x\, {\rm d}y\, f(x)g(y)\delta(z-xy)
                       \,.
\end{eqnarray}
In Eq.~(\ref{eq::sig_Gam}) the superscript ``bare'' refers to the
renormalization of the parton distribution functions, i.e.  the subtraction of
the $1/\epsilon$ poles in connection to radiation from incoming partons. It
does not refer to the renormalization of the ultraviolet divergences; here
we assume that all ultraviolet counterterms are already taken into account.
The kernels $\Gamma_{ij}$ in Eq.~(\ref{eq::sig_Gam})
are obtained from the splitting functions as follows
\begin{eqnarray} 
  \Gamma_{ij}(x) &=& \delta_{ij}\delta(1-x)
                     - \frac{\alpha_s^{(5)}(\mu)}{\pi}
                     \frac{P_{ij}^{(0)}}{\epsilon}
                     \nonumber\\&&
                     + \left(\frac{\alpha_s^{(5)}(\mu)}{\pi} \right)^{\!2}\!\!
                     \left\{\frac{1}{2\epsilon^2}\left[
                     \left(P_{ik}^{(0)}\otimes P_{kj}^{(0)}\right)(x)
                     + \beta_0 P_{ij}^{(0)}
                     \right]
                     -\frac{1}{2\epsilon} P_{ij}^{(1)}
                     \right\}
                     + \mathcal{O}\left(\alpha_s^3\right)
                     ,
                     \label{eq::Gamma_ij}
\end{eqnarray}
where $P_{ij}^{(0)}$ and
$P_{ij}^{(1)}$ are one- and two-loop splitting functions which can be found in
Refs.~\cite{Altarelli:1977zs,Curci:1980uw,Moch:2004pa,Vogt:2004mw}.  For
further details we refer to Refs.~\cite{Anastasiou:2002yz,Hoeschele:2013gga}
where the computation of the collinear counterterms is discussed for NNLO and
N$^3$LO single Higgs boson production.  The combinatorics for single and
double Higgs production is identical and thus the formalism outlined in detail
in~\cite{Anastasiou:2002yz} can be applied in the context of this calculation.
Additionally, this reference contains explicit formulae which solve
Eq.~(\ref{eq::sig_Gam}) for the renormalized cross section $\sigma_{ij}$
within perturbation theory. The resulting collinear counterterms are
computed from convolution integrals of the following form,
\begin{eqnarray}
  \sigma_{gg,\rm coll}^{(2)} 
  &=&
      - \frac{1}{\epsilon}\left(\frac{\mu^2}{\mu_f^2}\right)^\epsilon
      \int_{1-\delta}^1 \mathrm{d}z\, P^{(0)}_{gg}(z)\sigma_{gg}^{(1)}(x/z)
      + \ldots
      \,,
      \label{eq::collCT-example}
\end{eqnarray}
where the ellipses represent further contributions
to the collinear NNLO counterterm $\sigma_{gg,\rm coll}^{(2)}$.
Note that in Eq.~(\ref{eq::collCT-example}) we distinguish the
renormalization scale $\mu$ from the factorization scale $\mu_f$.
We compute the collinear counterterms in the $n_l$ flavour theory,
i.e., in Eq.~(\ref{eq::Gamma_ij}) the one-loop coefficient of the
QCD beta function is given by
\begin{eqnarray}
  \beta_{0} &=& \frac{11}{12} C_{A}-\frac{1}{3} T_f {n_l}
  \,,
\end{eqnarray}
where $n_l$ is the number of massless quarks.
We compute all contributions in terms of SU$(N_c)$ colour factors $C_A=N_c$ and
$C_F=(N_c^2-1)/(2N_c)$ and the trace normalization $T_f=1/2$. 

Inspection of Eq.~(\ref{eq::Gamma_ij}) shows that at NNLO the collinear
counterterm starts at $\mathcal{O}(1/\epsilon^{2})$ whereas the real and virtual
corrections contain poles up to $\mathcal{O}(1/\epsilon^{4})$.  In
Appendix~\ref{app::collCT} we provide useful formulae
for the analytic computation of the convolution integrals, such as the
one shown in Eq.~(\ref{eq::collCT-example}). These are obtained by expanding
in $\delta$; we have computed all contributions up to order
$\rho^3$ and $\delta^{30}$.


\subsection{\label{sec::virt}Virtual corrections}

\begin{figure}[t]
  \begin{center}
    \begin{tabular}{cccc}
      \includegraphics[width=0.22\textwidth]{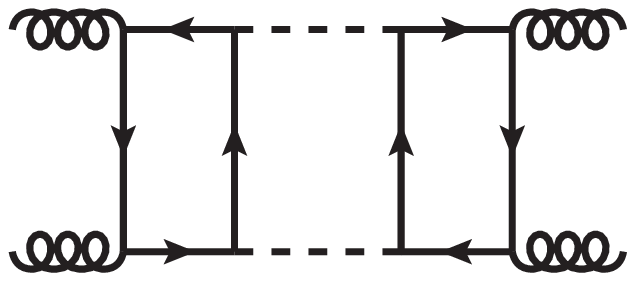}
      &
      \includegraphics[width=0.22\textwidth]{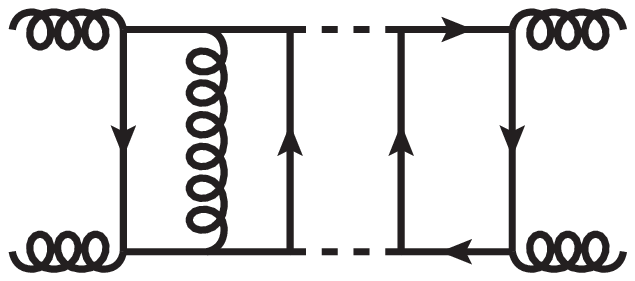}
      &
      \includegraphics[width=0.22\textwidth]{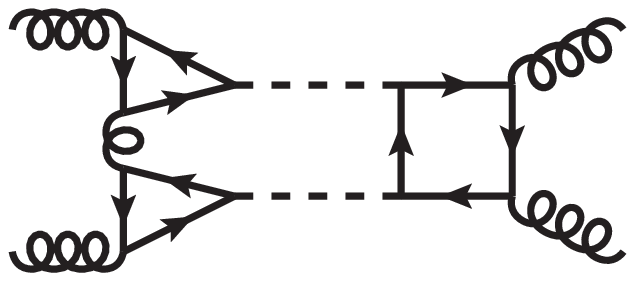}
      &
      \includegraphics[width=0.22\textwidth]{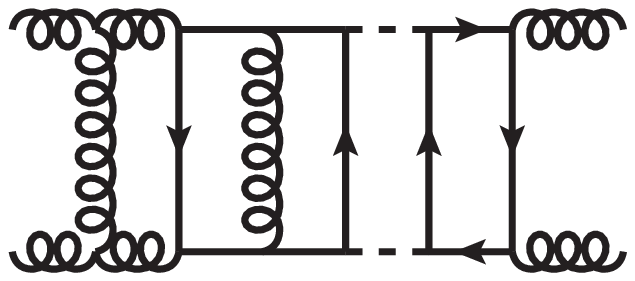}
      \\
      (a) LO & (b) NLO, $n_h^2$ & (c) NLO, $n_h^3$ & (d) NNLO, $n_h^2$
      \\
      \\
      \includegraphics[width=0.22\textwidth]{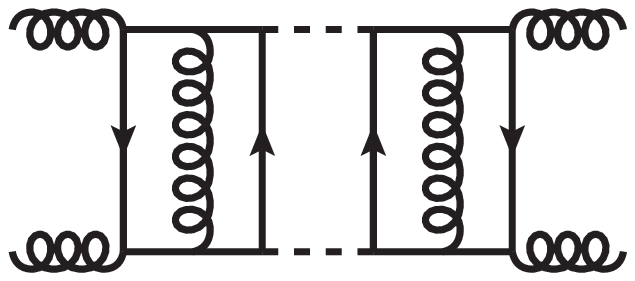}
      &
      \includegraphics[width=0.22\textwidth]{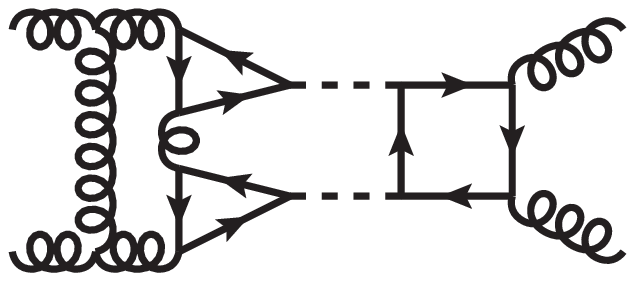}
      &
      \includegraphics[width=0.22\textwidth]{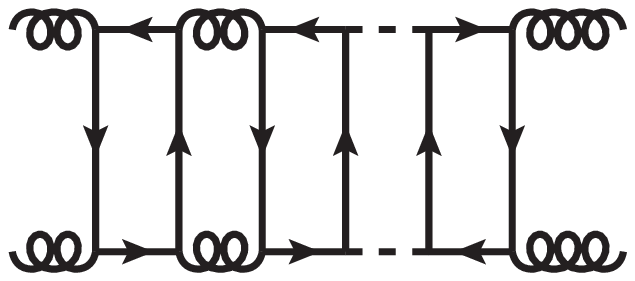}
      &
      \includegraphics[width=0.22\textwidth]{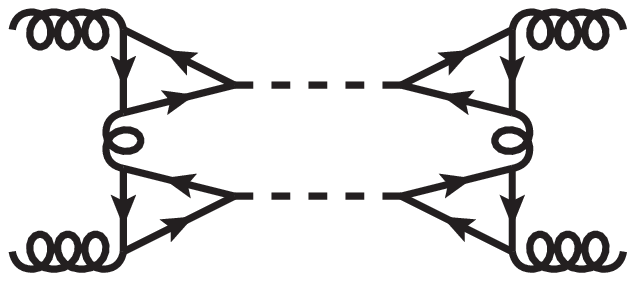}
      \\
      (e) NNLO, $n_h^2$ & (f) NNLO, $n_h^3$ & (g) NNLO, $n_h^3$ & (h) NNLO, $n_h^4$
    \end{tabular}
    \caption{\label{fig::diag_virt}
      Sample forward-scattering Feynman diagrams for the virtual corrections at
      LO, NLO and NNLO.
    }
  \end{center}
\end{figure}

Virtual corrections only exist for the $gg$ and $q\bar{q}$ channels.
In the latter case they contribute for the first time
at NNLO and are suppressed by $1/m_t^2$ at the level of the amplitude
(and so by $1/m_t^4$ at the level of the cross section).
We discuss them in Appendix~\ref{app::qqHH}. In this section we
concentrate on the channel $gg\to HH$. Sample Feynman diagrams for the forward-scattering
kinematics are shown in Fig.~\ref{fig::diag_virt}. At NNLO we distinguish 
contributions which have two, three and four closed top quark loops and we
denote the corresponding contributions by $n_h^2$, $n_h^3$ and $n_h^4$ respectively.
Note that all $n_h^3$ virtual corrections are already contained in
Ref.~\cite{Davies:2019xzc}, even those with a top quark loop which has
no coupling to the Higgs bosons (such as the diagram in Fig.~\ref{fig::diag_virt}(g)).
The $n_h^2$ terms develop poles up to
$1/\epsilon^4$ which cancel against those of the real-radiation corrections.
The $n_h^4$ contribution appears only in the virtual corrections and is therefore
finite. The corresponding analytic results are presented in Appendix~\ref{app::nh4}.

For the virtual corrections we do not use the optical theorem but
compute the contribution to the cross section from the form factors
obtained in Ref.~\cite{Davies:2019djw} in an expansion
up to order $1/m_t^8$. The triangle form factors are even
available up to order $1/m_t^{14}$.

The starting point is the amplitude for the process $gg\to HH$
which has two independent tensor structures. Expressed in terms
of the ``box'' and ``triangle'' form factors it is given by
\begin{eqnarray}
  \delta^{ab} {\cal M}
      &=&
      \delta^{ab}\:
      \varepsilon_{1,\mu}\varepsilon_{2,\nu}\:
      \frac{G_F}{\sqrt{2}} \frac{\alpha_s(\mu)}{2\pi} T_f n_h \, s \, 
      \left[
      \left(\frac{3 m_H^2}{s-m_H^2} F_{\rm tri} + F_{\rm box1}\right)
      A_1^{\mu\nu} + F_{\rm box2} A_2^{\mu\nu} 
      \right]
      ,
      \label{eq::virtamp}
\end{eqnarray}
where $G_F$ is the Fermi constant,
$a$ and $b$ are adjoint colour indices and the two Lorentz structures
are given by
\begin{eqnarray}
  A_1^{\mu\nu} &=& g^{\mu\nu} - {\frac{1}{q_{12}}q_1^\nu q_2^\mu
  }\,,\nonumber\\
  A_2^{\mu\nu} &=& g^{\mu\nu}
                   + \frac{1}{p_T^2 q_{12}}\left(
                   q_{33}    q_1^\nu q_2^\mu
                   - 2q_{23} q_1^\nu q_3^\mu
                   - 2q_{13} q_3^\nu q_2^\mu
                   + 2q_{12} q_3^\mu q_3^\nu \right)\,,
 \label{eq::A1A2}
\end{eqnarray}
with
\begin{eqnarray}
\label{eqn:qt2def}
  q_{ij} &=& q_i\cdot q_j\,,\qquad
  p_T^{\:2} \:\:\:=\:\:\: \frac{2q_{13}q_{23}}{q_{12}}-q_{33}
  \,.
\end{eqnarray}
Analytic results for the form factors $F_{\rm tri}$, $F_{\rm box1}$,
$F_{\rm box2}$ are available in the ancillary files of Ref.~\cite{Davies:2019djw}.

The differential cross section is constructed in terms
of the amplitude of Eq.~(\ref{eq::virtamp}) according to
\begin{eqnarray}
  \frac{{\rm d}\sigma}{{\rm d}{y}} 
  &=&
      \frac{N_A}{2 \cdot N_A^2 \cdot (2-2\epsilon)^2}
      {\frac{1}{2s}}
      f^{\rm 2PS}(\epsilon)
      \left|{\cal M}^\star{\cal M}\right|
\end{eqnarray}
where the factors $1/2$, $1/N_A^2=1/8^2$ and $1/(2-2\epsilon)^2$ are due to the
identical particles in the final state, and the gluon colour and spin averages,
respectively.  The factor $f^{\rm 2PS}$ comes from the $d$-dimensional
two-particle phase space and is given by
\begin{eqnarray}
	f^{\rm 2PS}(\epsilon) &=&
	\left(\frac{e^{\gamma_E}}{4\pi}\right)^\epsilon
	\left(\frac{\mu^2}{s}\right)^\epsilon
	\frac{2^{2\epsilon-3}\:\: \pi^{\epsilon-1}\:\: \delta^{1/2-\epsilon}}{\Gamma(1-\epsilon)}
	\left(y(1-y)\right)^{-\epsilon}.
	\label{eq::f2PS}
\end{eqnarray}

We perform the integration to obtain the cross section after expanding in
$\delta$. If we were to integrate over $t$, the integration boundaries would be
$\delta$ dependent. We therefore make the following change of variables:
\begin{eqnarray}
  t &\to & -\frac{s}{4}\left(1 + \delta - 4 y \sqrt{\delta} + 2 \sqrt{\delta}
        \right)
        \,.
        \label{eq::t2y}
\end{eqnarray}
In terms of the new integration variable $y$ the integration boundaries become
$[0,1]$ and the $\delta$ dependence is isolated to the integrand.
This allows us to expand in $\delta$ at the level of the integrand, and
thus integrate each term of the expansion individually.
In the supplementary material to this paper we
provide the results in such a way that the (divergent)
virtual corrections can be extracted.

We perform the construction described above for the ultraviolet 
renormalized form factors as published in Ref.~\cite{Davies:2019djw}.
The renormalization of the gluon wave function, top quark mass 
and strong coupling in the LO expression induces $1/\epsilon$
poles at NLO and $1/\epsilon^2$ poles at NNLO.
The insertion of the one-loop counterterms in the
virtual NLO expression (which has an $\epsilon$ expansion starting at 
$1/\epsilon^2$) induces $1/\epsilon^3$ poles at NNLO.



\section{\label{sec::masters}Phase-space master integrals}


\subsection{Transformation to $\epsilon$ form}

In this section, we discuss the computation of the 17 three-loop
three-particle cut master integrals depicted in Fig.~\ref{fig::3l3p}, as well
as the 57 three-loop four-particle cut master integrals depicted in
Figs.~\ref{fig::3l4p} and~\ref{fig::3l4p2}.  We follow the method of
differential equations \cite{Kotikov:1990kg,Gehrmann:1999as} and use
\texttt{LiteRed} to take the derivative of each master integral w.r.t.~$x$ and
reduce the resulting integrals back to master integrals.  Thus, we obtain two
systems of differential equations: one for the three-particle cut integrals
and one for the four-particle cut integrals. These systems have the form
\begin{flalign}
\partial_x \vec{L} = M\!\left(x,\epsilon\right)\,\vec{L}~.\label{eq::de1}
\end{flalign}
$\vec{L}$ is the vector of master integrals, and the matrix $M$ is block-triangular
with entries rational in both $x$ and $\epsilon$.
In the following we take two approaches to solve Eq.~(\ref{eq::de1}): in
Section~\ref{sec::canonical} we seek a canonical basis of master integrals
\cite{Henn:2013pwa} and in Section~\ref{sec::boundaries} we compute them in terms
of an asymptotic series in $\delta$.


\subsection{\label{sec::canonical}Finding a canonical basis}
A canonical basis $\vec{C}$ is a particularly useful basis of integrals in which the differential equations take the form
\begin{flalign}
\partial_x \vec{C} = \epsilon\sum_i \frac{ {\hat{M}_i }}{x-x_i}\vec{C}~,\label{eq::can}
\end{flalign}
where the $x_i$ are constants and the matrices {$\hat{M}_i$} do not depend
on $\epsilon$ or $x$. A system of the form of Eq.~(\ref{eq::can}) can be
integrated order-by-order in $\epsilon$ in terms of Goncharov Polylogarithms
(GPLs) \cite{Goncharov:1998kja} over the letters $x_i$.

To find a rational transformation from $\vec{L}$ to $\vec{C}$ we use the program \texttt{Libra} \cite{Lee:2020zfb} and adapt Lee's algorithm \cite{Lee:2014ioa} to the problem at hand. As
the matrix $M(x,\epsilon)$ contains rational functions with numerator and denominator degrees as high as $20$ in our case, a direct application of Lee's algorithm would lead to unmanageable intermediate expression swell.
We thus deviate from the standard algorithm by first bringing all diagonal blocks to Fuchsian form (that is, they have only simple poles in $x$), followed by bringing the off-diagonal blocks to Fuchsian form.
We then normalize the eigenvalues of all residues to be proportional to $\epsilon$ or take the form $a \epsilon \pm 1/2$, where $a$ is an integer.

Consequently, we arrive at an intermediate basis $\vec{I}$, related to $\vec{L}$ by a rational transformation, in which the differential equations are in Fuchsian form
and do not contain spurious poles:
\begin{flalign}
\partial_x \vec{I} = \sum_i \frac{\tilde{M}_i(\epsilon)}{x-x_i}\vec{I}~.\label{eq::fuchs}
\end{flalign}
For the three-particle cut master integrals the poles $x_i$ are given by $0$,
$1/4$ and $1$, corresponding to the kinematic limits $s \rightarrow \infty$,
$s \rightarrow 4 m_H^2$ and $s \rightarrow m_H^2$, respectively.  In the case
of the four-particle cut master integrals there are also poles at $-1/4$ and
$-1$, corresponding to the kinematic limits $s \rightarrow -4 m_H^2$ and
$s \rightarrow -m_H^2$ respectively.\footnote{Both of these poles only appear
  in the differential equations for master integrals such as $I^{(4)}_{30}$
  (see Fig.~\ref{fig::3l4p}), for which the momentum $(q_1 - q_2)^2 = -s$
  flows through the massive lines.}

Rational transformations render the eigenvalues of all poles proportional to $\epsilon$, apart from
$\pm 1/4$. Therefore to find a canonical basis, we need to rationalize the square roots
$\sqrt{1-4x}$ and $\sqrt{1 + 4x}$.
For the three-particle cut master integrals only the first square root plays a role. For this system, we perform the variable change
\begin{align}
x = \frac{z}{(z+1)^2}
\end{align}
mapping the physical interval $x \in [1/4,0)$ to $z \in [1,0)$.
In the case of the four-particle cut master integrals we need to rationalize both square roots
simultaneously, requiring the more complicated variable change
\begin{align}
x &= \frac{k^4 - 6 k^2 + 1}{4(1 + k^2)^2}~.
\end{align}
This change maps $x \in [1/4,0)$ to $k \in [0,\sqrt{2} - 1)$.

With all roots rationalized, we now proceed by normalizing the corresponding
eigenvalues to be proportional to integers, bringing the off-diagonal blocks
into Fuchsian form once more and factoring out $\epsilon$ to arrive at a
canonical basis in $z$ and $k$ for the three- and four-particle cut
systems, respectively. These resulting system can be integrated in terms of
GPLs and their boundary conditions can be fixed in the limit
$x \rightarrow 1/4$.

Once we have the leading-order term in the $\delta$ expansion for the master integrals
(see Section~\ref{sec::boundaries}), the exact expressions for the master integrals can
be computed in two ways. The first is to transform the leading-$\delta$ terms of all
master integrals into the canonical basis.
The differential system can then be integrated order-by-order in $\epsilon$ and the
integration constants are fixed by comparing the resulting expressions to the boundary
conditions in the limit $\delta \rightarrow 0$.

The second option is to first compute the path-ordered exponential $U$ of the
canonical differential equation matrix $M$ 
\begin{align}
U = \mathcal{P}\exp\left[ \epsilon\int\mathrm{d}x\: {\sum_i
  \frac{\hat{M}_i}{(x-x_i)} } \right]~,
\end{align}
in terms of which we can write the master integrals of the canonical basis as
\begin{align}
\vec{I}(x) = U \vec{I}_0~,
\end{align}
where $\vec{I}_0$ does not depend on $x$. \texttt{Libra} can decompose $\vec{I}_0$ as
\begin{align}
\vec{I}_0 = {K(\epsilon)}\: \vec{c}~,
\end{align}
where $K$ depends only on $\epsilon$ and $\vec{c}$ is the minimal number of necessary
boundary conditions. The entries of $\vec{c}$ are given by the leading-order terms
in the $\delta$ expansion of a certain subset of original master integrals, determined by
\texttt{Libra} to be sufficient to fix all boundary conditions.

The main difference between these two approaches is the apparent number of
necessary boundary conditions.  For example using the first approach, we need
the integrals $I_1^{(4)}$, $I_2^{(4)}$ and $I_3^{(4)}$ to determine
$I_1^{(4)},\ldots,I_{10}^{(4)}$, whereas \texttt{Libra} finds that
$I_1^{(4)}$ is sufficient.  Thus, it makes use of relations among the
leading-$\delta$ terms of different integrals.

In our calculation we have used both methods for the three-loop
three-particle cut master integrals. For the three-loop four-particle cut
master integrals we have only used the first method since our non-trivial
letters make the construction of the matrix $U$ difficult.

In the following section we describe how we compute the boundary conditions. We will
compute the leading-order term in the $\delta$ expansion for all master integrals; the
integrals which are not necessary to integrate the differential system serve as a cross
check.


\subsection{\label{sec::boundaries}Computation of the boundary conditions}

In the previous subsection we have presented the construction of the exact
results for the phase-space master integrals.  The purpose of this subsection
is to provide boundary conditions needed to compute the exact
expressions. In the next subsection we briefly describe a 
formalism which allows us
to compute the $\delta$ expansion of all master integrals to a high enough
order in $\delta$ to be sufficient for practical applications.
In fact, in the combination of the real-radiation contribution
with the virtual corrections and the collinear counterterm it is necessary to
use the $\delta$-expanded results in order check the cancellation of the
$\epsilon$ poles analytically.

We start with some definitions which are common to the three- and
four-particle phase-space integrals.
For the phase-space measures of the Higgs bosons we write
\begin{align}
  \mathcal{D}p_j
  \equiv
  \frac{\mathrm{d}^{d-1}p_j}{(2\pi)^{d-1}}
  \frac{1}{2E_j}
  =
  \frac{p_j^{d-2}\mathrm{d}p_j}{(2\pi)^{d-1}}
  \frac{1}{2E_j}
  \mathrm{d}\Omega_{d-1}^{(j)}
  \qquad \mathrm{with}\quad j=3,4\,,
\end{align}
and the phase-space measures of massless particles are given by
\begin{align}
  \mathcal{D}p_j
  \equiv
  \frac{1}{2}
  \frac{p_j^{d-3}\mathrm{d}p_j}{(2\pi)^{d-1}}
  \mathrm{d}\Omega_{d-1}^{(j)}
  \qquad \mathrm{with}\quad j=5,6
  \,.
\end{align}
When necessary, we use a parametrization for the
momenta such that the scalar products take the form
\begin{align}
  q_1 \cdot p_5=\frac{s}{4} \kappa_5 (1-\cos \theta_5),&\quad
  q_2 \cdot p_5=\frac{s}{4} \kappa_5 (1+\cos \theta_5),
  \nonumber\\
  q_1 \cdot p_6=\frac{s}{4} \kappa_6 (1-\cos \theta_6),&\quad
  q_2 \cdot p_6=\frac{s}{4} \kappa_6 (1+\cos \theta_6)
  \,,
  \nonumber\\
  q_1 \cdot (p_5+p_6)=\frac{s}{4}\kappa_{56}(1-\cos\theta_{56}),&\quad
  q_2 \cdot (p_5+p_6)=\frac{s}{4}\kappa_{56}(1+\cos\theta_{56})
  \,,
\end{align}
where $q_i$ and $p_j$ are momenta of the initial and final
state, respectively.

In the following, we use the symbol ``$\approx$'' to denote that the leading-order
term in $\delta$ expansion agrees on the the left- and right-hand sides of the
equation, but that the sub-leading terms may differ.

For the $(d-1)$-dimensional angular integration,
we use the following property:
\begin{align}
  \int \mathrm{d} \Omega_{d-1}^{(j)}
  &=\frac{2 \pi^{\frac{d-3}{2}}}{\Gamma\left(\frac{d-3}{2}\right)} 
    \int_{-1}^{1}\left(1-\cos ^{2} \theta_{j}\right)^{\frac{d-4}{2}} 
    \mathrm{~d} \cos \theta_{j} 
    \int_{-1}^{1}\left(1-\cos ^{2} \phi_{j}\right)^{\frac{d-5}{2}} 
    \mathrm{~d} \cos \phi_{j}
    =\frac{2 \pi^{\frac{d-1}{2}}}{\Gamma\left(\frac{d-1}{2}\right)} 
    \,.
\end{align}


\subsubsection{Three-particle phase space}

All 17 three-particle cut master integrals can be written as
\begin{align}
  I_{i}^{(3)}
  =
  \left(\frac{e^{\gamma_{E}} \mu^{2}}{4 \pi}\right)^{3 \epsilon} 
  \int \mathcal{D} p_{3} \mathcal{D} p_{4} \mathcal{D} p_{5}
  \:(2 \pi)^{d}\: \delta^{(d)}\!\left(q_{1}+q_{2}-p_{3}-p_{4}-p_{5}\right)
  \mathcal{Q}_{i}^{(3)} 
  \,,
  \label{eq:3ps}
\end{align}
where $\mathcal{Q}_i^{(3)}$ are one-loop integrals, which can easily be read
off from Fig.~\ref{fig::3l3p}. For convenience we provide explicit
expressions in Appendix~\ref{app::Q3_Q4}.
Note that the $\mathcal{Q}^{(3)}_i$ do not depend
on $p_3$ and $p_4$, since the external momenta of the loop diagrams are $q_1, q_2, p_5$
and the combination $p_3+p_4$.  The latter can be eliminated by using
momentum conservation.
By performing the integration over $p_4$ and
taking the leading order in $\delta$,
we obtain
\begin{align}
I_i^{(3)}
\approx
\left(\frac{e^{\gamma_{E}} \mu^{2}}{4 \pi}\right)^{3 \epsilon}
\int \mathcal{D} p_{3} \mathcal{D} p_{5}
\frac{4\pi}{s}\:
\delta\!\left(\delta-\frac{\vec{p}_{3}^{\,\,2}}{m_H^2}-\kappa_{5}\right)
\mathcal{Q}_{i}^{(3)}
\end{align}
which indicates
\begin{align}
 |\vec{p}_3| \leq m_H\sqrt{\delta},\quad
\kappa_5 \leq \delta
\,.
\label{eq:kappa5}
\end{align}
There are six one-loop integrals appearing in the 17 master integrals.
Three of them ($L_1, L_2, L_3$)
are massless one-loop two-point functions and are known for a general dimension $d$:
\begin{align}
L_1&=
\int \frac{\mathrm{d}^d \ell}{i(2\pi)^d}
\:\frac{1}{-(\ell+p_5)^2}
\:\frac{1}{-(\ell+q_1+q_2)^2}
\approx
\frac{1}{(4\pi)^{2-\epsilon}}
\frac{\Gamma(1-\epsilon)^2\Gamma(\epsilon)}{\Gamma(2-2\epsilon)}
s^{-\epsilon}
e^{i\epsilon\pi}
\,,\nonumber\\
L_2&=
\int \frac{\mathrm{d}^d \ell}{i(2\pi)^d}
\:\frac{1}{-\ell^2}
\:\frac{1}{-(\ell+q_1+q_2)^2}
=
\frac{1}{(4\pi)^{2-\epsilon}}
\frac{\Gamma(1-\epsilon)^2\Gamma(\epsilon)}{\Gamma(2-2\epsilon)}
s^{-\epsilon}
e^{i\epsilon\pi}
\approx L_1
\,,\nonumber\\
L_3&=
\int \frac{\mathrm{d}^d \ell}{i(2\pi)^d}
\:\frac{1}{-(\ell+p_5)^2}
\:\frac{1}{-(\ell+q_1)^2}
\approx
\frac{1}{(4\pi)^{2-\epsilon}}
\frac{\Gamma(1-\epsilon)^2\Gamma(\epsilon)}{\Gamma(2-2\epsilon)}
\frac{2^\epsilon}{(s(1-\cos\theta_5)\kappa_5)^\epsilon }
\,.\label{eq::L123}
\end{align}
Note that in the soft limit
there are only five independent one-loop integrals 
since $L_2\approx L_1$.
The remaining three integrals ($L_4, L_5, L_6$)
are either triangle or box diagrams,
and their soft limit
can be computed 
using expansion by regions~\cite{Beneke:1997zp,Pak:2010pt,Jantzen:2012mw}.
In particular, we use the implementation in the \texttt{Mathematica}
package \texttt{asy2.m}~\cite{Jantzen:2012mw}.
Applying this method is straightforward but the details of the computation are rather technical;
we do not discuss them here but simply refer to Ref.~\cite{Jantzen:2012mw}.
Rather, in this paper, we discuss an alternative way to obtain the same results
by making use of the Mellin-Barnes representation,
which serves as a cross-check of the calculation based on expansion by regions.
In the following, it is assumed that the integration contours 
of the Mellin-Barnes integrals are chosen properly,
(see, e.g., Ref.~\cite{Smirnov:2006ry})
and we abbreviate the product of $\Gamma$-functions as
\begin{align}
\Gamma\left[x_{1}, \ldots, x_{n}\right] \equiv \prod_{i=1}^{n} \Gamma\left(x_{i}\right)
\,.
\end{align}
We begin by expressing the integrals in the Mellin-Barnes representation:
\begin{align}
L_4&=
\int \frac{\mathrm{d}^d \ell}{i(2\pi)^d}
\:\frac{1}{-(\ell+p_5)^2}
\:\frac{1}{-(\ell+q_1)^2}
\:\frac{1}{-(\ell+q_1+q_2)^2}
\\
&=
\frac{-e^{i \pi \epsilon}}{(4\pi)^{2-\epsilon}}
\int \mathrm{d}z_1\mathrm{d}z_2
\frac{\kappa_5 ^{z_1} e^{i \pi  z_1}
\Gamma [-z_2, z_2+1, -\epsilon , z_2-z_1, z_1+\epsilon +1, -z_2-\epsilon ]}
{2^{z_2} s^{\epsilon +1} 
(1-\cos\theta_5)^{-z_2} 
\Gamma (1-2 \epsilon )}
\,,\nonumber
\\
L_5&=
\int \frac{\mathrm{d}^d \ell}{i(2\pi)^d}
\:\frac{1}{-\ell^2}
\:\frac{1}{-(\ell+p_5)^2}
\:\frac{1}{-(\ell+q_2)^2}
\:\frac{1}{-(\ell+q_1+q_2)^2}
\nonumber\\
&=
\frac{e^{i \pi \epsilon}}{(4\pi)^{2-\epsilon}}
\int \mathrm{d}z_1\mathrm{d}z_2
\frac{
\kappa_5 ^{z_1} e^{i \pi  z_1} 
\Gamma [z_1+1, -z_2, z_2+1, -\epsilon , z_2-z_1, z_1+\epsilon +2, -z_2-\epsilon -1]}
{2^{z_2+1}  s^{\epsilon +2} 
(1+\cos\theta_5)^{-z_2} 
\Gamma [z_1+2, -2 \epsilon ]}
\,,\nonumber
\\
L_6&=
\int \frac{\mathrm{d}^d \ell}{i(2\pi)^d}
\:\frac{1}{-\ell^2}
\:\frac{1}{-(\ell+q_1)^2}
\:\frac{1}{-(\ell-p_5+q_1)^2}
\:\frac{1}{-(\ell-p_5+q_1+q_2)^2}
\nonumber\\
&=
\frac{e^{i \pi \epsilon}}{(4\pi)^{2-\epsilon}}
\int \mathrm{d}z_1\mathrm{d}z_2
\frac{
\kappa_5 ^{z_1} e^{i \pi  z_1} 
\Gamma [-z_2, -\epsilon ,-\epsilon ,z_2-z_1, z_1+\epsilon +2, -z_2-\epsilon -1 ]}
{2^{z_1+1}s^{\epsilon +2}  
(1+\cos\theta_5)^{-z_2} 
(1-\cos\theta_5)^{-z_1+z_2}
\Gamma [-2 \epsilon , -z_2-\epsilon ]}
\nonumber\\
&
\qquad \times \frac{\Gamma(z_2-z_1-\epsilon -1)}{\Gamma(-z_1+z_2-\epsilon)}
\,.\nonumber\label{eq::L456}
\end{align}
Note that the Mellin-Barnes representation 
of a given Feynman integral is not unique;
here we show the representations used in our calculation.
We choose them in such a way that the exponent of $\kappa_5$ is $z_1$.
From Eq.~\eqref{eq:kappa5}, we see that the leading-order term in $\delta$
is obtained from the leading-order term in $\kappa_5$.
We solve $z_1$ integrals above
by closing the integration contour in the right half plane
and taking the residues of the poles of $\Gamma$-functions.
For example, in $L_5$
only the poles of $\Gamma (z_2-z_1)$ contribute
and their locations are $z_1=z_2-n$ where $n=0,1,2,...$.
Thus the leading order in $\kappa_5$ corresponds to $z_1=z_2$ pole.
Keeping this in mind, we first solve the $z_2$ integral also
by closing the integration contour in its right half plane.
This time, the poles from $\Gamma (-z_2)$ and $\Gamma (-z_2-\epsilon -1)$ contribute.
Among these, the pole at $z_2=-\epsilon -1$ gives the smallest value of $z_2$,
so we consider the residue at this pole.
The resulting expression is now a one-dimensional integral over $z_1$,
and its leading-order term in $\kappa_5$ is obtained by
taking the residue at $z_1 (=z_2) =-\epsilon -1$.
In this way, we extract the leading-order term of $L_5$ in $\kappa_5$.
The calculation of $L_4$ and $L_6$ proceeds in a similar way.
For $L_4$, the two poles at $z_2=0$ and $z_2=-\epsilon$
contribute to the leading order and we have to take both of them into account.
For $L_6$, the leading-order term is obtained from the pole at
$z_1=z_2-\epsilon -1$ instead of $z_1=z_2$.
The results are given by
\begin{align}
L_4&\approx
\frac{1}{(4\pi)^{2-\epsilon}}
\frac{
\Gamma (1+\epsilon ) \Gamma (-\epsilon )^2 
\left(2^{\epsilon } (1-\cos\theta_5)^{-\epsilon }  \kappa_5 ^{-\epsilon } 
-e^{i \pi  \epsilon }
\right)}
{s^{\epsilon +1}\Gamma (1-2 \epsilon ) \Gamma (\epsilon +1)}
\,,\nonumber\\
L_5&\approx
\frac{-1}{(4\pi)^{2-\epsilon}}
\frac{2^{\epsilon } (\cos\theta_5+1)^{-\epsilon -1}  \kappa_5 ^{-\epsilon -1} \Gamma (-\epsilon )^2 \Gamma (\epsilon )}{s^{\epsilon +2}\Gamma (-2 \epsilon )}
\,,\nonumber\\
L_6&\approx
\frac{1}{(4\pi)^{2-\epsilon}}
\frac{2^{2 \epsilon +1} e^{-i \pi  \epsilon } \left(1-\cos\theta_5^2\right)^{-\epsilon -1} 
\kappa_5 ^{-2 (\epsilon +1)} \Gamma (-\epsilon )^3 \Gamma (\epsilon +1)^2}
{s^{\epsilon +2} \Gamma (-2 \epsilon )}
\,.
\end{align}

We follow Ref.~\cite{Davies:2019xzc} to perform the three-particle phase-space
integration and obtain the following results for the leading-$\delta$ contribution
of the master integrals:
\begin{align}
I^{(3)}_{1}&\approx\delta ^{5/2}\left[\frac{s}{7680 \pi ^5 \epsilon }-\frac{s (15 \log \delta-56+20 \log 2)}{38400 \pi ^5}+\mathcal{O}(\epsilon^{1})\right]\,,\nonumber\\
I^{(3)}_{2}&\approx\delta ^{5/2}\left[-\frac{s^2}{10240 \pi ^5 \epsilon }+\frac{s^2 (15 \log \delta-56+20 \log 2)}{51200 \pi ^5}+\mathcal{O}(\epsilon^{1})\right]\,,\nonumber\\
I^{(3)}_{3}&\approx\delta ^{5/2}\left[\frac{s}{7680 \pi ^5 \epsilon }-\frac{s (15 \log \delta-56+20 \log 2)}{38400 \pi ^5}+\mathcal{O}(\epsilon^{1})\right]\,,\nonumber\\
I^{(3)}_{4}&\approx\delta ^{5/2}\left[-\frac{s^2}{10240 \pi ^5 \epsilon }+\frac{s^2 (15 \log \delta-56+20 \log 2)}{51200 \pi ^5}+\mathcal{O}(\epsilon^{1})\right]\,,\nonumber\\
I^{(3)}_{5}&\approx\delta ^{5/2}\left[\frac{s}{7680 \pi ^5 \epsilon }-\frac{s (30 \log \delta-107+45 \log 2)}{57600 \pi ^5}+\mathcal{O}(\epsilon^{1})\right]\,,\nonumber\\
I^{(3)}_{6}&\approx\delta ^{5/2}\left[-\frac{s^2}{10240 \pi ^5 \epsilon }+\frac{s^2 (30 \log \delta-107+45 \log 2)}{76800 \pi ^5}+\mathcal{O}(\epsilon^{1})\right]\,,\nonumber\\
I^{(3)}_{7}&\approx\delta ^{5/2}\left[\frac{-15 \log \delta+46-30 \log 2}{115200 \pi ^5 \epsilon }+\mathcal{O}(\epsilon^{0})\right]\,,\nonumber\\
I^{(3)}_{8}&\approx\sqrt{\delta }\left[-\frac{1}{2048 \pi ^5 s^2 \epsilon ^4}+\frac{2 \log \delta-4+3 \log 2}{1024 \pi ^5 s^2 \epsilon ^3}+\mathcal{O}(\epsilon^{-2})\right]\,,\nonumber\\
I^{(3)}_{9}&\approx\sqrt{\delta }\left[\frac{3}{8192 \pi ^5 s \epsilon ^4}-\frac{3 (2 \log \delta-4+3 \log 2)}{4096 \pi ^5 s \epsilon ^3}+\mathcal{O}(\epsilon^{-2})\right]\,,\nonumber\\
I^{(3)}_{10}&\approx\sqrt{\delta }\left[\frac{1}{4096 \pi ^5 s \epsilon ^4}+\frac{-5 \log \delta+10-8 \log 2}{4096 \pi ^5 s \epsilon ^3}+\mathcal{O}(\epsilon^{-2})\right]\,,\nonumber\\
I^{(3)}_{11}&\approx\sqrt{\delta }\left[-\frac{3}{16384 \pi ^5 \epsilon ^4}+\frac{3 (5 \log \delta-10+8 \log 2)}{16384 \pi ^5 \epsilon ^3}+\mathcal{O}(\epsilon^{-2})\right]\,,\nonumber\\
I^{(3)}_{12}&\approx\delta ^{5/2}\left[-\frac{1}{5760 \pi ^5 \epsilon }+\frac{15 \log \delta-56+20 \log 2}{28800 \pi ^5}+\mathcal{O}(\epsilon^{1})\right]\,,\nonumber\\
I^{(3)}_{13}&\approx\delta ^{5/2}\left[-\frac{1}{5760 \pi ^5 \epsilon }+\frac{15 \log \delta-56+20 \log 2}{28800 \pi ^5}+\mathcal{O}(\epsilon^{1})\right]\,,\nonumber\\
I^{(3)}_{14}&\approx\delta ^{5/2}\left[-\frac{1}{5760 \pi ^5 \epsilon }+\frac{30 \log \delta-107+45 \log 2}{43200 \pi ^5}+\mathcal{O}(\epsilon^{1})\right]\,,\nonumber\\
I^{(3)}_{15}&\approx\delta ^{5/2}\left[\frac{15 \log \delta-46+30 \log 2}{86400 \pi ^5 s \epsilon }+\mathcal{O}(\epsilon^{0})\right]\,,\nonumber\\
I^{(3)}_{16}&\approx\sqrt{\delta }\left[\frac{1}{1536 \pi ^5 s^3 \epsilon ^4}-\frac{2 \log \delta-4+3 \log 2}{768 \pi ^5 s^3 \epsilon ^3}+\mathcal{O}(\epsilon^{-2})\right]\,,\nonumber\\
I^{(3)}_{17}&\approx\sqrt{\delta }\left[-\frac{1}{3072 \pi ^5 s^2 \epsilon
              ^4}+\frac{5 \log \delta-10+8 \log 2}{3072 \pi ^5 s^2 \epsilon
              ^3}+\mathcal{O}(\epsilon^{-2})\right]\,,
\label{eq:I3_1_17}
\end{align}
where for brevity for some integrals we have truncated the $\epsilon$ expansion
at $1/\epsilon^2$ and do not display the finite term,
and we set $\mu^2=s$. Note that the $\mu$ dependence can easily be
reconstructed by multiplying $(\mu^2/s)^{3\epsilon}$.
The complete set of $\epsilon$ orders needed
for our calculation and also additional terms in the $\delta$ expansion can be found
in the supplementary material of this paper~\cite{progdata}.
In order to fix the boundary
conditions of the differential equations discussed in Section~\ref{sec::canonical},
only the results for $I_1^{(3)}, \ldots, I_6^{(3)}$ and
$I_8^{(3)}, \ldots, I_{11}^{(3)}$ are needed. Therefore in practice, only
$L_1, L_2, L_3, L_5$ and $L_6$ need to be computed.  All other master
integrals in Eq.~(\ref{eq:I3_1_17}) and also the higher-order terms of the $\delta$
expansion serve as cross-check.


\subsubsection{Four-particle phase space}

The 57 four-particle phase-space master integrals can be parametrized as
\begin{align}
  I_{i}^{(4)}
  =
  \left(\frac{e^{\gamma_{E}} \mu^{2}}{4 \pi}\right)^{3 \epsilon} 
  \int \mathcal{D} p_{3} \mathcal{D} p_{4} \mathcal{D} p_{5} \mathcal{D} p_{6}
  (2 \pi)^{d} \delta^{(d)}\left(q_{1}+q_{2}-p_{3}-p_{4}-p_{5}-p_{6}\right) \mathcal{Q}_{i}^{(4)}
  \,,
  \label{eq:4ps}
\end{align}
where the soft limit all $Q_i^{(4)}$ are given in Appendix~\ref{app::Q3_Q4}.
We group the Higgs momenta ($p_3, p_4$) into $p_{HH} \equiv p_3+p_4$ and the parton momenta
($p_5, p_6$) into $p_{gg} \equiv p_5+p_6$ by inserting
\begin{align}
  &1=\int {\rm d}(m_{HH}^2)\: {\rm d}^{d-1}p_{HH}\: \frac{1}{2\sqrt{(\vec{p}_{HH})^2+m_{HH}^2}}
    \delta^{(d)} (p_{HH} -p_3-p_4)\,,
  \\
  &1=\int {\rm d}(m_{gg}^2)\: {\rm d}^{d-1}p_{gg}\: \frac{1}{2\sqrt{(\vec{p}_{gg})^2+m_{gg}^2}}
    \delta^{(d)} (p_{gg} -p_5-p_6)
    \,,
\end{align}
where $m_{HH}^2=p_{HH}^2$ and $m_{gg}^2=p_{gg}^2$.
The momentum assignments of $Q_i^{(4)}$ are chosen such that they do not depend
on $p_3, p_4$ or $p_{HH}$, which is possible because of the fact that $p_3$
and $p_4$ appear only in the combination $p_{HH}$, which can be
eliminated using momentum conservation.
The dependence of
$p_5, p_6$ and $p_{gg}$ is chosen such that $Q_i^{(4)}$ factorizes in these
variables, i.e., when a propagator contains $p_5+p_6$ we replace it with
$p_{gg}$ and in the remaining cases we keep $p_5$ and $p_6$.

With the help of this parametrisation, it is straightforward to perform the
phase-space integration over $p_3, p_4, p_5$ and $p_6$ using the well-known
trick of boosting to the centre-of-mass frames of $p_{HH}$ and $p_{gg}$.  The
resulting integrals can be evaluated in an expansion in $\delta$. The
leading-$\delta$ terms read
\begin{align}
I^{(4)}_{1}&\approx\delta ^{9/2}\left[\frac{s^2}{120960 \pi ^5}-\frac{s^2 \epsilon  (315 \log \delta-1126+504 \log 2)}{7620480 \pi ^5}+\mathcal{O}(\epsilon^{2})\right]\,,\nonumber\\
I^{(4)}_{2}&\approx\delta ^{9/2}\left[-\frac{s^3}{161280 \pi ^5}+\frac{s^3 \epsilon  (315 \log \delta-1126+504 \log 2)}{10160640 \pi ^5}+\mathcal{O}(\epsilon^{2})\right]\,,\nonumber\\
I^{(4)}_{3}&\approx\delta ^{9/2}\left[\frac{s^3}{483840 \pi ^5}-\frac{s^3 \epsilon  (315 \log \delta-1126+504 \log 2)}{30481920 \pi ^5}+\mathcal{O}(\epsilon^{2})\right]\,,\nonumber\\
I^{(4)}_{4}&\approx\delta ^{9/2}\left[-\frac{s}{90720 \pi ^5}+\frac{s \epsilon  (315 \log \delta-1126+504 \log 2)}{5715360 \pi ^5}+\mathcal{O}(\epsilon^{2})\right]\,,\nonumber\\
I^{(4)}_{5}&\approx\delta ^{9/2}\left[-\frac{s^2}{362880 \pi ^5}+\frac{s^2 \epsilon  (315 \log \delta-1126+504 \log 2)}{22861440 \pi ^5}+\mathcal{O}(\epsilon^{2})\right]\,,\nonumber\\
I^{(4)}_{6}&\approx\delta ^{7/2}\left[-\frac{1}{26880 \pi ^5}+\frac{\epsilon  (105 \log \delta-394+168 \log 2)}{564480 \pi ^5}+\mathcal{O}(\epsilon^{2})\right]\,,\nonumber\\
I^{(4)}_{7}&\approx\delta ^{7/2}\left[\frac{s}{35840 \pi ^5}-\frac{s \epsilon  (105 \log \delta-394+168 \log 2)}{752640 \pi ^5}+\mathcal{O}(\epsilon^{2})\right]\,,\nonumber\\
I^{(4)}_{8}&\approx\delta ^{9/2}\left[-\frac{s}{120960 \pi ^5}+\frac{s \epsilon  (315 \log \delta-1189+504 \log 2)}{7620480 \pi ^5}+\mathcal{O}(\epsilon^{2})\right]\,,\nonumber\\
I^{(4)}_{9}&\approx\delta ^{7/2}\left[\frac{1}{20160 \pi ^5 s}+\frac{\epsilon  (-105 \log \delta+394-168 \log 2)}{423360 \pi ^5 s}+\mathcal{O}(\epsilon^{2})\right]\,,\nonumber\\
I^{(4)}_{10}&\approx\delta ^{9/2}\left[\frac{1}{90720 \pi ^5}+\frac{\epsilon  (-315 \log \delta+1189-504 \log 2)}{5715360 \pi ^5}+\mathcal{O}(\epsilon^{2})\right]\,,\nonumber\\
I^{(4)}_{11}&\approx\sqrt{\delta }\left[\frac{1}{8192 \pi ^5 s^2 \epsilon ^4}+\frac{-5 \log \delta+10-8 \log 2}{8192 \pi ^5 s^2 \epsilon ^3}+\mathcal{O}(\epsilon^{-2})\right]\,,\nonumber\\
I^{(4)}_{12}&\approx\sqrt{\delta }\left[-\frac{3}{32768 \pi ^5 s \epsilon ^4}+\frac{3 (5 \log \delta-10+8 \log 2)}{32768 \pi ^5 s \epsilon ^3}+\mathcal{O}(\epsilon^{-2})\right]\,,\nonumber\\
I^{(4)}_{13}&\approx\sqrt{\delta }\left[-\frac{1}{6144 \pi ^5 s^3 \epsilon ^4}+\frac{5 \log \delta-10+8 \log 2}{6144 \pi ^5 s^3 \epsilon ^3}+\mathcal{O}(\epsilon^{-2})\right]\,,\nonumber\\
I^{(4)}_{14}&\approx\delta ^{5/2}\left[\frac{1}{7680 \pi ^5}+\frac{\epsilon  (-15 \log \delta+58-24 \log 2)}{23040 \pi ^5}+\mathcal{O}(\epsilon^{2})\right]\,,\nonumber\\
I^{(4)}_{15}&\approx\delta ^{5/2}\left[-\frac{s}{10240 \pi ^5}+\frac{s \epsilon  (15 \log \delta-58+24 \log 2)}{30720 \pi ^5}+\mathcal{O}(\epsilon^{2})\right]\,,\nonumber\\
I^{(4)}_{16}&\approx\delta ^{5/2}\left[-\frac{1}{5760 \pi ^5 s}+\frac{\epsilon  (15 \log \delta-58+24 \log 2)}{17280 \pi ^5 s}+\mathcal{O}(\epsilon^{2})\right]\,,\nonumber\\
I^{(4)}_{17}&\approx\sqrt{\delta }\left[\frac{1}{8192 \pi ^5 s^2 \epsilon ^4}+\frac{-5 \log \delta+10-8 \log 2}{8192 \pi ^5 s^2 \epsilon ^3}+\mathcal{O}(\epsilon^{-2})\right]\,,\nonumber\\
I^{(4)}_{18}&\approx\sqrt{\delta }\left[-\frac{3}{32768 \pi ^5 s \epsilon ^4}+\frac{3 (5 \log \delta-10+8 \log 2)}{32768 \pi ^5 s \epsilon ^3}+\mathcal{O}(\epsilon^{-2})\right]\,,\nonumber\\
I^{(4)}_{19}&\approx\sqrt{\delta }\left[-\frac{1}{6144 \pi ^5 s^3 \epsilon ^4}+\frac{5 \log \delta-10+8 \log 2}{6144 \pi ^5 s^3 \epsilon ^3}+\mathcal{O}(\epsilon^{-2})\right]\,,\nonumber\\
I^{(4)}_{20}&\approx\delta ^{5/2}\left[\frac{1}{7680 \pi ^5 \epsilon ^2}+\frac{-15 \log \delta+46-24 \log 2}{23040 \pi ^5 \epsilon }+\mathcal{O}(\epsilon^{0})\right]\,,\nonumber\\
I^{(4)}_{21}&\approx\delta ^{5/2}\left[-\frac{1}{5760 \pi ^5 s \epsilon ^2}+\frac{15 \log \delta-46+24 \log 2}{17280 \pi ^5 s \epsilon }+\mathcal{O}(\epsilon^{0})\right]\,,\nonumber\\
I^{(4)}_{22}&\approx\sqrt{\delta }\left[\frac{1}{2048 \pi ^5 s^2 \epsilon ^4}+\frac{-5 \log \delta+10-8 \log 2}{2048 \pi ^5 s^2 \epsilon ^3}+\mathcal{O}(\epsilon^{-2})\right]\,,\nonumber\\
I^{(4)}_{23}&\approx\sqrt{\delta }\left[-\frac{3}{8192 \pi ^5 s \epsilon ^4}+\frac{3 (5 \log \delta-10+8 \log 2)}{8192 \pi ^5 s \epsilon ^3}+\mathcal{O}(\epsilon^{-2})\right]\,,\nonumber\\
I^{(4)}_{24}&\approx\sqrt{\delta }\left[-\frac{1}{1536 \pi ^5 s^3 \epsilon ^4}+\frac{5 \log \delta-10+8 \log 2}{1536 \pi ^5 s^3 \epsilon ^3}+\mathcal{O}(\epsilon^{-2})\right]\,,\nonumber\\
I^{(4)}_{25}&\approx\delta ^{9/2}\left[\frac{1}{120960 \pi ^5}+\frac{\epsilon  (-315 \log \delta+1126-504 \log 2)}{7620480 \pi ^5}+\mathcal{O}(\epsilon^{2})\right]\,,\nonumber\\
I^{(4)}_{26}&\approx\delta ^{9/2}\left[-\frac{s}{161280 \pi ^5}+\frac{s \epsilon  (315 \log \delta-1126+504 \log 2)}{10160640 \pi ^5}+\mathcal{O}(\epsilon^{2})\right]\,,\nonumber\\
I^{(4)}_{27}&\approx\delta ^{7/2}\left[\frac{1}{26880 \pi ^5 \epsilon }+\frac{-105 \log \delta+352-168 \log 2}{564480 \pi ^5}+\mathcal{O}(\epsilon^{1})\right]\,,\nonumber\\
I^{(4)}_{28}&\approx\delta ^{9/2}\left[-\frac{1}{90720 \pi ^5 s}+\frac{\epsilon  (315 \log \delta-1126+504 \log 2)}{5715360 \pi ^5 s}+\mathcal{O}(\epsilon^{2})\right]\,,\nonumber\\
I^{(4)}_{29}&\approx\delta ^{7/2}\left[-\frac{1}{20160 \pi ^5 s \epsilon }+\frac{105 \log \delta-352+168 \log 2}{423360 \pi ^5 s}+\mathcal{O}(\epsilon^{1})\right]\,,\nonumber\\
I^{(4)}_{30}&\approx\delta ^{5/2}\left[\frac{1}{7680 \pi ^5 s^2 \epsilon ^2}+\frac{-15 \log \delta+46-24 \log 2}{23040 \pi ^5 s^2 \epsilon }+\mathcal{O}(\epsilon^{0})\right]\,,\nonumber\\
I^{(4)}_{31}&\approx\delta ^{5/2}\left[-\frac{1}{10240 \pi ^5 s \epsilon ^2}+\frac{15 \log \delta-46+24 \log 2}{30720 \pi ^5 s \epsilon }+\mathcal{O}(\epsilon^{0})\right]\,,\nonumber\\
I^{(4)}_{32}&\approx\delta ^{5/2}\left[-\frac{1}{5760 \pi ^5 s^3 \epsilon ^2}+\frac{15 \log \delta-46+24 \log 2}{17280 \pi ^5 s^3 \epsilon }+\mathcal{O}(\epsilon^{0})\right]\,,\nonumber\\
I^{(4)}_{33}&\approx\delta ^{5/2}\left[\frac{1}{7680 \pi ^5 s^2 \epsilon ^2}+\frac{-15 \log \delta+46-24 \log 2}{23040 \pi ^5 s^2 \epsilon }+\mathcal{O}(\epsilon^{0})\right]\,,\nonumber\\
I^{(4)}_{34}&\approx\delta ^{5/2}\left[-\frac{1}{5760 \pi ^5 s^3 \epsilon ^2}+\frac{15 \log \delta-46+24 \log 2}{17280 \pi ^5 s^3 \epsilon }+\mathcal{O}(\epsilon^{0})\right]\,,\nonumber\\
I^{(4)}_{35}&\approx\delta ^{7/2}\left[\frac{1}{26880 \pi ^5 \epsilon }+\frac{-105 \log \delta+352-168 \log 2}{564480 \pi ^5}+\mathcal{O}(\epsilon^{1})\right]\,,\nonumber\\
I^{(4)}_{36}&\approx\delta ^{9/2}\left[\frac{1}{120960 \pi ^5}+\frac{\epsilon  (-315 \log \delta+1126-504 \log 2)}{7620480 \pi ^5}+\mathcal{O}(\epsilon^{2})\right]\,,\nonumber\\
I^{(4)}_{37}&\approx\delta ^{9/2}\left[-\frac{s}{161280 \pi ^5}+\frac{s \epsilon  (315 \log \delta-1126+504 \log 2)}{10160640 \pi ^5}+\mathcal{O}(\epsilon^{2})\right]\,,\nonumber\\
I^{(4)}_{38}&\approx\delta ^{7/2}\left[-\frac{1}{20160 \pi ^5 s \epsilon }+\frac{105 \log \delta-352+168 \log 2}{423360 \pi ^5 s}+\mathcal{O}(\epsilon^{1})\right]\,,\nonumber\\
I^{(4)}_{39}&\approx\delta ^{9/2}\left[-\frac{1}{90720 \pi ^5 s}+\frac{\epsilon  (315 \log \delta-1126+504 \log 2)}{5715360 \pi ^5 s}+\mathcal{O}(\epsilon^{2})\right]\,,\nonumber\\
I^{(4)}_{40}&\approx\delta ^{5/2}\left[\frac{1}{7680 \pi ^5 s^2 \epsilon ^2}+\frac{-15 \log \delta+46-24 \log 2}{23040 \pi ^5 s^2 \epsilon }+\mathcal{O}(\epsilon^{0})\right]\,,\nonumber\\
I^{(4)}_{41}&\approx\delta ^{5/2}\left[-\frac{1}{10240 \pi ^5 s \epsilon ^2}+\frac{15 \log \delta-46+24 \log 2}{30720 \pi ^5 s \epsilon }+\mathcal{O}(\epsilon^{0})\right]\,,\nonumber\\
I^{(4)}_{42}&\approx\delta ^{5/2}\left[-\frac{1}{5760 \pi ^5 s^3 \epsilon ^2}+\frac{15 \log \delta-46+24 \log 2}{17280 \pi ^5 s^3 \epsilon }+\mathcal{O}(\epsilon^{0})\right]\,,\nonumber\\
I^{(4)}_{43}&\approx\delta ^{5/2}\left[\frac{1}{7680 \pi ^5 s^2 \epsilon ^2}+\frac{-15 \log \delta+46-24 \log 2}{23040 \pi ^5 s^2 \epsilon }+\mathcal{O}(\epsilon^{0})\right]\,,\nonumber\\
I^{(4)}_{44}&\approx\delta ^{5/2}\left[-\frac{1}{5760 \pi ^5 s^3 \epsilon ^2}+\frac{15 \log \delta-46+24 \log 2}{17280 \pi ^5 s^3 \epsilon }+\mathcal{O}(\epsilon^{0})\right]\,,\nonumber\\
I^{(4)}_{45}&\approx\delta ^{7/2}\left[\frac{1}{26880 \pi ^5 \epsilon }+\frac{-105 \log \delta+352-168 \log 2}{564480 \pi ^5}+\mathcal{O}(\epsilon^{1})\right]\,,\nonumber\\
I^{(4)}_{46}&\approx\delta ^{5/2}\left[-\frac{1}{15360 \pi ^5 s \epsilon ^2}+\frac{15 \log \delta-46+24 \log 2}{46080 \pi ^5 s \epsilon }+\mathcal{O}(\epsilon^{0})\right]\,,\nonumber\\
I^{(4)}_{47}&\approx\delta ^{5/2}\left[\frac{1}{20480 \pi ^5 \epsilon ^2}+\frac{-15 \log \delta+46-24 \log 2}{61440 \pi ^5 \epsilon }+\mathcal{O}(\epsilon^{0})\right]\,,\nonumber\\
I^{(4)}_{48}&\approx\delta ^{7/2}\left[-\frac{1}{20160 \pi ^5 s \epsilon }+\frac{105 \log \delta-352+168 \log 2}{423360 \pi ^5 s}+\mathcal{O}(\epsilon^{1})\right]\,,\nonumber\\
I^{(4)}_{49}&\approx\delta ^{5/2}\left[\frac{1}{11520 \pi ^5 s^2 \epsilon ^2}+\frac{-15 \log \delta+46-24 \log 2}{34560 \pi ^5 s^2 \epsilon }+\mathcal{O}(\epsilon^{0})\right]\,,\nonumber\\
I^{(4)}_{50}&\approx\delta ^{3/2}\left[\frac{1}{6144 \pi ^5 s^2 \epsilon ^3}+\frac{-15 \log \delta+40-24 \log 2}{18432 \pi ^5 s^2 \epsilon ^2}+\mathcal{O}(\epsilon^{-1})\right]\,,\nonumber\\
I^{(4)}_{51}&\approx\delta ^{3/2}\left[-\frac{1}{4608 \pi ^5 s^3 \epsilon ^3}+\frac{15 \log \delta+8 (3 \log 2-5)}{13824 \pi ^5 s^3 \epsilon ^2}+\mathcal{O}(\epsilon^{-1})\right]\,,\nonumber\\
I^{(4)}_{52}&\approx\sqrt{\delta }\left[\frac{1}{8192 \pi ^5 s^2 \epsilon ^4}+\frac{-5 \log \delta+10-8 \log 2}{8192 \pi ^5 s^2 \epsilon ^3}+\mathcal{O}(\epsilon^{-2})\right]\,,\nonumber\\
I^{(4)}_{53}&\approx\sqrt{\delta }\left[-\frac{3}{32768 \pi ^5 s \epsilon ^4}+\frac{3 (5 \log \delta-10+8 \log 2)}{32768 \pi ^5 s \epsilon ^3}+\mathcal{O}(\epsilon^{-2})\right]\,,\nonumber\\
I^{(4)}_{54}&\approx\sqrt{\delta }\left[-\frac{1}{6144 \pi ^5 s^3 \epsilon ^4}+\frac{5 \log \delta-10+8 \log 2}{6144 \pi ^5 s^3 \epsilon ^3}+\mathcal{O}(\epsilon^{-2})\right]\,,\nonumber\\
I^{(4)}_{55}&\approx\sqrt{\delta }\left[-\frac{3}{8192 \pi ^5 s \epsilon ^4}-\frac{3 (-5 \log \delta+10-8 \log 2)}{8192 \pi ^5 s \epsilon ^3}+\mathcal{O}(\epsilon^{-2})\right]\,,\nonumber\\
I^{(4)}_{56}&\approx\sqrt{\delta }\left[\frac{9}{32768 \pi ^5 \epsilon ^4}+\frac{9 (-5 \log \delta+10-8 \log 2)}{32768 \pi ^5 \epsilon ^3}+\mathcal{O}(\epsilon^{-2})\right]\,,\nonumber\\
I^{(4)}_{57}&\approx\sqrt{\delta }\left[\frac{1}{2048 \pi ^5 s^2 \epsilon ^4}+\frac{-5 \log \delta+10-8 \log 2}{2048 \pi ^5 s^2 \epsilon ^3}+\mathcal{O}(\epsilon^{-2})\right]\,.
\label{eq::BC_4PC}
\end{align}
As in Eq.~(\ref{eq:I3_1_17}) the $\epsilon$ expansion of some integrals has been
truncated before the finite term can be displayed,
and we set $\mu^2=s$.
We provide higher order $\epsilon$ and additional terms
in the $\delta$ expansion in the ancillary files~\cite{progdata}.
Of the 57 boundary conditions given in Eq.~(\ref{eq::BC_4PC}) we only require
the information from the following integrals, and a dedicated calculation is only
necessary for one integral of each set:
$\{I^{(4)}_{1}, I^{(4)}_{2}, I^{(4)}_{3}\}$, 
$\{I^{(4)}_{11}, I^{(4)}_{12}\}$, 
$\{I^{(4)}_{17}, I^{(4)}_{18}\}$, 
$\{I^{(4)}_{22}, I^{(4)}_{23}\}$, 
$\{I^{(4)}_{52}, I^{(4)}_{53}\}$, 
$\{I^{(4)}_{55}, I^{(4)}_{56}\}$.
The remaining integrals of the sets only differ
by a trivial factor which is obtained from a numerator factor in the soft limit.
The remaining 44 integrals of Eq.~(\ref{eq::BC_4PC}) serve as consistency and cross check.



\subsection{\label{sec::expansion}Deep $\delta$ expansion of the master integrals}

In this subsection, we briefly mention a method to obtain for the master
integrals higher order terms in the $\delta$ expansion with the help of
differential equations.  This method is used also in
Refs.~\cite{Melnikov:2016qoc,Davies:2018ood,Mishima:2018olh,Davies:2019xzc}.
After changing from $x$ to $\delta$ the differential equations in
Eq~\eqref{eq::de1} can be written as
\begin{flalign}
  \partial_\delta \vec{L} = M\!\left(\delta,\epsilon\right)\,\vec{L}~.\label{eq::dedel1}
\end{flalign}
We are interested in $\vec{L}$ as a series expansion in $\delta$ and
$\epsilon$ which has the form
\begin{flalign}
  \vec{L} = \sum_{n_1,n_2,n_3}
  \vec{c}_{n_1,n_2,n_3}
  \delta^{n_1} (\log \delta)^{n_2} \epsilon^{n_3}\,,
\label{eq::ansatz}
\end{flalign}
where the coefficients $\vec{c}_{n_1,n_2,n_3}$ are constants containing
$\zeta_2, \zeta_3, \zeta_4$ and $\log 2$.  The matrix
$M\!\left(\delta,\epsilon\right)$ in Eq.~\eqref{eq::dedel1} consists of
rational functions of $\delta$ and $\epsilon$ and it can be expanded in both
variables and truncated at a certain order.  By substituting the ansatz in
Eq.~\eqref{eq::ansatz} into Eq.~\eqref{eq::dedel1} we obtain a set of
recurrence relations for the coefficients $\vec{c}_{n_1,n_2,n_3}$ where the
initial conditions are obtained by the leading terms calculated in the
previous subsection.  Thus, by solving the recurrence relations, it is
possible to obtain the deep $\delta$ expansion of the master integrals
efficiently.


\subsection{Comparison of exact and $\delta$-expanded phase-space master integrals}

\begin{figure}[t]
  \begin{center}
    \begin{tabular}{cc}
      \hspace{-4mm}
      \includegraphics[width=0.51\textwidth]{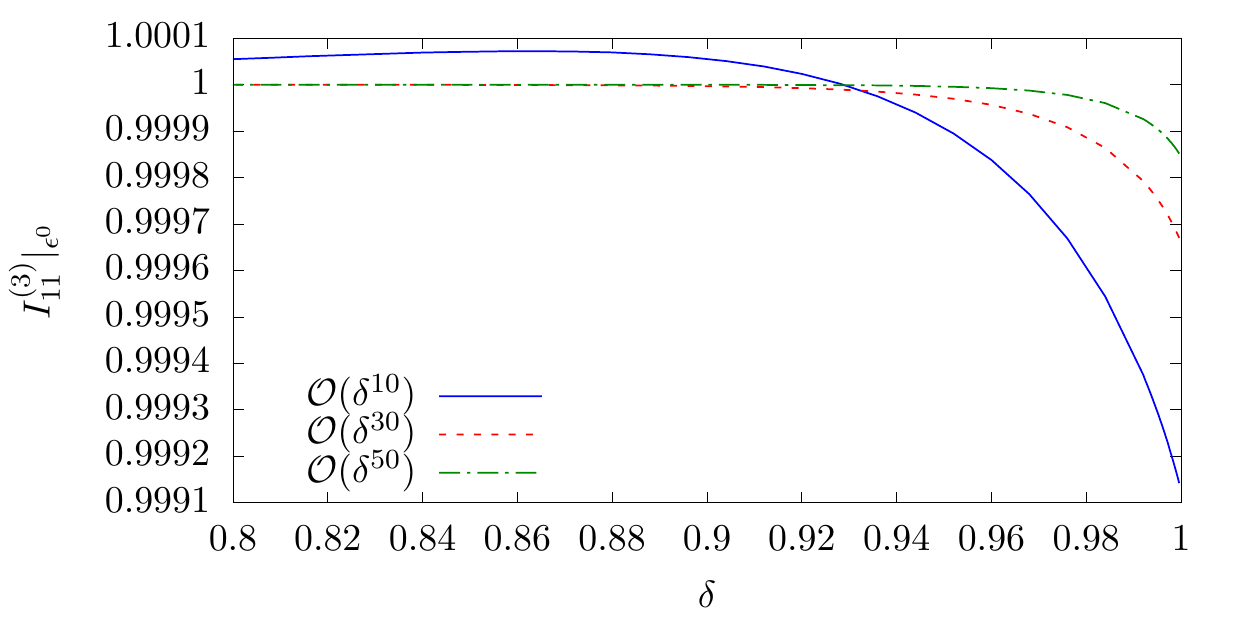}
      &
      \hspace{-6mm}
      \includegraphics[width=0.51\textwidth]{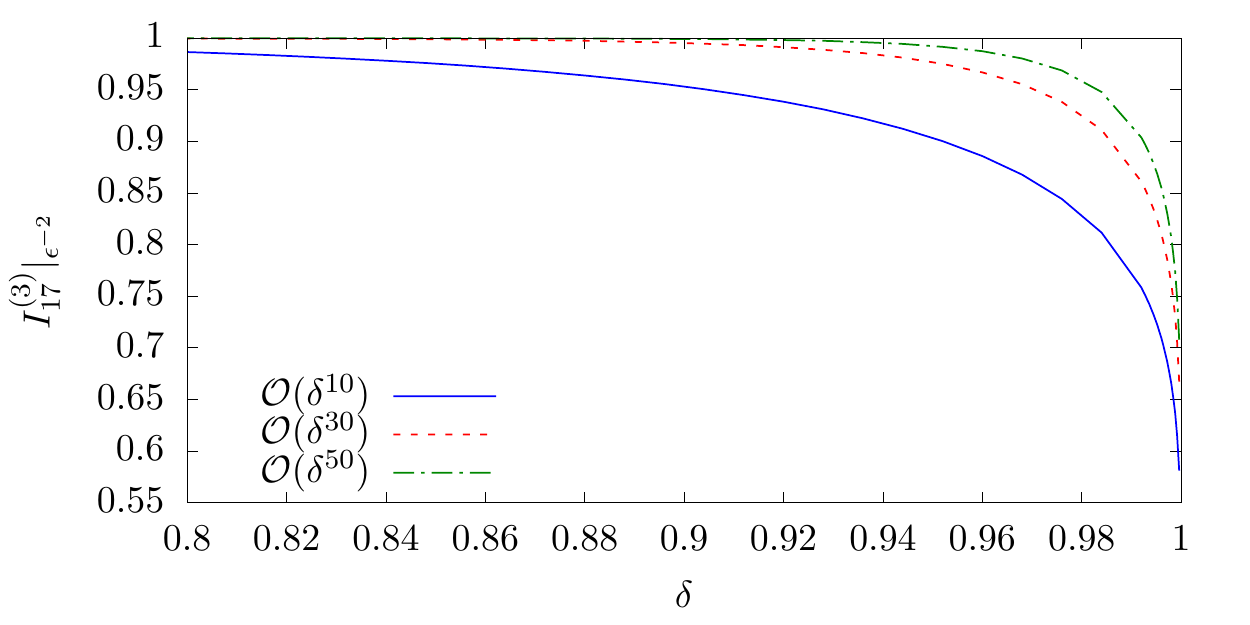}
      \\
      \hspace{-4mm}
      \includegraphics[width=0.51\textwidth]{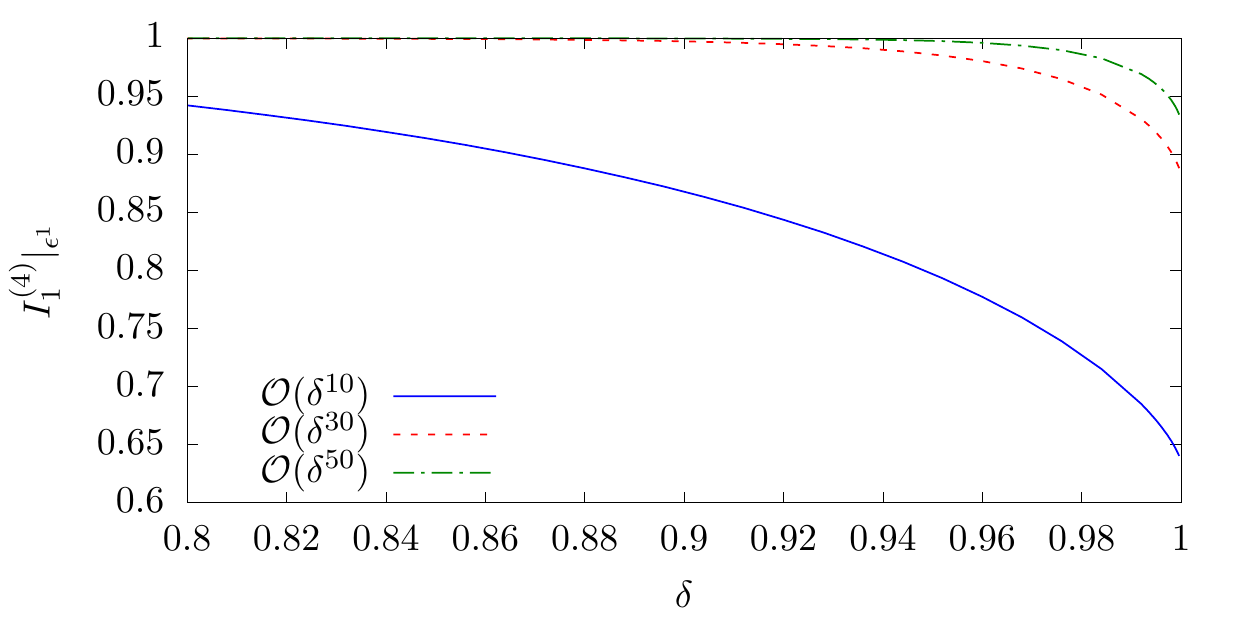}
      &
      \hspace{-6mm}
      \includegraphics[width=0.51\textwidth]{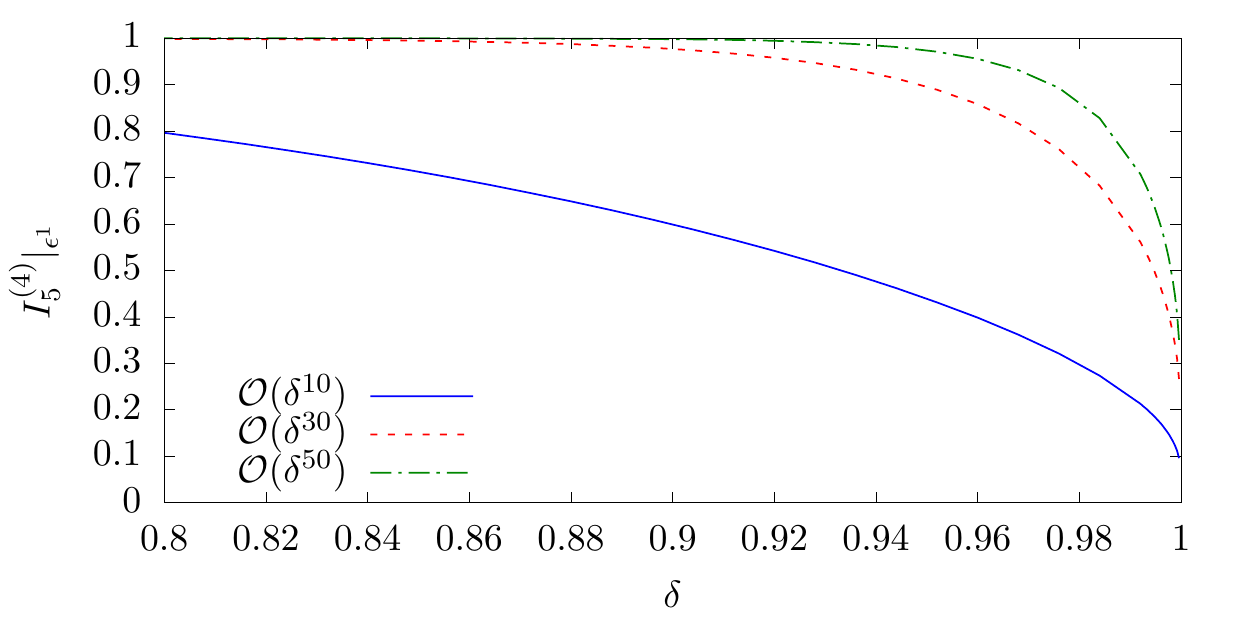}
    \end{tabular}
    \caption{\label{fig::MIexp}Comparison of exact and expanded phase-space
      master integrals for four typical examples, where the results are
      normalized to the exact expressions.  The panel in the top row
      correspond to three-particle cut master integrals and the corresponding
      pictures can be found in Fig.~\ref{fig::3l3p}.  The pictures for the
      four-particle cut master integrals in the bottom row can found in
      Fig.~\ref{fig::3l4p}.  The considered $\epsilon$ order is indicated on
      the $y$-axis labels.}
  \end{center}
\end{figure}  

Following the approach described in the previous subsections we were able to
obtain analytic results for all three- and four-particle cut master integrals,
which are exact in $s$ and $m_H^2$. However, the expressions are quite large (in
some cases of the order of several MB) and contain GPLs which have to
be evaluated at some fixed value of their argument. We did not make any
effort to obtain minimal expressions, since for our application it
is sufficient to consider an expansion in the limit $\delta\to0$. This can be
obtained in a straightforward way by either expanding the exact results or
by using the boundary conditions of Section~\ref{sec::boundaries} in
combination with the differential equations.

In Fig.~\ref{fig::MIexp}, for four typical examples of
$\epsilon$ coefficients of master integrals, we compare the $\delta$-expanded and exact
results. We normalize the curves to the exact expressions and plot the expansions
as functions of $\delta$. In each panel three curves are
shown, which correspond to approximations including 10, 30 and 50 terms of
the $\delta$ expansion. All curves tend to 1 for $\delta\to0$, so the plots
focus on the region $0.8<\delta<1$ where the expansions start to diverge from
the exact results.

The $\epsilon^0$ term of the three-particle cut integral $I^{(11)}$
(the first panel of Fig.~\ref{fig::MIexp})
is an example for which as few as 10 $\delta$-expansion terms provide
an excellent approximation.
Even for $\delta=1$ the deviation from the exact result is at the level of 0.01\%.
Including additional expansion terms leads to further improvement.
For the other three examples shown, we observe that the $\delta^{10}$ curves show
deviations from the exact result at the percent level, even for $\delta$ values
as small as 0.8.
The approximations including expansion terms to order $\delta^{30}$ and $\delta^{50}$
show percent-level agreement with the exact results for $\delta$ values up to around
0.9 and 0.95, respectively.

In our final expression for the cross sections, we use the expansions up to $\delta^{30}$;
from a practical point of view this is sufficient, since our large-$m_t$ approximation
is only valid below the top threshold at $s=4m_t^2$. This corresponds to
$\delta = 1-4m_H^2/m_t^2 \approx 0.48$, 
where the deviation of the expansions up to $\delta^{30}$ from the exact results
is below $0.0001\%$ for all master integrals. On the other hand, the
expansions up to $\delta^{10}$ show deviations of several percent.



\section{Results\label{sec::res}}

We are now in a position to combine the virtual, real-radiation and
collinear counterterm corrections according to Eq.~(\ref{eq::sig_tot}).  It is
interesting to check their contributions to the cancellation of the poles.  In
the $gg$ channel the highest-order poles, i.e. the $1/\epsilon^4$ and $1/\epsilon^3$
terms, receive contributions only from the virtual and real-radiation
corrections. Starting from $1/\epsilon^2$ the collinear counterterm
also contributes. In fact, there are two sources which can be identified from
Eq.~(\ref{eq::Gamma_ij}): either the term proportional to $1/\epsilon^2$ is
convoluted with the LO cross section, or there are two convolutions of the
one-loop splitting functions in the ${\cal O}(\alpha_s)$ term, again with the
LO cross section.  The convolutions involving the two-loop splitting function and
the convolutions of the one-loop splitting function with NLO partonic cross
sections develop only $1/\epsilon$ poles.

For the partonic channels with quarks or anti-quarks in the initial state
there are no virtual corrections.\footnote{With the exception of the finite
  virtual correction in the $q\bar{q}$ channel discussed in
  Appendix~\ref{app::qqHH}.}  As a consequence the renormalized real-radiation
contribution develops at most $1/\epsilon^2$ poles which cancel against those
of the collinear counterterm contributions.

In the following we present analytic results for the renormalized partonic
NNLO cross sections for all five channels. We restrict the expressions to the two
leading terms in the $\delta$ expansion and we present only results up to
order $1/m_t^2$ (or, $\rho^1$). To obtain compact expressions we set $\mu^2=m_H^2$.
The top quark mass is renormalized in the on-shell scheme.
In the supplementary material to this paper~\cite{progdata}
we provide expansions for general renormalization scale $\mu$ 
up to $\delta^{30}$ and $1/m_t^6$ ($\rho^3$) (and for the quark-induced
channels up to $1/m_t^8$ ($\rho^4$)) for both on-shell and $\overline{\rm MS}$ top
quark masses.
For the $gg$ channel our result is as follows:
\begin{align}
  \sigma_{gg}^{(2),n_h^2} &= \frac{a_s^4 G_F^2
                            m_H^2}{\pi}\Bigg\{\delta^{3/2}\Bigg(\frac{19}{6912}
                            + \frac{n_l}{1296}\Bigg) +
                            \rho\Bigg[\sqrt{\delta}\Bigg(\frac{133}{552960} +
                            \frac{7}{103680}n_l\Bigg)
                            \nonumber\\&{}\vphantom{\Bigg(}
                            + \delta^{3/2}\Bigg(\frac{613614937}{398131200} -
                            \frac{695}{1492992}\log\rho
                            - \frac{121217}{51840}\log 2
                            + \frac{1964}{1215}\log^2(2)
                            \nonumber\\&{}\vphantom{\Bigg(}
                            - \frac{17129}{25920}\log^3(2)
                            + \frac{7}{45}\log^4(2)
                            - \frac{7753\pi^2}{103680}
                            + \frac{18515\pi^2}{186624}\log 2 -
                            \frac{7\pi^2}{135}\log^2(2)
                            \nonumber\\&{}\vphantom{\Bigg(}
                            - \frac{931\pi^4}{829440}  -
                            \frac{171070393}{477757440}\zeta_3
                            + \frac{1211}{2880}\zeta_3\log 2 + \log\delta
                            \Bigg\{-\frac{363953}{466560} +
                            \frac{29009}{25920}\log 2
                            \nonumber\\&{}\vphantom{\Bigg(}
                            -\frac{12341}{17280}\log^2(2) +
                            \frac{7}{30}\log^3(2)
                            + \frac{3479\pi^2}{103680} -
                            \frac{203\pi^2}{5760}\log 2 +
                            \frac{539}{3840}\zeta_3\Bigg\}
                            \nonumber\\&{}\vphantom{\Bigg(}
                            +\log^2(\delta)\Bigg\{\frac{29009}{155520} -
                            \frac{1421}{5760}\log 2 +
                            \frac{91}{720}\log^2(2)
                            - \frac{203\pi^2}{34560}\Bigg\}
                            + \log^3(\delta)\Bigg\{-\frac{1421}{51840}
                           \nonumber\\&{}\vphantom{\Bigg(}
                            + \frac{7}{240}\log 2\Bigg\} +
                            \frac{7}{2880}\log^4(\delta) +
                            n_l\Bigg\{-\frac{1067831}{111974400}
                            - \frac{533}{22394880}\log\rho
                           \nonumber\\&{}\vphantom{\Bigg(}
                            + \frac{50359}{3732480}\log 2 -
                            \frac{161}{19440}\log^2(2) +
                            \frac{91}{38880}\log^3(2) +
                            \frac{119\pi^2}{311040}
                            - \frac{49\pi^2}{155520}\log 2
                           \nonumber\\&{}\vphantom{\Bigg(}
                            + \frac{371}{311040}\zeta_3 +
                            \log\delta\Bigg[\frac{1057}{233280} -
                            \frac{49}{8640}\log 2 + \frac{7}{2880}\log^2(2) -
                            \frac{7\pi^2}{51840}\Bigg]
                           \nonumber\\&{}\vphantom{\Bigg(}
                            + \log^2(\delta)\Bigg[-\frac{49}{51840} +
                            \frac{7}{8640}\log 2\Bigg] +
                            \frac{7}{77760}\log^3(\delta) \Bigg\}\Bigg)\Bigg]
                            + \mathcal{O}(\delta^{5/2}) +
                            \mathcal{O}(\rho^2)\Bigg\} 
                            \label{eq::sig2ggnh2}
                          \,.
\end{align}
Note that the subset of the $n_h^3$ real-radiation terms with a closed top quark loop which
has no coupling to Higgs bosons only contribute to higher-order $\delta$ terms not shown in
Eq.~(\ref{eq::sig2ggnh2}).
The $n_h^4$ terms (see Eq.~(\ref{eq::ggnh4}) in Appendix~\ref{app::nh4}) are not included
in this expression.

The result for the $gq$ channel is given by
\begin{align}
\sigma_{gq}^{(2),n_h^2} &= \frac{a_s^4 G_F^2 m_H^2}{\pi}\Bigg\{\rho\delta^{5/2}\Bigg(-\frac{259448903}{7873200000} + \frac{3096503}{65610000}\log 2 - \frac{170443}{5832000}\log^2(2)
                        \nonumber\\&\vphantom{\Bigg)}+ \frac{7777}{874800}\log^3(2) + \frac{553\pi^2}{432000} - \frac{1099\pi^2}{874800}\log 2 + \frac{4151}{777600}\zeta_3 + \log\delta\Bigg[\frac{8090963}{524880000}
                        \nonumber\\&\vphantom{\Bigg)}- \frac{346577}{17496000}\log 2 + \frac{623}{64800}\log^2(2) - \frac{973\pi^2}{2332800}\Bigg] + \log^2(\delta)\Bigg[-\frac{227423}{69984000}
                        \nonumber\\&\vphantom{\Bigg)}+ \frac{3871}{1166400}\log 2\Bigg] + \frac{2569}{6998400}\log^3(\delta) + n_l\Bigg[\frac{42287}{262440000} - \frac{973}{4374000}\log 2
                        \nonumber\\&\vphantom{\Bigg)}+
  \frac{49}{583200}\log^2(2) { - \frac{7\pi^2}{6998400} }+ \log\delta\Bigg\{-\frac{119}{2187000} + \frac{7}{194400}\log 2 \Bigg\}
                        \nonumber\\&\vphantom{\Bigg)}+ \frac{7}{2332800}\log^2(\delta)\Bigg]\Bigg) + \rho^2\Bigg[\delta^{3/2}\Bigg(-\frac{1446053}{755827200} + \frac{2829953}{1007769600}\log 2
                        \nonumber\\&\vphantom{\Bigg)}- \frac{207221}{111974400}\log^2(2) + \frac{54439}{83980800}\log^3(2) + \frac{18179 \pi^2}{223948800} - \frac{7693\pi^2}{83980800}\log 2
                        \nonumber\\&\vphantom{\Bigg)}+ \frac{29057}{74649600}\zeta_3 + \log\delta\Bigg\{\frac{369229}{403107840} - \frac{16807}{13436928}\log 2 + \frac{4361}{6220800}\log^2(2)
                        \nonumber\\&\vphantom{\Bigg)}- \frac{6811\pi^2}{223948800}\Bigg\} + \log^2(\delta)\Bigg\{-\frac{275233}{1343692800} + \frac{27097}{111974400}\log 2\Bigg\} + \frac{17983}{671846400}\log^3(\delta)
                        \nonumber\\&\vphantom{\Bigg)}+ n_l\Bigg\{\frac{10241}{1007769600} - \frac{637}{41990400}\log 2 + \frac{343}{55987200}\log^2(2) - \frac{49\pi^2}{671846400}
                        \nonumber\\&\vphantom{\Bigg)}+
  \log\delta\Bigg[-\frac{637}{167961600} + \frac{49}{18662400} { \log 2}\Bigg] + \frac{49}{223948800}\log^2(\delta)\Bigg\}\Bigg)
                        \nonumber\\&\vphantom{\Bigg)}+ \delta^{5/2}\Bigg(-\frac{189454926667}{5290790400000} - \frac{10087}{279936000}\log\rho + \frac{8996488777}{176359680000}\log 2
                        \nonumber\\&\vphantom{\Bigg)}- \frac{1218467}{39191040}\log^2(2) + \frac{2780833}{293932800}\log^3(2) + \frac{5301953\pi^2}{3919104000} - \frac{392971\pi^2}{293932800}\log 2
                        \nonumber\\&\vphantom{\Bigg)}+ \frac{1484279}{261273600}\zeta_3 + \log\delta\Bigg\{\frac{5876601277}{352719360000} - \frac{2211991}{104976000}\log 2 + \frac{222767}{21772800}\log^2(2)
                        \nonumber\\&\vphantom{\Bigg)}- \frac{347917\pi^2}{783820800}\Bigg\} + \log^2(\delta)\Bigg\{-\frac{40606843}{11757312000} + \frac{197737}{55987200}\log 2\Bigg\}
                        \nonumber\\&\vphantom{\Bigg)}+ \frac{918601}{2351462400}\log^3(\delta) + n_l\Bigg\{\frac{1138687}{6298560000} - \frac{1435327}{5878656000}\log 2 + \frac{2503}{27993600}\log^2(2)
                        \nonumber\\&\vphantom{\Bigg)}- \frac{2503\pi^2}{2351462400} + \log\delta\Bigg[-\frac{702131}{11757312000} + \frac{2503}{65318400}\log 2 \Bigg]
                        \nonumber\\&\vphantom{\Bigg)}+ \frac{2503}{783820800}\log^2(\delta)\Bigg\}\Bigg)
                        \Bigg] + \mathcal{O}(\delta^{7/2}) + \mathcal{O}(\rho^3)\Bigg\}
\,.
\end{align}
Here we display the $\rho\delta^{5/2}$, $\rho^2\delta^{3/2}$ and
$\rho^2\delta^{5/2}$ terms. Note that the leading $\rho^0$ term
starts at $\delta^{7/2}$ and so does not appear here. The channels 
$q\bar{q}$, $qq$ and $qq^\prime$ are further suppressed,
as can be seen from their analytic expressions:
\begin{align}
\sigma_{qq'}^{(2),n_h^2} &= \frac{a_s^4 G_F^2 m_H^2}{\pi}\Bigg\{\rho\delta^{7/2}\Bigg(\frac{565457}{1607445000} - \frac{599}{1275750}\log 2 + \frac{\log^2(2)}{4050} - \frac{\pi^2}{72900}
                         \nonumber\\&\vphantom{\Bigg)}+ \log\delta\Bigg[-\frac{599}{3827250} + \frac{\log 2}{6075}\Bigg] + \frac{\log^2(\delta)}{36450}\Bigg) + \rho^2\Bigg[\delta^{5/2}\Bigg(\frac{493577}{15746400000}
                         \nonumber\\&\vphantom{\Bigg)}- \frac{3773}{87480000}\log 2 + \frac{49}{1944000}\log^2(2) - \frac{49\pi^2}{34992000} + \log\delta\Bigg\{-\frac{3773}{262440000}
                         \nonumber\\&\vphantom{\Bigg)}+ \frac{49}{2916000}\log 2\Bigg\} + \frac{49}{17496000}\log^2(\delta)\Bigg) + \delta^{7/2}\Bigg(\frac{344611787}{840157920000} - \frac{181147}{333396000}\log 2
                         \nonumber\\&\vphantom{\Bigg)}+ \frac{1783}{6350400}\log^2(2) - \frac{1783\pi^2}{114307200} + \log\delta\Bigg[-\frac{181147}{1000188000} + \frac{1783}{9525600}\log 2\Bigg]
                         \nonumber\\&\vphantom{\Bigg)}+ \frac{1783}{{ 57153600}}\log^2(\delta) \Bigg)
                        \Bigg] + \mathcal{O}(\delta^{9/2}) + \mathcal{O}(\rho^3)\Bigg\}\,,
\\\vphantom{\Bigg)}
\Delta^{(2),n_h^2}_{qq} &= { \frac{a_s^4 G_F^2 m_H^2}{\pi}\Bigg\{ } -
                          \frac{7}{104976000}\rho^2\delta^{7/2} +
                          \mathcal{O}(\delta^{9/2}) + \mathcal{O}(\rho^3) \Bigg\}\,,
\\\vphantom{\Bigg)}
\Delta^{(2),n_h^2}_{q\overline{q}} &= { \frac{a_s^4 G_F^2
                                     m_H^2}{\pi}\Bigg\{ }\frac{31}{1020600}\rho^2\delta^{7/2} + \mathcal{O}(\delta^{9/2}) + \mathcal{O}(\rho^3)
\Bigg\}\,,
\end{align}
where $\sigma_{xy} = \Delta_{xy} + \sigma_{qq'}$.

In the ancillary files~\cite{progdata} we provide the renormalized cross
sections with tags for the virtual, real-radiation and collinear counterterm
contributions, which allows one to extract the individual contributions if desired.
This means that these expressions contain poles in $\epsilon$, which cancel upon
removing the tags.

\begin{figure}[t]
  \begin{center}
    \begin{tabular}{cc}
      \hspace*{-2mm}\includegraphics[width=0.51\textwidth]{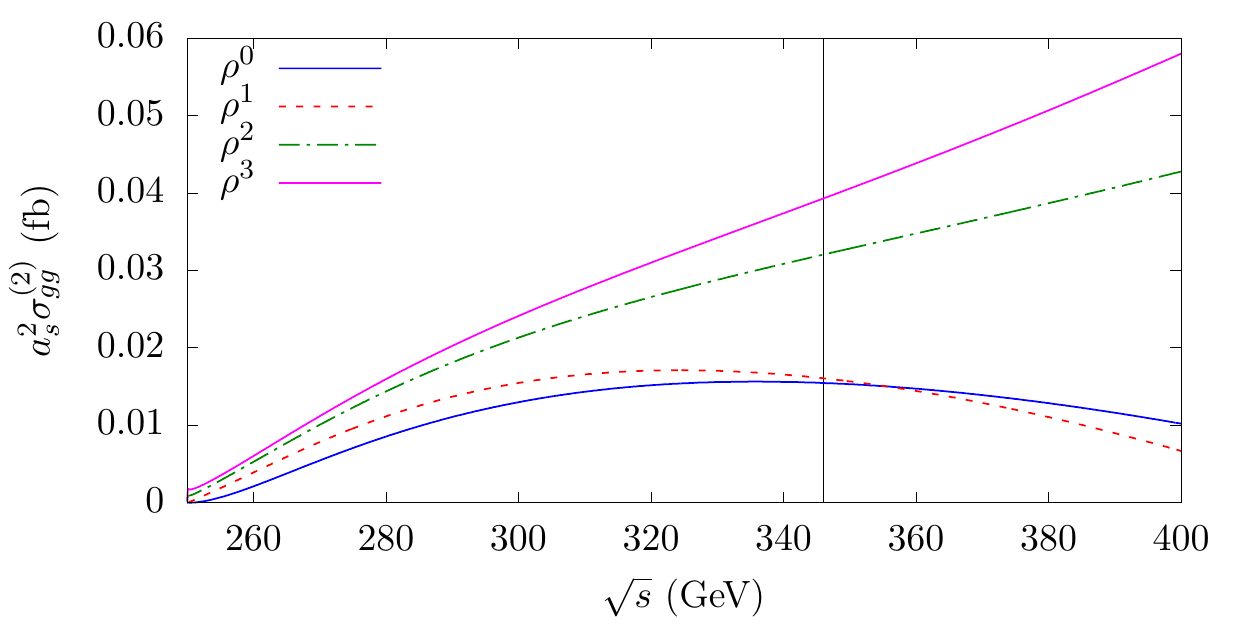} &
      \hspace*{-8mm}\includegraphics[width=0.51\textwidth]{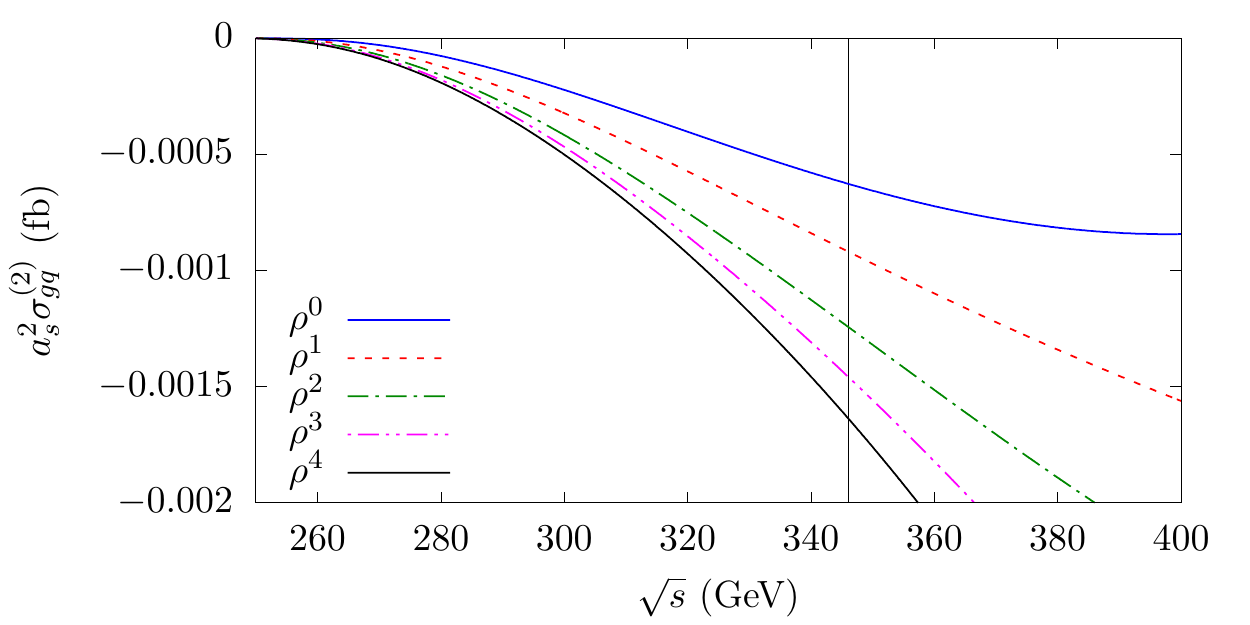}
      \\
      (a) & (b) \\ 
      \hspace*{-2mm}\includegraphics[width=0.51\textwidth]{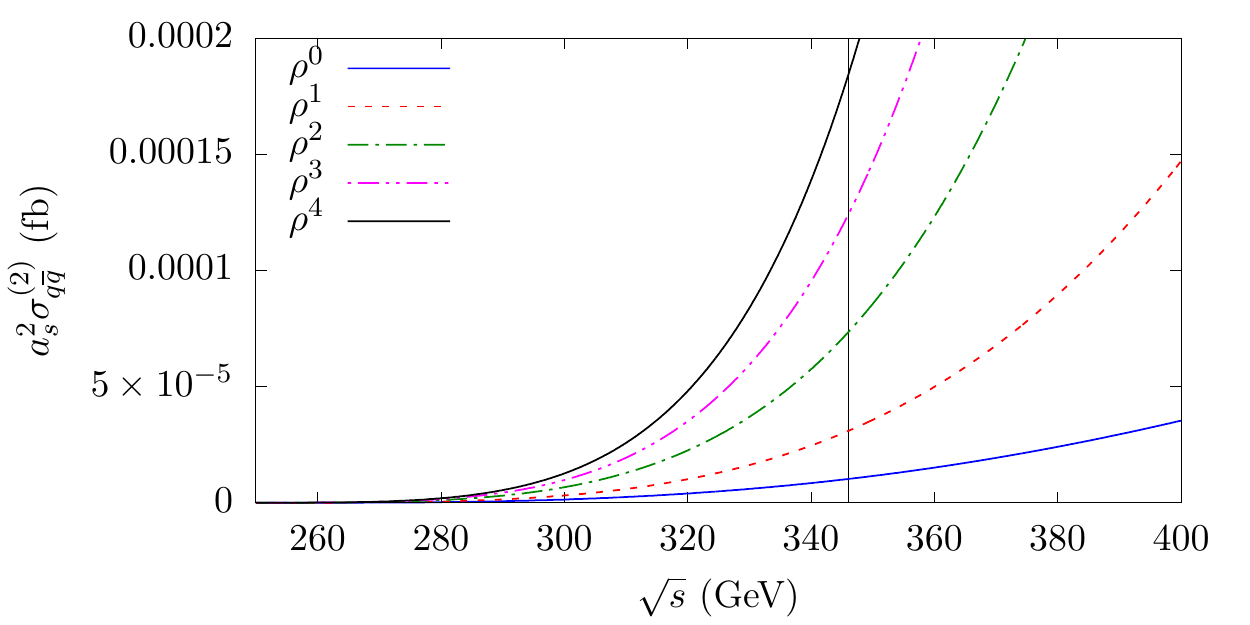} &
      \hspace*{-8mm}\includegraphics[width=0.51\textwidth]{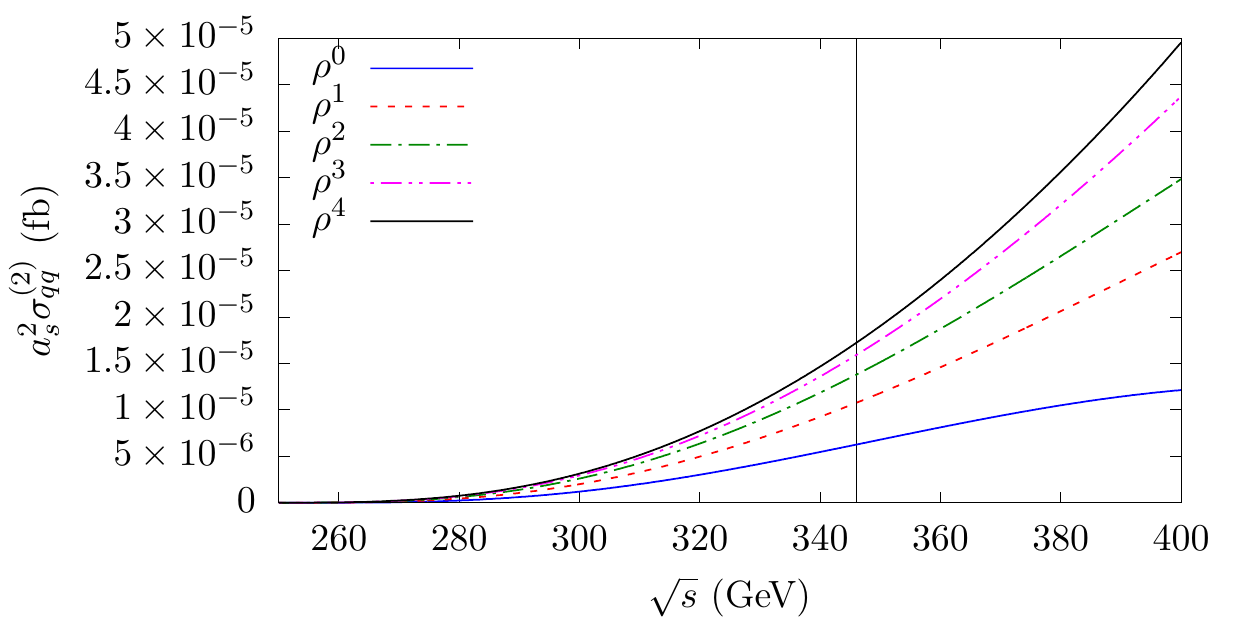} \\
      (c) & (d) \\
      \hspace*{-2mm}\includegraphics[width=0.51\textwidth]{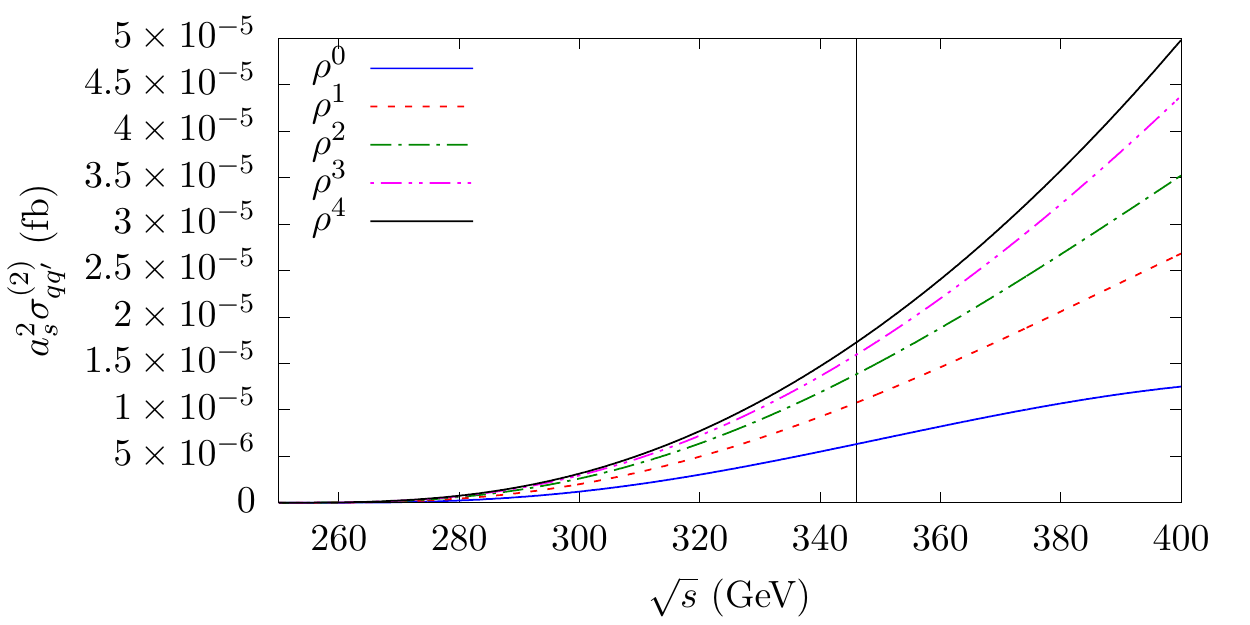} \\
      (e)
    \end{tabular}
    \caption{\label{fig::sig2} The five NNLO partonic contributions as a
      function of $\sqrt{s}$, with the renormalization scale $\mu^2=m_H^2$.
      The colours and line styles denote the inclusion
      of higher orders in the $\rho=m_H^2/m_t^2$ expansion.
      Note that the cross sections have been
      multiplied by two powers of $a_s=\alpha_s^{(5)}/\pi$. The vertical
      black line indicates the top quark threshold at $\sqrt{s}=2 m_t$.}
  \end{center}
\end{figure}

In Fig.~\ref{fig::sig2} we plot the partonic cross section for all five
channels as a function of $\sqrt{s}$, with the renormalization scale $\mu^2=m_H^2$.
We combine all corrections, including
the $n_h^3$ terms from Ref.~\cite{Davies:2019xzc},
the $n_h^4$ terms from Appendix~\ref{app::nh4} and the
$q\bar{q}\to HH$ virtual corrections from Appendix~\ref{app::qqHH}.

The overall picture is similar to the LO, NLO and NNLO $n_h^3$
contributions~\cite{Grigo:2013rya,Davies:2019xzc}; a reasonable convergence
pattern is observed for $\sqrt{s}\lsim 320$~GeV, which rapidly
deteriorates for values of $\sqrt{s}$ above the top quark threshold.
For the $qg$, $qq$ and $qq^\prime$ channels one observes a better convergence
behaviour than for the $gg$ channel. Similar to NLO, the
$q\bar{q}$ channel shows the worst behaviour.
Note that there is a strong hierarchy in the numerical size of the various
contributions, ranging from ${\cal O}(10^{-2})$~fb for the
$gg$ channel to ${\cal O}(10^{-5})$~fb for the $qq$ and $qq^\prime$ channels.

By construction all curves have to approach zero in a continuous way for
$\sqrt{s}\to 2m_H$. This is clearly visible from the plots for all channels
with quarks in the initial state, however for the $gg$ channel the plot gives
the impression that this is not the case for the $\rho^2$ and $\rho^3$ curves.
Inspecting the analytic result reveals that there are terms containing
$\rho^2\sqrt{\delta}\log^4(\delta)$ and $\rho^3\sqrt{\delta}\log^4(\delta)$
which are responsible for a rapid rise in the cross section close to the
threshold.

Although the radius of convergence of the large-$m_t$ expansion is quite
limited here, it contains useful information for the construction of
approximated NNLO results. For example, in Ref.~\cite{Grober:2017uho}
NLO virtual corrections to double Higgs boson production are considered.
The analytic structure of the form factors 
close to the top quark threshold is combined with results
obtained in the large-$m_t$ limit. Stable approximations are only obtained
after including power-suppressed $1/m_t$ terms.
This approach has been applied in Refs.~\cite{Davies:2019nhm,Davies:2019roy}
to the Higgs-gluon form factor and also there it was found that
it is important to include higher-order $1/m_t$ terms to obtain a
stable result in the high-energy region.


\section{Conclusions\label{sec::conclusion}}

The main achievement of this paper is the computation of the NNLO
real-radiation contribution to the total cross section of the process
$gg\to HH$ in an expansion for large top quark mass. This improves
the description beyond the infinite top quark mass limit and provides
important information for NNLO approximation procedures.

We provide a detailed description of the various steps of our
calculation and in particular discuss the computation of the
three- and four-particle phase-space master integrals. We provide
various intermediate results which might be useful for other
calculations.  In the ancillary files of this paper we provide
separate results for the virtual corrections, real-radiation
contribution and the collinear counterterm.

The inclusion of the power-suppressed terms significantly increases the
cross section below the top quark threshold. For example, at NNLO, for a partonic
center-of-mass energy $\sqrt{s}\approx 300$~GeV the cross section
increases almost by a factor of two after including the $1/m_t^4$ and
$1/m_t^6$ terms.



\section*{Acknowledgements}

We thank R.~Lee for discussions regarding the use of \texttt{Libra}.
This research was supported by the Deutsche Forschungsgemeinschaft (DFG,
German Research Foundation) under grant 396021762 --- TRR 257
``Particle Physics Phenomenology after the Higgs Discovery''.
F.H.~acknowledges the support of the Alexander von Humboldt Foundation.
This document was prepared using the resources of the Fermi National Accelerator
Laboratory (Fermilab), a U.S. Department of Energy, Office of Science, HEP User
Facility. Fermilab is managed by Fermi Research Alliance, LLC (FRA), acting
under Contract No. DE-AC02-07CH11359.
The work of G.M.~was in part supported by JSPS KAKENHI (No. JP20J00328).
The work of J.D.~was in part supported by the Science and Technology
Facilities Council (STFC) under the Consolidated Grant ST/T00102X/1.


\appendix


\begin{appendix}

\section*{Appendix}


\section{\label{app::collCT}Auxiliary formulae for the collinear counterterm}

In this appendix we provide some details on the analytic computation
of the convolution integrals in an expansion in $\delta$.

All integrals contributing to Eq.~(\ref{eq::sig_Gam}) have the form
\begin{align}
  \int ^1_{1-\delta}\mathrm{d}z\:
  Q(z) \sigma_X (x/z)
  \,,
  \label{eq::conv_int}
\end{align}
where $Q(z)$ is either a single splitting function or the convolution
of two splitting functions and $\sigma_X$ is one of the following
LO or NLO cross sections,
\begin{align}
  \sigma_X \in \left\{
  \sigma_{gg}^{(0)}
  ,\quad
  \sigma_{gg}^{(1)}
  ,\quad
  \sigma_{gq}^{(1)}
  ,\quad
  \sigma_{q\bar q}^{(1)}
  \right\}
  \,.
\end{align}
These are renormalized finite quantities
which we need up to ${\cal O}(\epsilon^2)$ at LO
and ${\cal O}(\epsilon)$ at NLO.

We introduce the variables
\begin{align}
  x=\frac{m_H^2}{s}
  ,\quad
  \delta =1-4x
  ,\quad
  z=\frac{1-\delta}{1-\delta\nu}
  \,,
  \label{eq::z2nu}
\end{align}
and obtain 
\begin{align}
  \int ^1_{1-\delta}\mathrm{d}z
  =
  \int^1_0\frac{\delta(1-\delta)}{(1-\delta\nu)^2}
  \mathrm{d}\nu
  \,,
  \label{eq::intz2intnu-1}
\end{align}
and
\begin{align}
  \frac{x}{z}=\frac{1-\delta\nu}{4}
  \,.
  \label{eq:x/z}
\end{align}
This allows us to rewrite
Eq.~(\ref{eq::conv_int}) as
\begin{align}
  \int^1_{1-\delta} \mathrm{d}z\:
  Q(z) \sigma_X (x/z) 
  = \int^1_0\mathrm{d}\nu\:
  Q\!\left(\frac{1-\delta}{1-\delta\nu}\right)
  \frac{\delta(1-\delta)}{(1-\delta\nu)^2}
  \sigma_X\left({\frac{1-\delta\nu}{4}}\right) 
  \,,
  \label{eq::intz2intnu}
\end{align}
where $Q(z)$ can be rational functions of $z$, a plus distribution
or a delta function. It may also contain logarithms or dilogarithms.
In the latter case it is convenient to rewrite the
arguments using the identities
\begin{align}
  \mathrm{Li}_2(-z)
  &=
    \frac{1}{2}
    \mathrm{Li}_2(z^2)
    -\mathrm{Li}_2(z)
    \,,\nonumber\\
  \mathrm{Li}_2(z)
  &=
    -\mathrm{Li}_2(1-z)
    -\log(z)\log(1-z)
    +\frac{\pi^2}{6}
    \,,\nonumber\\
  \mathrm{Li}_2(z^2)
  &=
    -\mathrm{Li}_2(1-z^2)
    -2\log(z)\log(1-z)
    -2\log(z)\log(1+z)
    +\frac{\pi^2}{6}
    \,.
\end{align}
In this way, we obtain $Q(z)$ in terms of $\mathrm{Li}_2(1-z)$
and $\mathrm{Li}_2(1-z^2)$ and the expansion around 
$\delta=0$, which corresponds to $z=1$, is straightforward.

\subsection{$\sigma_{gg}^{(0)}$}

Let us first consider the contributions involving the LO cross
section which has a simple series expansion in $\delta$
\begin{align}
  \sigma_{gg}^{(0)}(x)\Big|_{x=\frac{1-\delta}{4}}
  =
  \sum _{n=0}^\infty \delta^{\frac{1}{2}+n-\epsilon}
  \sum_{u=0}^{N_\epsilon^\mathrm{max}}
  \epsilon^u
  c^{(n,u),(0)}_{gg}
  \,,
  \label{eq::sigLO_delta}
\end{align}
where for our application we have $N_\epsilon^\mathrm{max}=2$.
Note the the coefficients $c^{(n,u),(0)}_{gg}$ are available from
an explicit calculation of $\sigma_{gg}^{(0)}(x)$.
In the following we discuss the various cases for $Q(z)$ which appear.

For the case of $Q(z)=z^r$, where in our case only $r\geq -1$ is needed, we
have
\begin{align}
  \int^1_{1-\delta} \mathrm{d}z
  ~\sigma_{gg}^{(0)} (x/z) ~z^r
  &=
    \sum_{n=0}^\infty 
    \delta^{3/2+n} (1-\delta)^{r+1}
    \sum _{j=0}^n
    \sum_{u=0}^{N_\epsilon^\mathrm{max}}
    \frac{(n-j+r+1)!}{(r+1)!(n-j)!}
    \frac{\delta^{-\epsilon}}{\frac{3}{2}+n-\epsilon}
    c^{(j,u),(0)}_{gg}
    \epsilon^u
    \,,
\end{align}
and for $Q(z)=1/(1+z)$,
\begin{align}
  &\int^1_{1-\delta} \mathrm{d}z
  ~\frac{\sigma_{gg}^{(0)} (x/z)}{1+z}
    =
    \nonumber\\
  &
    \qquad
    \sum_{n=0}^\infty 
    \delta^{\frac{3}{2}+n}(1-\delta)
    \sum _{k=0}^n
    \sum_{j=0}^k
    \sum _{s=0}^{n-k} 
    \sum_{u=0}^{N_\epsilon^\mathrm{max}}
    \frac{1}{2^{n-k+1}}
    \frac{(n-k)!}{s!(n-k-s)!} 
    \frac{\delta^{-\epsilon}}{\frac{3}{2}+k+s-\epsilon}
    c^{(j,u),(0)}_{gg}
    \epsilon^u
    \,.
\end{align}
These formulae are obtained with the help of Eqs.~(\ref{eq::z2nu})
and~(\ref{eq::intz2intnu-1}) which are used to replace $z$ by $\nu$.
Afterwards we insert Eq.~(\ref{eq::sigLO_delta}),
expand in $\delta$ and integrate each term individually.
This approach is also used for the results which we present below.
Due to the more complicated functions $Q(z)$ some of the formulae
are more involved. Although the formulae we present are exact,
we apply an upper limit to the $\delta$ expansion when we use
them in our calculations.

Now we consider
$Q(z)=\left[\frac{(\log(1-z))^r}{1-z}\right]_+$. 
Using the definition of the plus distribution we obtain
\begin{align}
  &  \int^1_{1-\delta}\mathrm{d}z
    \left[\frac{(\log(1-z))^r}{1-z}\right]_+
    \sigma_{gg}^{(0)}(x/z)
    =
    \nonumber\\&\qquad
    \int^1_{1-\delta}
    \mathrm{d}z
    \frac{(\log(1-z))^r}{1-z}
    \left( \sigma_{gg}^{(0)}\left(\frac{x}{z}\right) -\sigma_{gg}^{(0)}(x) \right)
    +
    \frac{(\log\delta)^{r+1}}{r+1}
    \sigma_{gg}^{(0)}(x)
    \,.
    \label{plus}
\end{align}
In order to compute the finite integral on the r.h.s., we consider
$Q(z) = (1-z)^{-1+\eta}$ which leads to
\begin{align}
  &\int^1_{1-\delta} \mathrm{d}z
  ~\sigma_{gg}^{(0)} (x/z)~(1-z)^{-1+\eta}
  =
  \nonumber\\&\qquad 
    \sum_{l=0}^\infty
    \sum _{m=0}^\infty 
    \sum_{u=0}^{N_\epsilon^\mathrm{max}}
    \frac{\Gamma(1+\eta+l)}{\Gamma(1+\eta)l!}
    \delta^{\frac{1}{2}+m+l+\eta-\epsilon}
    (1-\delta)
    \epsilon^u
    c^{(m,u),(0)}_{gg}
    \frac{
    \Gamma(\eta)\Gamma(\frac{3}{2}+m+l-\epsilon)
    }
    {\Gamma( \frac{3}{2}+m+l+\eta-\epsilon )}
    \,.
\end{align}
Note that this formula is exact in $\eta$. 
Using $\sigma_{gg}^{(0)} (x)$ instead of $\sigma_{gg}^{(0)} (x/z)$ we have
\begin{align}
  &\int^1_{1-\delta} \mathrm{d}z
  ~\sigma_{gg}^{(0)} (x)~(1-z)^{-1+\eta}
  =
  \nonumber\\&\qquad 
    \sum_{l=0}^\infty
    \sum _{m=0}^\infty 
    \sum_{u=0}^{N_\epsilon^\mathrm{max}}
    \frac{\Gamma(1+\eta+l)}{\Gamma(1+\eta)l!}
    \delta^{\frac{1}{2}+m+l+\eta-\epsilon}
    (1-\delta)
    \epsilon^u
    c^{(m,u),(0)}_{gg}
    \frac{
    \Gamma(\eta)\Gamma(1+l)
    }
    {\Gamma( 1+l+\eta )}
    \,.
\end{align}
After combining both expressions we obtain
\begin{align}
  &\int^1_{1-\delta} \mathrm{d}z
  ~(1-z)^{-1+\eta}
  \left[ \sigma_{gg}^{(0)}\left(\frac{x}{z}\right)
  -\sigma_{gg}^{(0)} (x) \right]=
    \sum_{n=0}^\infty
    \delta^{\frac{1}{2}+n-\epsilon}
    (1-\delta)
    \sum _{j=0}^n
    \sum_{u=0}^{N_\epsilon^\mathrm{max}}
    \epsilon^u
    c^{(j,u),(0)}_{gg}
  \nonumber\\
  &\mbox{}\qquad   \left[
    \Psi_{j,n}
    +
    \eta \left(
    (\log\delta + \psi (n-j+1) ) \Psi_{j,n}
    +\frac{\Psi^{(1)}_{j,n}-\Psi^{(2)}_{j,n}}{2}
    \right)
    +\mathcal{O}(\eta^2)
    \right]
    \,,
    \label{eq::eta2}
\end{align}
where
\begin{align}
  \Psi_{j,n}
  &=
    \psi\left(n-j+1\right)
    -
    \psi\left(\frac{3}{2}+n-\epsilon\right)\,,
  \nonumber\\
  \Psi^{(1)}_{j,n}
  &=
    \psi'\left(n-j+1\right)
    -
    \psi'\left(\frac{3}{2}+n-\epsilon\right)\,,
  \nonumber\\
  \Psi^{(2)}_{j,n}
  &=
    \left[ \psi \left(n-j+1\right)\right]^2
    -
    \left[ \psi \left(\frac{3}{2}+n-\epsilon\right)\right]^2\,.
\end{align}
Here $\psi(x)$ is the polygamma function and $\psi'(x)\equiv\psi^{(1)}$ its
derivative.
The results for Eq.~(\ref{plus}) are obtained by expanding the l.h.s.
of Eq.~(\ref{eq::eta2}) in $\eta$ according to
\begin{align}
  (1-z)^{-1+\eta} = \frac{1}{1-z} \left[ 1+ \eta \log (1-z) +\frac{\eta^2}{2}
  \left( \log(1-z) \right) ^2 +\cdots \right] 
  \,.
\end{align}
Note that the r.h.s~of Eq.~(\ref{eq::eta2}) is already expanded in $\eta$.
The comparison of the $\eta^0$ and $\eta^1$ terms provides
results for Eq.~(\ref{plus}) for $r=0$ and $r=1$, which we need for
our calculations.

Next we treat cases such as
\begin{align}
  Q(z)=\frac{\log(z)\log(1-z)}{1-z},\quad Q(z) = z^2\log(1-z), \ldots
  \,,
  \label{eq::Qz_log}
\end{align}
i.e., $Q(z)$ contains factors of $z$ and $1-z$ with positive or negative
exponents, and/or logarithms with $z$ or $1-z$ as argument. 
Here it is convenient first to consider
\begin{align}
  Q(z)=z^{r_1}(1-z)^{r_2}\,,
  \label{eq::Qz_r1r2}
\end{align}
with non-integer exponents $r_1$ and $r_2$.
The resulting formula can then be expanded around $r_1,r_2=-1,0,\ldots$
in order to obtain the result for the desired $Q(z)$ of Eq.~(\ref{eq::Qz_log}).
The generic result for $Q(z)$ from Eq.~(\ref{eq::Qz_r1r2}) is given by
\begin{align}
  &
    \int^1_{1-\delta} \mathrm{d}z
    ~\sigma_{gg}^{(0)} (x/z)~z^{r_1}(1-z)^{r_2}
    =
    \sum_{n=0}^\infty
    \delta ^{\frac{3}{2}+n-\epsilon+r_2} (1-\delta)^{1+r_1} 
    \nonumber\\&\qquad
    \sum _{j=0}^n
    \sum_{u=0}^{N_\epsilon^\mathrm{max}}
    \frac{
    \Gamma(r_1+r_2+2+n-j)
    \Gamma(r_2+1)
    \Gamma(\frac{3}{2}+n-\epsilon)
    }{
    \Gamma(r_1+r_2+2)
    \Gamma(\frac{5}{2}+n-\epsilon+r_2)
    (n-j)!
    }
    \epsilon^u
    c^{(j,u),(0)}_{gg}
    \,.
    \label{eq::formulae_r1r2}
\end{align}
To obtain the result for $Q(z)=z^2\log(1-z)$ we set $r_1=2$
and $r_2=\eta$. Afterwards we expand Eq.~(\ref{eq::formulae_r1r2})
in $\eta$ and take the coefficient of the linear term, on both sides.
In a similar way one can treat $Q(z)=\log(z)$ by setting $r_1=\eta$ and $r_2=0$.

In case $Q(z)$ (from Eq.~(\ref{eq::Qz_log})) contains a $\log(z)$ term
a non-integer value for $r_1$ must be chosen.

Cases such as
\begin{align}
  Q(z)=\frac{\log(z)\log(1+z)}{1+z},\quad z^2\log(1+z),\ldots
\end{align}
can be treated in a similar way. Here a convenient auxiliary function is given by
\begin{align}
  Q(z)=z^{r_1}(1+z)^{r_2}\,,
\end{align}
where the desired result is again obtained by a
suitable choice of $r_1$ and $r_2$ supplemented by 
proper expansions. The generic integral is given by
\begin{align}
  &
    \int^1_{1-\delta} \mathrm{d}z
    ~\sigma_{gg}^{(0)} (x/z)~z^{r_1}(1+z)^{r_2}
    =
    \sum_{n=0}^\infty
    \sum _{k=0}^n
    \sum_{j=0}^k
    \sum_{s=0}^{n-k}
    \sum_{u=0}^{N_\epsilon^\mathrm{max}}
    {2^{r_2-n+k}}
    \delta ^{\frac{3}{2}+n-\epsilon} (1-\delta)^{1+r_1} 
    \nonumber\\&\qquad
    \frac{\Gamma(-r_2+n-k)\Gamma(r_1+r_2+2+k-j)}
    {\Gamma(-r_2)(n-k-s)!s!\Gamma(r_1+r_2+{2})(k-j)!}
    \epsilon^u
    c^{(j,u),(0)}_{gg}
    \frac{1}{\frac{3}{2}+k+s-\epsilon}
    \,.
\end{align}
Note that the expression on the r.h.s.~is significantly more
complex than Eq.~(\ref{eq::formulae_r1r2}). This is
due to the fact that the factor $(1+z)^{r_2}$ introduces a new
type of denominator whereas $(1-z)\sim\delta$ and thus
no new structure is introduced.

Having dealt with logarithms, we now turn to functions $Q(z)$ which
involve dilogarithms. For $Q(z)=z^r\mathrm{Li}_2(1-z)$ we have
\begin{align}
  &
    \int^1_{1-\delta} \mathrm{d}z
    ~z^r \mathrm{Li}_2 (1-z)
    ~\sigma_{gg}^{(0)} (x/z)
    =
    \sum_{n=1}^\infty
    \delta ^{\frac{3}{2}+n-\epsilon} (1-\delta)^{1+r} 
    \nonumber\\&\qquad
    \sum _{k=0}^{n-1}
    \sum_{j=0}^k
    \sum_{u=0}^{N_\epsilon^\mathrm{max}}
    \frac{1}{(n-k)^2}
    \frac{(r+n-j+1)!}{(r+n-k+1)!(k-j)!}
    \frac{
    \Gamma(n-k+1)\Gamma(\frac{3}{2}+k-\epsilon)}
    {\Gamma(\frac{5}{2}+n-\epsilon)}
    \epsilon^u
    c^{(j,u),(0)}_{gg}
    \,,
\end{align}
and for $Q(z)=\mathrm{Li}_2(1-z)/(1-z)$
the corresponding integral has the form
\begin{align}
  &
    \int^1_{1-\delta} \mathrm{d}z
    ~\frac{\mathrm{Li}_2 (1-z)}{1-z} 
    ~\sigma_{gg}^{(0)} (x/z)
    =
    \nonumber\\&\qquad
    \sum_{n=1}^\infty
    \delta ^{\frac{1}{2}+n-\epsilon} (1-\delta)
    \sum _{k=0}^{n-1}
    \sum_{j=0}^k
    \sum_{u=0}^{N_\epsilon^\mathrm{max}}
    \frac{
    (n-j)! \Gamma(\frac{3}{2}+k-\epsilon)}
    {(n-k)^3 (k-j)! \Gamma(\frac{3}{2}+n-\epsilon)}
    \epsilon^u
    c^{(j,u),(0)}_{gg}
    \,.
\end{align}
For $Q(z)=\mathrm{Li}_2(1-z)/(1+z)$ we have
\begin{align}
  &
    \int^1_{1-\delta} \mathrm{d}z
    ~\frac{\mathrm{Li}_2 (1-z)}{1+z} 
    ~\sigma_{gg}^{(0)} (x/z)
    =
    \sum_{n=1}^\infty
    \delta ^{\frac{3}{2}+n-\epsilon} (1-\delta)
    \sum _{h=0}^{n-1}
    \sum_{k=0}^h
    \sum_{j=0}^k
    \sum_{w=0}^{h-k}
    \sum_{u=0}^{N_\epsilon^\mathrm{max}}
    \frac{
    \epsilon^u
    c^{(j,u),(0)}_{gg}
    }{2^{h-k+1}}
    \nonumber\\&\qquad
    \frac{(h-k)!(n-h+k-j)!}{(h-k-w)!w!(k-j)!(n-h)^2}
    \frac{\Gamma(\frac{3}{2}+k+w-\epsilon)}
    {\Gamma(\frac{5}{2}+n-h+k+w-\epsilon)}
    \,,
\end{align}
and for $Q(z)=z^r\mathrm{Li}_2(1-z^2)$
the integral is given by
\begin{align}
  &
    \int^1_{1-\delta} \mathrm{d}z
    ~z^r \mathrm{Li}_2 (1-z^2)
    ~\sigma_{gg}^{(0)} (x/z)
    =
    \sum_{n=1}^\infty
    \delta^{\frac{3}{2}+n-\epsilon}
    (1-\delta)^{1+r}
    \sum_{k=0}^{n-1}
    \sum_{j=0}^k
    \sum_{w=0}^{[\frac{n-k}{2}]}
    \sum_{h=0}^{w}
    \sum_{u=0}^{N_\epsilon^\mathrm{max}}
    \epsilon^u
    c^{(j,u),(0)}_{gg}
    \nonumber\\&\qquad
    \frac{(-1)^w 
    ((n-k-w)!)^2(r+2n-k-j-2w+1)!
    2^{n-k-2w}
    }{(n-k-w)^2(n-k-2w)!(w-h)!h!(r+2n-2k-2w+1)!(k-j)!
    }
    \times
    \nonumber\\&\qquad
    \frac{ \Gamma(\frac{3}{2}+k+h-\epsilon) } {  \Gamma(\frac{5}{2}+n+h-w-\epsilon) }
    \,.
\end{align}
Finally, for $Q(z)=\mathrm{Li}_2(1-z^2)/(1+z)$
we have
\begin{align}
  &
    \int^1_{1-\delta} \mathrm{d}z
    ~\frac{1}{1+z} \mathrm{Li}_2 (1-z^2)
    ~\sigma_{gg}^{(0)} (x/z)
    =
    \sum_{n=1}^\infty
    \delta^{\frac{3}{2}+n-\epsilon}
    (1-\delta)
    \sum_{k=0}^{n-1}
    \sum_{j=0}^k
    \sum_{w=0}^{[\frac{n-k-1}{2}]}
    \sum_{h=0}^{w}
    \sum_{u=0}^{N_\epsilon^\mathrm{max}}
    \epsilon^u
    c^{(j,u),(0)}_{gg}
    \nonumber\\&\qquad
    \frac{(-1)^w 
    ((n-k-w-1)!)^2(2n-k-j-2w)!
    2^{n-k-2w-1}
    }{(n-k-w)(n-k-2w-1)!(w-h)!h!(2n-2k-2w)!(k-j)!
    }
    \times
    \nonumber\\&\qquad
    \frac{ \Gamma(\frac{3}{2}+k+h-\epsilon)  } {  \Gamma(\frac{5}{2}+n+h-w-\epsilon) }
    \,.
\end{align}


\subsection{$\sigma_{gg}^{(1)}, \sigma_{gq}^{(1)}, \sigma_{q\bar q}^{(1)}$}

In this subsection we consider the NLO partonic cross sections
$\sigma_{gg}^{(1)}, \sigma_{gq}^{(1)}$ and $\sigma_{q\bar q}^{(1)}$
which have the following expansion in $\delta$
\begin{align}
  \sigma_X(x)\Big|_{x=\frac{1-\delta}{4}}
  =\sum _{n=0}^\infty \delta^{\frac{1}{2}+n}
  \sum_{u=0}^{N_\epsilon^\mathrm{max}}
  \sum_{v=0}^{u+2}
  \epsilon^u
  (\log \delta)^v
  c^{(n,u,v)}_X
  \,,
  \label{sigxz}
\end{align}
with known coefficients $c^{(n,u,v)}_X$. For our application
$N_\epsilon^\mathrm{max}=1$ is sufficient.  Note the explicit presence of
$\log\delta$ terms which cannot be factorized as at LO in
Eq.~(\ref{eq::sigLO_delta}).  With the help of Eq.~(\ref{eq:x/z}) we also
introduce
\begin{align}
  \sigma_X(x/z)\Big|_{x=\frac{1-\delta}{4}}
  =\sum _{n=0}^\infty (\delta\nu)^{\frac{1}{2}+n}
  \sum_{u=0}^{N_\epsilon^\mathrm{max}}
  \sum_{v=0}^{u+2}
  \epsilon^u
  (\log \nu)^v
  \tilde{c}^{(n,u,v)}_X
  \,,
  \label{sigxz-1}
\end{align}
where the coefficients $\tilde{c}^{(n,u,v)}_X$ are functions of $\log\delta$,
which for the relevant values of $u$ and $v$ are given in terms of $c^{(n,u,v)}_X$
by:
\begin{align}
  \tilde{c}^{(n,0,0)}_X
  &=
    c^{(n,0,0)}_X
    +c^{(n,0,1)}_X\log\delta
    +c^{(n,0,2)}_X(\log\delta)^2
    \,,\nonumber\\
  \tilde{c}^{(n,0,1)}_X
  &=
    c^{(n,0,1)}_X
    +2c^{(n,0,2)}_X\log\delta
    \,,\nonumber\\
  \tilde{c}^{(n,0,2)}_X
  &=
    c^{(n,0,2)}_X
    \,,\nonumber\\
  \tilde{c}^{(n,1,0)}_X
  &=
    c^{(n,1,0)}_X
    +c^{(n,1,1)}_X\log\delta
    +c^{(n,1,2)}_X(\log\delta)^2
    +c^{(n,1,3)}_X(\log\delta)^3
    \,,\nonumber\\
  \tilde{c}^{(n,1,1)}_X
  &=
    c^{(n,1,1)}_X
    +2c^{(n,1,2)}_X\log\delta
    +3c^{(n,1,3)}_X(\log\delta)^2
    \,,\nonumber\\
  \tilde{c}^{(n,1,2)}_X
  &=
    c^{(n,1,2)}_X
    +3c^{(n,1,3)}_X\log\delta
    \,,\nonumber\\
  \tilde{c}^{(n,1,3)}_X
  &=
    c^{(n,1,3)}_X
    \,.
    \label{eq::ctil}
\end{align}
Note that the coefficients in Eq.~(\ref{eq::ctil}) can be used to
write Eq.~(\ref{sigxz}) in the form
\begin{align}
  \sigma_X(x)\Big|_{x=\frac{1-\delta}{4}}
  =\sum _{n=0}^\infty \delta^{\frac{1}{2}+n}
  \sum_{u=0}^{N_\epsilon^\mathrm{max}}
  \epsilon^u
  \tilde{c}^{(n,u,0)}_X
  \,.
\end{align}

We are now in the position to insert the relevant expressions
for $Q(z)$ in Eq.~(\ref{eq::intz2intnu}) and perform an expansion
in $\delta$. The NLO cross sections are only convoluted
with one-loop splitting functions which leaves us with
$Q(z)=z^r$ and $Q(z)=\left[\frac{1}{1-z}\right]_+$.
For $Q(z)=z^r$ we have
\begin{align}
  &\int^1_{1-\delta} \mathrm{d}z
  ~\sigma_X (x/z) ~z^r
  =
    \nonumber\\&\qquad
  (1-\delta)^{r+1}
  \sum_{n=0}^\infty
  \delta^{\frac{3}{2}+n}
  \sum _{j=0}^n
  \frac{(r+n-j+1)!}{(r+1)!(n-j)!}
  \sum_{u=0}^{N_\epsilon^\mathrm{max}}
  \sum_{v=0}^{u+2}
  \epsilon^u
  \tilde{c}^{(j,u,v)}_X
  \frac{(-1)^vv!}{(3/2+n)^{v+1}}
  \,.
\end{align}
For $Q(z)=\left[\frac{1}{1-z}\right]_+$ we use Eq.~\eqref{plus}
for $r=0$ and obtain for the integral on the r.h.s.
\begin{align}
  &
    \int^1_{1-\delta}
    \mathrm{d}z
    \frac{1}{1-z}
    \left( \sigma_X\left(\frac{x}{z}\right) -\sigma_X(x) \right)
    =
    (1-\delta)\times
    \nonumber\\&\qquad
    \sum _{n=0}^\infty 
    \sum _{j=0}^n
    \delta^{\frac{1}{2}+n}
    \sum_{u=0}^{N_\epsilon^\mathrm{max}}
    \epsilon^u
    \left[
    \psi(n-j+1)
    \tilde{c}^{(j,u,0)}_X
    -
    \sum_{v={0}}^{u+2}
    \psi^{(v)}
    \!\left(\!\frac{3}{2}+n\!\right)
    \tilde{c}^{(j,u,v)}_X
    \right] 
    .
  \label{eq::NLO_plus}
\end{align}
For $N_\epsilon^\mathrm{max}=1$, terms with up to 
three derivatives of the polygamma function appear
on the r.h.s.~of Eq.~(\ref{eq::NLO_plus}).



\section{\label{app::h2hh}Cross-check from $gg\to H$}

In this appendix we describe how the leading terms in the $1/m_t$ expansion
(i.e.~the $m_t^0$ terms)
of the Higgs pair cross section can be cross-checked using known results from
single-Higgs boson production. The approach described below can be used
to cross-check the divergent building blocks entering the individual channels.

At leading order in $1/m_t$ the Higgs-pair cross section
is obtained by interpreting the Higgs boson mass in the single-Higgs
cross section as the invariant mass of the Higgs boson pair, $m_{HH}$,
and subsequently integrating over $m_{HH}$ (see also Ref.~\cite{Chen:2019fhs}).
Our master formula reads
\begin{align}
  \sigma_{HH}^{[x]} =
  \int ^{\sqrt{s}}_{2m_H} {\rm d} m_{HH}
  \frac{d\sigma^\mathrm{[x]}_{HH}}{dm_{HH}}
  &=
    \int ^{\sqrt{s}}_{2m_H} {\rm d} m_{HH}
    f^\epsilon _{H\to HH}
    \left(
    \frac{C_{HH}}{C_H}
    -\frac{6\lambda v^2}{m_{HH}^2-m_H^2}
    \right)^2
    \sigma^\mathrm{[x]}_H \Big|_{m_H\to m_{HH}}
    \,,
    \label{eq::h2hh}
\end{align}
where the superscript ``[x]'' of $\sigma^\mathrm{[x]}_{HH}$ and
$\sigma^\mathrm{[x]}_H$ represent the (divergent) individual pieces.  For
example, $\sigma^\mathrm{[rr]}_{HH}$ denotes the real-real
and $\sigma^\mathrm{[coll]}_{HH}$ the collinear counterterm contribution.
In Eq.~(\ref{eq::h2hh}) the function $f^\epsilon _{H\to HH}$ takes into
account the $d$-dimensional two-particle phase space and is given by
\begin{align}
  f^\epsilon _{H\to HH}
  =
  e^{\epsilon \gamma_E}
  \frac{\Gamma(1-\epsilon)}{\Gamma(2-2\epsilon)}
  \frac{(m_{HH}^2-4m_H^2)^{1/2-\epsilon}}{16\pi^2 v^2}
  \left(\frac{\mu^2}{m_H^2}\right)^\epsilon
  \,.
\end{align}
The ratio of the matching coefficients $C_{HH}$ and $C_H$, including higher
order terms in $\epsilon$, reads~\cite{Gerlach:2018hen}
\begin{align}
  \frac{C_{HH}}{C_H} =
  &1+2\epsilon
    + \frac{\alpha_s^{(5)}(\mu)}{\pi} \left[
    \frac{3  \epsilon}{2}
    + \left(\frac{3}{2}\log\frac{\mu^2}{m_t^2} - \frac{11}{3}\right) \epsilon ^2
    \right]
    +\left( \frac{\alpha_s^{(5)}(\mu)}{\pi} \right)^2 \Bigg\{
    \frac{19}{8} -\frac{11 n_h}{12}+\frac{2 n_l}{3}
    \nonumber\\&\qquad
  +\epsilon  \left[
  -\frac{95}{16}
  +\frac{91 n_h}{144}
  -\frac{13 n_l}{8}
  + \log\frac{\mu^2}{m_t^2} \left(
  \frac{71}{8}
  -\frac{11 n_h}{6}
  +\frac{13 n_l}{12}
  \right)
  \right]
  \Bigg\}
  \,,
\end{align}
where $m_t\equiv m_t(\mu)$ is the top quark mass renormalized in the $\overline{\rm MS}$ scheme.

Our aim is to compute the integral in Eq.~(\ref{eq::h2hh}) in an expansion in
$\delta$. To do so it is convenient to use the following change of variable,
\begin{align}
  m_{HH}^2 =\frac{4m_H^2(1-\delta \nu)}{1-\delta }
  \,.
\end{align}
This allows us to rewrite $f^\epsilon _{H\to HH}$ as
\begin{align}
  f^\epsilon _{H\to HH}
  =
  e^{\epsilon \gamma_E}
  \frac{\Gamma(1-\epsilon)}{\Gamma(2-2\epsilon)}
  \frac{
  (4\delta)^{1/2-\epsilon}\:
  (1-\delta)^{-1/2+\epsilon}\:
  (1-\nu)^{1/2-\epsilon}
  }{16\pi^2 v^2}
  \left(\frac{\mu^2}{m_H^2}\right)^\epsilon
\end{align}
and the integral in Eq.~(\ref{eq::h2hh}) is then written as
\begin{align}
   & \int ^{\sqrt{s}}_{2m_H} {\rm d} m_{HH}
    f^\epsilon _{H\to HH}
    \left(
    \frac{C_{HH}}{C_H}
    -\frac{6\lambda v^2}{m_{HH}^2-m_H^2}
    \right)^2
    \sigma^\mathrm{[x]}_H \Big|_{m_H\to m_{HH}}
  =
  \nonumber\\&\qquad
     \int^1_0
     {\rm d}\nu 
     \frac{2m_H\delta}{\sqrt{1-\delta}\sqrt{1-\delta\nu}}
     f^\epsilon _{H\to HH}
     \left(
     \frac{C_{HH}}{C_H}
     -\frac{6\lambda v^2(\delta-1)}
     { m_H^2 (-3 - \delta + 4 \delta \nu)  }
     \right)^2
     \sigma^\mathrm{[x]}_H \Big|_{m_H\to m_{HH}}
     \,.
\end{align}
In this form, we can expand the integrand in terms of $\delta$ before integration,
considerably simplifying the computation.  Note that the single Higgs boson
cross sections $\sigma^\mathrm{[x]}_H$, which are expressed in terms of
$x=m_H^2/s$, now depend on
\begin{align}
  x \to \frac{m_{HH}^2}{s} 
  = 1 - \delta \nu
  \,.
\end{align}

We have used this method to check the $m_t^0$ terms of the LO, NLO and NNLO
cross sections, including terms of order $\epsilon^2$, $\epsilon^1$ and $\epsilon^0$,
respectively. The results for the partonic cross sections of single Higgs production,
expanded to the proper orders in $\epsilon$, are taken from
Ref.~\cite{Hoschele:2013pvt}.



\section{\label{app::qqHH}Virtual corrections to $q\bar{q}\to HH$}

\begin{figure}[t]
  \begin{center}
    \begin{tabular}{cc}
      \includegraphics[width=0.3\textwidth]{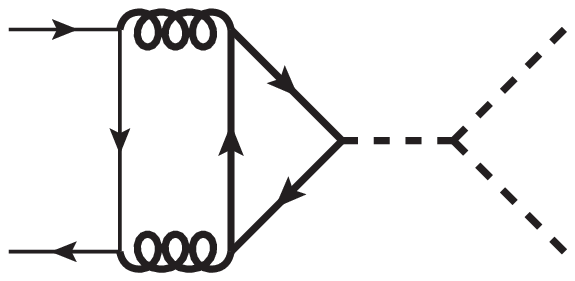}
      &
      \includegraphics[width=0.3\textwidth]{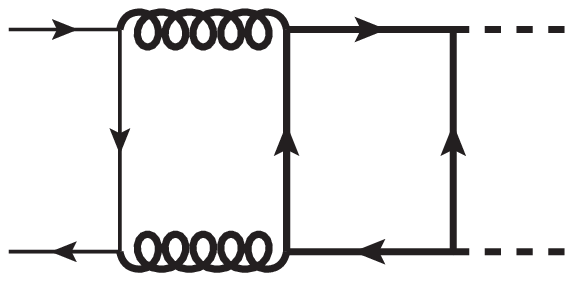}
      \\
      (a) & (b) 
    \end{tabular}
    \caption{\label{fig::diag_qqHH}Sample Feynman diagrams
      for the process $q\bar{q}\to HH$.}
  \end{center}
\end{figure}  

In this appendix we discuss the virtual corrections to the process
$q\bar{q}\to HH$. The coupling of the initial-state quark anti-quark pair
to the Higgs bosons in the final state has to be mediated by top quarks. Thus
at the lowest order it proceeds via two-loop diagrams such as those shown in
Fig.~\ref{fig::diag_qqHH}.  As a consequence such virtual corrections
contribute for the first time at NNLO.

Diagrams involving a triple-Higgs boson coupling (cf.~Fig.~\ref{fig::diag_qqHH}(a)) vanish 
for massless quarks $q$ since the (effective) coupling of
the $q\bar{q}$ pair to one Higgs boson requires a helicity flip.  However,
diagrams such as the one in Fig.~\ref{fig::diag_qqHH}(b) can provide non-vanishing
contributions. We have shown by an explicit calculation that such diagrams
vanish in the $m_t\to\infty$ limit. However, non-vanishing contributions exist
starting from $1/m_t^2$.  In the following we describe their calculation
and present the results.

In principle there are two methods to compute the total cross section.
We can either compute the two-loop amplitude $q\bar{q}\to HH$
and integrate over the phase space, or we can consider five-loop
forward-scattering $q\bar{q}\to q\bar{q}$ diagrams and take the imaginary
part. We have used both approaches and have obtained identical results.
Note that the real corrections to the  $q\bar{q}$ channel discussed above
are finite. Thus we expect the two-loop virtual corrections
$q\bar{q}\to HH$ also to be finite.

We first describe the computation of the amplitude ${\cal M}(q\bar{q} \to HH)$.
This is conveniently done using the projector
\begin{eqnarray}
  P_{q\bar{q} \to HH} &=&
  \frac{\delta^{ij}}{N_c}\frac{\slashed{q}_1\slashed{q}_3\slashed{q}_2}{2 u t - 2 m_H^4}
  \,,
\end{eqnarray}
where $N_c=3$ and $q_1$ and $q_2$ are the incoming momenta of the quark and anti-quark,
respectively, and $q_3$ is the incoming momentum of one of the Higgs bosons.
$t=(q_1+q_3)^2$ and $u=(q_2+q_3)^2$ are Mandelstam variables and
$i$ and $j$ are colour indices of the quark and anti-quark, respectively. 
Asymptotic expansion in the large-$m_t$ limit, which we apply with the
help of {\rm exp}~\cite{Harlander:1997zb,Seidensticker:1999bb}, leads to 
two-loop vacuum integrals and products of one-loop vacuum and
one-loop massless form factor integrals. We perform the calculation with a general
QCD gauge parameter and check that it drops out in the sum of all contributing diagrams.
Next we square the amplitude obtain the differential cross section
\begin{eqnarray}
  \frac{{\rm d}\sigma_{q\bar{q} \to HH}}{{\rm d}{y}}  &=&
      \frac{ 
       f^{\rm 2PS}(\epsilon)
       }{2 \cdot N_c^2 \cdot 2^2 {\cdot 2s}}
       N_c (2 u t - {2}m_H^4 )
      \left|{\cal M}^\star_{q\bar{q} \to HH}{\cal M}_{q\bar{q} \to HH}\right|
       \,,
\end{eqnarray}
where the denominators in the first factor are due to the two identical Higgs
bosons in the final state, the colour and the spin averages, and the flux factor.
The function
$f^{\rm 2PS}(\epsilon)$ can be found in Eq.~(\ref{eq::f2PS}) and $y$ is defined in
Eq.~(\ref{eq::t2y}).
The integrand is expanded in $\delta$, and we then perform the integration over $y$
for each expansion term separately. Our
result for the total cross section up to terms of order $1/m_t^6$ ($\rho^3$) is given by
\begin{eqnarray}
	\sigma_{q\bar{q} \to HH} &=& \frac{a_s^4 G_F^2 m_H^2}{\pi} \Bigg\{
  		\delta^{5/2} \rho^2 \Bigg(
  			   \frac{22201}{2952450000}
  			- \frac{1639 \log 2}{73811250}
  			+ \frac{121 \log^2 2}{7381125}
  			- \frac{1639 \log\rho}{147622500}
\nonumber\\&&  			
  			{}+ \frac{121 \log 2 \log\rho}{7381125}
  			+ \frac{121 \log^2\rho}{29524500}
  			+ \frac{121 \pi^2}{29524500}
  		\Bigg)
  		+ \delta^{5/2} \rho^3 \Bigg(
  			- \frac{588401}{124002900000}
\nonumber\\&&
  			{}+ \frac{5891 \log 2}{6200145000}
  			+ \frac{22 \log^2 2}{2460375}
  			+ \frac{5891 \log\rho}{12400290000}
  			+ \frac{22 \log 2 \log\rho}{2460375}
  			+ \frac{11 \log^2\rho}{4920750}
\nonumber\\&&
  			{}+ \frac{11 \pi^2}{4920750}
  		\Bigg)
  	+{\cal O}\left(\delta^{7/2}\right) + {\cal O}\left(\rho^4\right)
	\Bigg\}\,.
\label{eq::sig_qqHH}
\end{eqnarray}
The ancillary files of this paper~\cite{progdata} contain the
$\sigma_{q\bar{q} \to HH}$ expansion to $\rho^4$ and $\delta^{30}$.

An alternative approach to obtain the cross section $\sigma_{q\bar{q} \to HH}$
is based on the computation of the imaginary part of the forward scattering
amplitude, which has been applied to the real corrections described in the
main part of this paper. From the computational point of view this is much
more demanding since we have to consider five-loop amplitudes which factorize
into one- and two-loop vacuum contributions and one-loop form factor
integrals. We have cross-checked Eq.~(\ref{eq::sig_qqHH}) up to order $1/m_t^6$
($\rho^3$) using this approach.

\begin{figure}
  \begin{center}
     \includegraphics[width=0.9\textwidth]{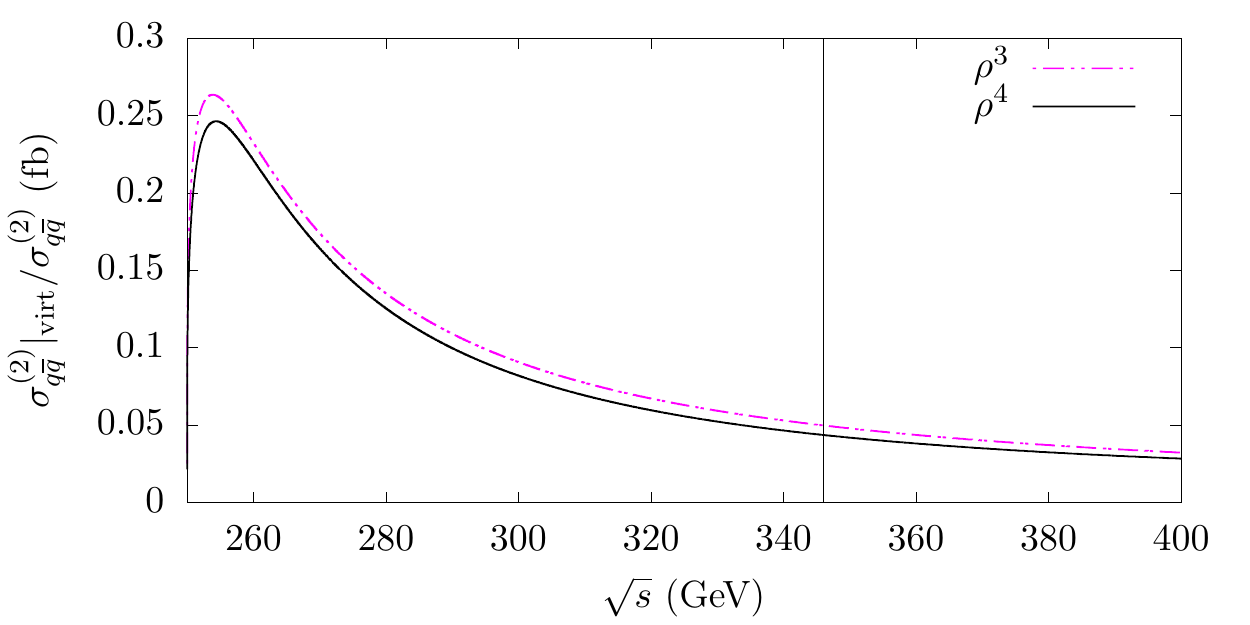}
     \caption{\label{fig::sigqQHHv}The ratio between the $q\bar{q}\to HH$ virtual
     corrections and the full $q\bar{q}$ cross section, as plotted in
     Fig.~\ref{fig::sig2}(c).}
  \end{center}
\end{figure}  

In Fig.~\ref{fig::sigqQHHv} we show the size of $\sigma_{q\bar{q} \to HH}$ compared
to $\sigma_{q\bar{q}}$. The virtual corrections form a significant part of the
total cross section, particularly near the production threshold. Therefore they
should be included for a proper NNLO description of the $q\bar{q}$ channel.



\section{\label{app::nh4}Virtual NNLO corrections proportional to $n_h^4$}

In this appendix we provide analytic results for the 
NNLO corrections which involve (in the forward-scattering kinematics)
four closed top quark loops. Such $n_h^4$ terms are only
present in the virtual corrections, see
Fig.~\ref{fig::diag_virt}(h).
The leading terms in the $\rho$ and $\delta$ expansion are given by
\begin{align}
  \sigma_{gg}^{(2),n_h^4} &= \frac{a_s^4 G_F^2
                            m_H^2}{\pi}\Bigg\{\frac{\sqrt{\delta}}{20736} +
                            \frac{\delta^{3/2}}{20736} +
                            \rho\Bigg(-\frac{\sqrt{\delta}}{155520} - \frac{{
                            13}\delta^{3/2}}{311040}\Bigg) + \mathcal{O}(\delta^{5/2}) + \mathcal{O}(\rho^2)\Bigg\}\,,
                             \label{eq::ggnh4}
\end{align}
and all terms up to $\rho^4$ and $\delta^{30}$ can be found
in the ancillary files of this paper~\cite{progdata}.



\section{\label{app::Q3_Q4}Explicit expressions for $\mathcal{Q}_i^{(3)}$ and $\mathcal{Q}_i^{(4)}$}

In Tabs.~\ref{tab::Q3} and~\ref{tab::Q4} we provide explicit expressions for $Q_i^{(3)}$ and
$Q_i^{(4)}$ introduced in Section~\ref{sec::boundaries}, where the
denominators as given by
\begin{align}
D_{1}&=-(-p_5+q_1+q_2)^2\approx-s\,,\nonumber\\
D_{2}&=-(p_5-q_1)^2\approx\frac{\kappa_5 (1-\cos\theta_5) s}{2}\,,\nonumber\\
D_{3}&=m_H^2-(p_5+p_6-q_1-q_2)^2\approx-\frac{3 s}{4}\,,\nonumber\\
D_{4}&=-(p_4+p_5+p_6-q_1)^2\approx\frac{s}{4}\,,\nonumber\\
D_{5}&=-(p_5+p_6)^2\approx-m_{gg}^2\,,\nonumber\\
D_{6}&=-(p_6-q_1)^2\approx\frac{\kappa_6 (1-\cos\theta_6) s}{2}\,,\nonumber\\
D_{7}&=-(p_5+p_6-q_1)^2\approx\frac{\kappa_{56} (1-\cos\theta_{56}) s}{2}\,,\nonumber\\
D_{8}&=-(p_5-q_2)^2\approx\frac{\kappa_5 (1+\cos\theta_5) s}{2}\,,\nonumber\\
D_{9}&=-(p_6-q_2)^2\approx\frac{\kappa_6 (1+\cos\theta_6) s}{2}\,,\nonumber\\
D_{10}&=-(p_5+p_6-q_2)^2\approx\frac{\kappa_{56} (1+\cos\theta_{56}) s}{2}\,,\nonumber\\
D_{11}&=-(-p_6+q_1+q_2)^2\approx-s
                               \,.
                               \label{eq::den}
\end{align}
After the symbol ``$\approx$'' the leading-order term in $\delta$ is given.

\begin{table}[ht]
\centering
{\scalefont{0.8}
\begin{tabular}{|r|c|r|c|r|c|r|c|r|c|}
\hline
$i$&$Q^{(3)}_i$&$i$&$Q^{(3)}_i$&$i$&$Q^{(3)}_i$&$i$&$Q^{(3)}_i$&$i$&$Q^{(3)}_i$\\\hline\hline
1&$L_{1}$&
2&$ (D_{1}+m_H^2)L_{1}$&
3&$L_{2}$&
4&$ (D_{1}+m_H^2)L_{2}$&
5&$L_{3}$\\
6&$ (D_{1}+m_H^2)L_{3}$&
7&$L_{4}$&
8&$\frac{1}{D_{2}}L_{5}$&
9&$\frac{ (D_{1}+m_H^2)}{D_{2}}L_{5}$&
10&$L_{6}$\\
11&$ (D_{1}+m_H^2)L_{6}$&
12&$\frac{1}{D_{1}+m_H^2}L_{1}$&
13&$\frac{1}{D_{1}+m_H^2}L_{2}$&
14&$\frac{1}{D_{1}+m_H^2}L_{3}$&
15&$\frac{1}{D_{1}+m_H^2}L_{4}$\\
16&$\frac{1}{D_{2} (D_{1}+m_H^2)}L_{5}$&
17&$\frac{1}{D_{1}+m_H^2}L_{6}$&
&&&&&\\
\hline
\end{tabular}
}
\caption{\label{tab::Q3}$Q_i^{(3)}$ expressed in terms of the denominator
  factors from Eq.~(\ref{eq::den}) and the integrals $L_i$ from Eqs.~(\ref{eq::L123}) and (\ref{eq::L456}).}
\end{table}

\begin{table}[!h]
\centering
{\scalefont{0.8}
\begin{tabular}{|r|c|r|c|r|c|r|c|r|c|}
\hline
$i$&$Q^{(4)}_i$&$i$&$Q^{(4)}_i$&$i$&$Q^{(4)}_i$&$i$&$Q^{(4)}_i$&$i$&$Q^{(4)}_i$\\\hline\hline
1&$1$&
2&$D_{3}$&
3&$D_{4}$&
4&$\frac{1}{D_{3}}$&
5&$\frac{D_{4}}{D_{3}}$\\
6&$\frac{1}{D_{1} D_{7}}$&
7&$\frac{D_{3}}{D_{1} D_{7}}$&
8&$\frac{D_{8}}{D_{1} D_{7}}$&
9&$\frac{1}{D_{1} D_{3} D_{7}}$&
10&$\frac{D_{8}}{D_{1} D_{3} D_{7}}$\\
11&$\frac{1}{D_{1} D_{5} D_{7} D_{8}}$&
12&$\frac{D_{3}}{D_{1} D_{5} D_{7} D_{8}}$&
13&$\frac{1}{D_{1} D_{3} D_{5} D_{7} D_{8}}$&
14&$\frac{1}{D_{10} D_{7}}$&
15&$\frac{D_{3}}{D_{10} D_{7}}$\\
16&$\frac{1}{D_{10} D_{3} D_{7}}$&
17&$\frac{1}{D_{10} D_{2} D_{7} D_{8}}$&
18&$\frac{D_{3}}{D_{10} D_{2} D_{7} D_{8}}$&
19&$\frac{1}{D_{10} D_{2} D_{3} D_{7} D_{8}}$&
20&$\frac{1}{D_{2} D_{9}}$\\
21&$\frac{1}{D_{2} D_{3} D_{9}}$&
22&$\frac{1}{D_{2} D_{6} D_{8} D_{9}}$&
23&$\frac{D_{3}}{D_{2} D_{6} D_{8} D_{9}}$&
24&$\frac{1}{D_{2} D_{3} D_{6} D_{8} D_{9}}$&
25&$\frac{1}{D_{1} D_{11}}$\\
26&$\frac{D_{3}}{D_{1} D_{11}}$&
27&$\frac{1}{D_{11} D_{8}}$&
28&$\frac{1}{D_{1} D_{11} D_{3}}$&
29&$\frac{1}{D_{11} D_{3} D_{8}}$&
30&$\frac{1}{D_{1} D_{11} D_{8} D_{9}}$\\
31&$\frac{D_{3}}{D_{1} D_{11} D_{8} D_{9}}$&
32&$\frac{1}{D_{1} D_{11} D_{3} D_{8} D_{9}}$&
33&$\frac{1}{D_{1} D_{11} D_{6} D_{8}}$&
34&$\frac{1}{D_{1} D_{11} D_{3} D_{6} D_{8}}$&
35&$\frac{1}{D_{1} D_{6}}$\\
36&$\frac{1}{D_{1} D_{11}}$&
37&$\frac{D_{3}}{D_{1} D_{11}}$&
38&$\frac{1}{D_{1} D_{3} D_{6}}$&
39&$\frac{1}{D_{1} D_{11} D_{3}}$&
40&$\frac{1}{D_{1} D_{11} D_{2} D_{6}}$\\
41&$\frac{D_{3}}{D_{1} D_{11} D_{2} D_{6}}$&
42&$\frac{1}{D_{1} D_{11} D_{2} D_{3} D_{6}}$&
43&$\frac{1}{D_{1} D_{11} D_{6} D_{8}}$&
44&$\frac{1}{D_{1} D_{11} D_{3} D_{6} D_{8}}$&
45&$\frac{1}{D_{11} D_{8}}$\\
46&$\frac{1}{D_{10} D_{11} D_{8}}$&
47&$\frac{D_{3}}{D_{10} D_{11} D_{8}}$&
48&$\frac{1}{D_{11} D_{3} D_{8}}$&
49&$\frac{1}{D_{10} D_{11} D_{3} D_{8}}$&
50&$\frac{1}{D_{10} D_{11} D_{6} D_{8}}$\\
51&$\frac{1}{D_{10} D_{11} D_{3} D_{6} D_{8}}$&
52&$\frac{1}{D_{10} D_{6} D_{7} D_{8}}$&
53&$\frac{D_{3}}{D_{10} D_{6} D_{7} D_{8}}$&
54&$\frac{1}{D_{10} D_{3} D_{6} D_{7} D_{8}}$&
55&$\frac{1}{D_{5} D_{6} D_{8}}$\\
56&$\frac{D_{3}}{D_{5} D_{6} D_{8}}$&
57&$\frac{1}{D_{3} D_{5} D_{6} D_{8}}$&
&&&&&\\
\hline\end{tabular}
}
\caption{\label{tab::Q4}$Q_i^{(4)}$ expressed in terms of the denominator
  factors from Eq.~(\ref{eq::den}).}
\end{table}


\end{appendix}


\clearpage




\begin{thebibliography}{99}


%
%

\bibitem{ATLAS:2012yve}
G.~Aad \textit{et al.} [ATLAS],
Phys. Lett. B \textbf{716} (2012), 1-29
%
[arXiv:1207.7214 [hep-ex]].

\bibitem{CMS:2012qbp}
S.~Chatrchyan \textit{et al.} [CMS],
Phys. Lett. B \textbf{716} (2012), 30-61
%
[arXiv:1207.7235 [hep-ex]].

\bibitem{ParticleDataGroup:2020ssz}
P.~A.~Zyla \textit{et al.} [Particle Data Group],
PTEP \textbf{2020} (2020) no.8, 083C01.
%

\bibitem{ATLAS:2020jgy}
G.~Aad \textit{et al.} [ATLAS],
JHEP \textbf{07} (2020), 108
[erratum: JHEP \textbf{01} (2021), 145; erratum: JHEP \textbf{05} (2021), 207]
%
[arXiv:2001.05178 [hep-ex]].

\bibitem{CMS:2020tkr}
A.~M.~Sirunyan \textit{et al.} [CMS],
JHEP \textbf{03} (2021), 257
%
[arXiv:2011.12373 [hep-ex]].

\bibitem{Glover:1987nx}
  E.~W.~N.~Glover and J.~J.~van der Bij,
  Nucl.\ Phys.\ B {\bf 309} (1988) 282.

\bibitem{Plehn:1996wb}
  T.~Plehn, M.~Spira and P.~M.~Zerwas,
  Nucl.\ Phys.\ B {\bf 479} (1996) 46
   Erratum: [Nucl.\ Phys.\ B {\bf 531} (1998) 655]
  [hep-ph/9603205].
%

\bibitem{Dawson:1998py}
  S.~Dawson, S.~Dittmaier and M.~Spira,
  Phys.\ Rev.\ D {\bf 58} (1998) 115012
  [hep-ph/9805244].

\bibitem{Grigo:2013rya}
  J.~Grigo, J.~Hoff, K.~Melnikov and M.~Steinhauser,
  Nucl.\ Phys.\ B {\bf 875} (2013) 1
  [arXiv:1305.7340 [hep-ph]].
%

\bibitem{Degrassi:2016vss}
G.~Degrassi, P.~P.~Giardino and R.~Gr\"ober,
Eur.\ Phys.\ J.\ C {\bf 76} (2016) no.7,  411
%
[arXiv:1603.00385 [hep-ph]].

\bibitem{Davies:2018ood}
  J.~Davies, G.~Mishima, M.~Steinhauser and D.~Wellmann,
  JHEP {\bf 1803} (2018) 048
  [arXiv:1801.09696 [hep-ph]].

\bibitem{Davies:2018qvx}
  J.~Davies, G.~Mishima, M.~Steinhauser and D.~Wellmann,
  JHEP {\bf 1901} (2019) 176
  [arXiv:1811.05489 [hep-ph]].
%

\bibitem{Bonciani:2018omm}
  R.~Bonciani, G.~Degrassi, P.~P.~Giardino and R.~Gr\"ober,
  Phys.\ Rev.\ Lett.\  {\bf 121} (2018) no.16,  162003
  [arXiv:1806.11564 [hep-ph]].
%

\bibitem{Grober:2017uho}
  R.~Gr\"ober, A.~Maier and T.~Rauh,
  JHEP {\bf 1803} (2018) 020
  [arXiv:1709.07799 [hep-ph]].
%

\bibitem{Maltoni:2014eza}
  F.~Maltoni, E.~Vryonidou and M.~Zaro,
      JHEP {\bf 1411} (2014) 079
%
          [arXiv:1408.6542 [hep-ph]].

\bibitem{Borowka:2016ehy}
  S.~Borowka, N.~Greiner, G.~Heinrich, S.~P.~Jones, M.~Kerner, J.~Schlenk, U.~Schubert and T.~Zirke,
  Phys.\ Rev.\ Lett.\  {\bf 117} (2016) no.1,  012001
   Erratum: [Phys.\ Rev.\ Lett.\  {\bf 117} (2016) no.7,  079901]
  [arXiv:1604.06447 [hep-ph]].

\bibitem{Borowka:2016ypz}
  S.~Borowka, N.~Greiner, G.~Heinrich, S.~P.~Jones, M.~Kerner, J.~Schlenk and T.~Zirke,
  JHEP {\bf 1610} (2016) 107
  [arXiv:1608.04798 [hep-ph]].

\bibitem{Baglio:2018lrj}
J.~Baglio, F.~Campanario, S.~Glaus, M.~M\"uhlleitner, M.~Spira and
J.~Streicher,
Eur. Phys. J. C \textbf{79} (2019) no.6, 459
%
[arXiv:1811.05692 [hep-ph]].
%

\bibitem{Davies:2019dfy}
J.~Davies, G.~Heinrich, S.~P.~Jones, M.~Kerner, G.~Mishima, M.~Steinhauser and
D.~Wellmann,
JHEP \textbf{11} (2019), 024
[arXiv:1907.06408 [hep-ph]].

\bibitem{Baglio:2020wgt}
J.~Baglio, F.~Campanario, S.~Glaus, M.~M\"uhlleitner, J.~Ronca and M.~Spira,
Phys. Rev. D \textbf{103} (2021) no.5, 056002
[arXiv:2008.11626 [hep-ph]].

\bibitem{deFlorian:2013jea}
  D.~de Florian and J.~Mazzitelli,
  Phys.\ Rev.\ Lett.\  {\bf 111} (2013) 201801
  [arXiv:1309.6594 [hep-ph]].

\bibitem{deFlorian:2013uza}
  D.~de Florian and J.~Mazzitelli,
  Phys.\ Lett.\ B {\bf 724} (2013) 306
  [arXiv:1305.5206 [hep-ph]].

\bibitem{Grigo:2014jma}
  J.~Grigo, K.~Melnikov and M.~Steinhauser,
  Nucl.\ Phys.\ B {\bf 888} (2014) 17
  [arXiv:1408.2422 [hep-ph]].

\bibitem{Grigo:2015dia}
  J.~Grigo, J.~Hoff and M.~Steinhauser,
  Nucl.\ Phys.\ B {\bf 900} (2015) 412
  [arXiv:1508.00909 [hep-ph]].
%

\bibitem{Davies:2019xzc}
J.~Davies, F.~Herren, G.~Mishima and M.~Steinhauser,
JHEP \textbf{05} (2019), 157
%
[arXiv:1904.11998 [hep-ph]].

\bibitem{Grazzini:2018bsd}
  M.~Grazzini, G.~Heinrich, S.~Jones, S.~Kallweit, M.~Kerner, J.~M.~Lindert
  and J.~Mazzitelli,
  JHEP {\bf 1805} (2018) 059
%
  [arXiv:1803.02463 [hep-ph]].
%

\bibitem{Chen:2019lzz}
L.~B.~Chen, H.~T.~Li, H.~S.~Shao and J.~Wang,
Phys. Lett. B \textbf{803} (2020), 135292
%
[arXiv:1909.06808 [hep-ph]].

\bibitem{Chen:2019fhs}
L.~B.~Chen, H.~T.~Li, H.~S.~Shao and J.~Wang,
JHEP \textbf{03} (2020), 072
%
[arXiv:1912.13001 [hep-ph]].

\bibitem{Anastasiou:2016cez}
C.~Anastasiou, C.~Duhr, F.~Dulat, E.~Furlan, T.~Gehrmann, F.~Herzog,
A.~Lazopoulos and B.~Mistlberger,
JHEP \textbf{05} (2016), 058
%
[arXiv:1602.00695 [hep-ph]].

\bibitem{Mistlberger:2018etf}
B.~Mistlberger,
JHEP \textbf{05} (2018), 028
%
[arXiv:1802.00833 [hep-ph]].

\bibitem{Spira:2016zna}
  M.~Spira,
  JHEP {\bf 1610} (2016) 026
  [arXiv:1607.05548 [hep-ph]].

\bibitem{Gerlach:2018hen}
  M.~Gerlach, F.~Herren and M.~Steinhauser,
  JHEP {\bf 1811} (2018) 141
  [arXiv:1809.06787 [hep-ph]].

\bibitem{Banerjee:2018lfq}
  P.~Banerjee, S.~Borowka, P.~K.~Dhani, T.~Gehrmann and V.~Ravindran,
  JHEP {\bf 1811} (2018) 130
  [arXiv:1809.05388 [hep-ph]].

\bibitem{Nogueira:1991ex}
  P.~Nogueira,
  J.\ Comput.\ Phys.\  {\bf 105} (1993) 279.

\bibitem{Pak_gen}
A.~Pak, unpublished.

\bibitem{Harlander:1997zb}
  R.~Harlander, T.~Seidensticker and M.~Steinhauser,
  Phys.\ Lett.\ B {\bf 426} (1998) 125
  [hep-ph/9712228].

\bibitem{Seidensticker:1999bb}
  T.~Seidensticker,
  hep-ph/9905298.

\bibitem{Ruijl:2017dtg}
  B.~Ruijl, T.~Ueda and J.~Vermaseren,
  arXiv:1707.06453 [hep-ph].

\bibitem{Steinhauser:2000ry}
  M.~Steinhauser,
  Comput.\ Phys.\ Commun.\  {\bf 134} (2001) 335
  [arXiv:hep-ph/0009029].

\bibitem{Davies:2019esq}
J.~Davies, F.~Herren, G.~Mishima and M.~Steinhauser,
PoS \textbf{RADCOR2019} (2019), 022
[arXiv:1912.01646 [hep-ph]].

\bibitem{Herren:2020ccq}
F.~Herren,
``Precision Calculations for Higgs Boson Physics at the LHC - Four-Loop
Corrections to Gluon-Fusion Processes and Higgs Boson Pair-Production at
NNLO,'' PhD thesis, 2020, KIT.

\bibitem{Pak:2011xt}
A.~Pak,
J. Phys. Conf. Ser. \textbf{368} (2012), 012049
%
[arXiv:1111.0868 [hep-ph]].

\bibitem{Lee:2012cn}
  R.~N.~Lee,
  arXiv:1212.2685 [hep-ph].

\bibitem{Lee:2013mka}
  R.~N.~Lee,
  J.\ Phys.\ Conf.\ Ser.\  {\bf 523} (2014) 012059
  [arXiv:1310.1145 [hep-ph]].

\bibitem{Altarelli:1977zs}
G.~Altarelli and G.~Parisi,
Nucl. Phys. B \textbf{126} (1977), 298-318

\bibitem{Curci:1980uw}
G.~Curci, W.~Furmanski and R.~Petronzio,
Nucl. Phys. B \textbf{175} (1980), 27-92

\bibitem{Moch:2004pa}
S.~Moch, J.~A.~M.~Vermaseren and A.~Vogt,
Nucl. Phys. B \textbf{688} (2004), 101-134
[arXiv:hep-ph/0403192 [hep-ph]].

\bibitem{Vogt:2004mw}
A.~Vogt, S.~Moch and J.~A.~M.~Vermaseren,
Nucl. Phys. B \textbf{691} (2004), 129-181
[arXiv:hep-ph/0404111 [hep-ph]].

\bibitem{Anastasiou:2002yz}
C.~Anastasiou and K.~Melnikov,
Nucl. Phys. B \textbf{646} (2002), 220-256
%
[arXiv:hep-ph/0207004 [hep-ph]].

\bibitem{Hoeschele:2013gga}
M.~H\"oschele, J.~Hoff, A.~Pak, M.~Steinhauser and T.~Ueda,
Comput. Phys. Commun. \textbf{185} (2014), 528-539
%
[arXiv:1307.6925 [hep-ph]].

\bibitem{Davies:2019djw}
J.~Davies and M.~Steinhauser,
JHEP \textbf{10} (2019), 166
%
[arXiv:1909.01361 [hep-ph]].

\bibitem{Kotikov:1990kg}
A.~V.~Kotikov,
Phys. Lett. B \textbf{254} (1991), 158-164
%

\bibitem{Gehrmann:1999as}
T.~Gehrmann and E.~Remiddi,
Nucl. Phys. B \textbf{580} (2000), 485-518
[arXiv:hep-ph/9912329 [hep-ph]].

\bibitem{Henn:2013pwa}
J.~M.~Henn,
Phys. Rev. Lett. \textbf{110} (2013), 251601
%
[arXiv:1304.1806 [hep-th]].

\bibitem{Goncharov:1998kja}
A.~B.~Goncharov,
Math. Res. Lett. \textbf{5} (1998), 497-516
%
[arXiv:1105.2076 [math.AG]].

\bibitem{Lee:2020zfb}
R.~N.~Lee,
%
[arXiv:2012.00279 [hep-ph]].

\bibitem{Lee:2014ioa}
R.~N.~Lee,
JHEP \textbf{04} (2015), 108
%
[arXiv:1411.0911 [hep-ph]].

\bibitem{Beneke:1997zp}
M.~Beneke and V.~A.~Smirnov,
Nucl. Phys. B \textbf{522} (1998), 321-344
%
[arXiv:hep-ph/9711391 [hep-ph]].

\bibitem{Pak:2010pt}
A.~Pak and A.~Smirnov,
Eur. Phys. J. C \textbf{71} (2011), 1626
%
[arXiv:1011.4863 [hep-ph]].

\bibitem{Jantzen:2012mw}
B.~Jantzen, A.~V.~Smirnov and V.~A.~Smirnov,
Eur. Phys. J. C \textbf{72} (2012), 2139
%
[arXiv:1206.0546 [hep-ph]].

\bibitem{Smirnov:2006ry}
V.~A.~Smirnov,
``Feynman integral calculus,'' Springer (2006) 

\bibitem{progdata}
\verb|https://www.ttp.kit.edu/preprints/2021/ttp21-034/|.

\bibitem{Melnikov:2016qoc}
K.~Melnikov, L.~Tancredi and C.~Wever,
JHEP \textbf{11} (2016), 104
%
[arXiv:1610.03747 [hep-ph]].

\bibitem{Mishima:2018olh}
G.~Mishima,
JHEP \textbf{02} (2019), 080
%
[arXiv:1812.04373 [hep-ph]].

\bibitem{Davies:2019nhm}
J.~Davies, R.~Gr\"ober, A.~Maier, T.~Rauh and M.~Steinhauser,
Phys. Rev. D \textbf{100} (2019) no.3, 034017
[erratum: Phys. Rev. D \textbf{102} (2020) no.5, 059901]
%
[arXiv:1906.00982 [hep-ph]].

\bibitem{Davies:2019roy}
J.~Davies, R.~Gr\"ober, A.~Maier, T.~Rauh and M.~Steinhauser,
PoS \textbf{RADCOR2019} (2019), 079
%
[arXiv:1912.04097 [hep-ph]].

\bibitem{Hoschele:2013pvt}
M.~H\"oschele, J.~Hoff, A.~Pak, M.~Steinhauser and T.~Ueda,
Comput. Phys. Commun. \textbf{185} (2014), 528-539
[arXiv:1307.6925 [hep-ph]].


\end{thebibliography}
\end{document}